\documentclass[12pt,3p]{llncs}

\usepackage[sectionbib]{natbib}

\usepackage{color}
\usepackage{eufrak}
\usepackage{bm}
\usepackage{comment}
\usepackage{amsmath}
\usepackage{mathrsfs,amsmath} 
\usepackage{amsfonts}
\usepackage{graphicx} 
\usepackage{epstopdf}
\usepackage{enumitem}
\usepackage{xcolor,colortbl}
\usepackage{subfigure}
\usepackage{setspace}
\usepackage{bm}
\definecolor{Gray}{gray}{0.85}
\usepackage[top=25mm, bottom=25mm, left=20mm, right=20mm]{geometry}
\allowdisplaybreaks

\usepackage{amssymb}
\usepackage{xcolor,colortbl}
  \usepackage{colortbl}
  \definecolor{Lightgray}{RGB}{235,235,235}
\definecolor{Gray2}{gray}{0.9}
\definecolor{Gray}{gray}{0.85}
\definecolor{Gray3}{gray}{0.8}
\usepackage{enumitem}
\usepackage{amsbsy}
 \usepackage{soul}
\usepackage{float}
\usepackage{enumitem}
\usepackage{mathrsfs}
\usepackage{amsmath}
\usepackage{amssymb}
\usepackage{graphicx}
\usepackage{mdwlist}
\usepackage{color}
\usepackage{soul}
\usepackage{url}
\usepackage[]{framed}
  \usepackage[percent]{overpic}
  
\usepackage{caption}

\usepackage{graphicx}
\usepackage{amsmath}
\usepackage{amsbsy}
\usepackage{amssymb}
\usepackage{amscd}
\usepackage{amsfonts}
\usepackage{graphics}
\usepackage{graphicx}
\usepackage{subfigure}
\usepackage{epsfig}
\usepackage{float}
\usepackage{times}
\usepackage{epstopdf}
\usepackage{fancyhdr}
\usepackage{fancyheadings}
\usepackage{rotating}
\usepackage{pstricks}
\usepackage{color}
\usepackage{soul}
\begin{document}

\title{Characterization of hybrid piezoelectric nanogenerators through dynamic asymptotic homogenization }

\author{ Maria Laura De Bellis\inst{1}, Andrea Bacigalupo\inst{2}, Giorgio Zavarise\inst{3}}
\institute{University of Chieti-Pescara, Department INGEO, Viale Pindaro 42, Pescara, Italy  \and IMT School for Advanced Studies, Piazza S. Francesco 19, 55100 Lucca, Italy \and Polytechnic University of Turin, Department DISEG, Corso Duca degli Abruzzi 24, Torino 10129, Italy }

\maketitle

\begin{abstract}
In the framework of energy scavenging for applications in flexible/strechable electronics,
hybrid piezoelectric nanogenerators, made up with Zinc oxyde nanorods, embedded in a polymeric matrix, and growth on a flexible polymeric support, are investigated. The ZnO nanorods are arranged in clusters, forming nearly regular distributions, so that periodic topologies can be realistically assumed. Focus is on a dynamic multi-field asymptotic homogenization approach, proposed to grasp the overall constitutive behaviour of such complex microstrutcures.
A set of applications, both in static and dynamic regime, is proposed to explore different design paradigms, related to nanogenerators based on three working principles. Both extension and bending nanogenerators are, indeed, analysed, considering either extension along the nanorods axis, or orthogonally to it.  The study of the wave propagation is, also, exploited to comprehend the main features of such piezoelectric devices in the dynamic regime.
\end{abstract}


\section{Introduction}
\noindent In the last decades, energy harvesting is increasingly becoming a topic of great interest in different engineering fields. The key idea is that the energy, naturally available in the environment (for instance in the form of heat or kinetic energy) is captured and converted into electrical energy, used to power small devices, such as wireless sensors and micro electronics, no longer requiring electro-chemical batteries.
Among others, piezoelectric generators, exploiting their intrinsic electro-mechanical coupling, are competitive solutions in energy harvesting, as witnessed by the growing number of applications, ranging from structural monitoring, to automoviles, up to Internet of Things.\\
More recently particular attention has been drawn 
to emerging applications, such as biomedical monitoring, wearable technology, pervasive computing, micro and nano robotics, tire condition monitoring and extreme technology. To this aim,
cutting-edge research has been devoted to design  piezoelectric devices characterized by 
smaller and smaller size and high performances.
Starting from the pioneering work by \cite{wang2006piezoelectric}, piezoelectric nanogenerators 
have been successfully proposed, based on different electro-active materials and working principles \citep{BRISCOE201515,MCCARTHY2016355,
jin2016self,zhang2016wearable,Ahmed2017,LI2017,saadatnia2017modeling,wang2017toward,
liu2018shape,Askari2019}. 
Relevant examples concern either the use of 
piezoelectric zinc oxide (ZnO) nanowire arrays grown on  conductive rigid supports \citep{yi2005zno,wang2006piezoelectric}, or the adoption of  lead zirconate
titanate (PZT), polyvinylidene fluoride (PVDF) and barium titanate (BT).\\
An important improvement in the design of ZnO nanorods-based 
piezoelectric generators has been achieved by adopting flexible substrates made of electro-active polymeric materials. The main advantage is, indeed, the possibility of exploiting relevant flexural mechanisms, besides the standard direct compression of the device. In 
\cite{FAN2016} different typologies of flexible nanogenerators are presented and critically commented. 
Also patterned growth of ZnO nanowires can be exploited to enhance the performances of flexible nanogenerators, as discussed in \cite{Dechao2017}.\\
Further benefits can be obtained by resorting to so-called hybrid nanogenerators, made up by embedding the ZnO nanorods within a polymeric matrix. More specifically, \cite{STASSI2015} propose highly oriented ZnO nanotubes in a porous polycarbonate (PC) matrix. The result is an efficient nanogenerator based on such a highly flexible ZnO–PC composite.  Moreover, in \cite{CHOI2017462}
a hybrid piezoelectric
structure made of ZnO nanowires and a matrix of PVDF polymer is investigated in order to obtain a power enhancement. The authors, indeed, find that 
the ZnO nanowires are able to deliver internal strain to the PVDF, which increase the electrical power output of the hybrid nanogenerator.\\
Based on the aforementioned considerations,
with the aim of energy harvesting from green and sustainable energy
resources, our focus is on hybrid flexible nanogenerators, made up with clusters of ZnO nanorods embedded into a polymeric matrix and growth on a flexible support. The choice of ZnO nanorods, is motivated by their relatively simple forming processes using low temperature. In particular, 
by exploiting innovative growth techniques, it is possible to synthesize ZnO nanorods clamped on top of a flexible support, which is typically realized by sandwiching a polymer base and a layer of   conductive fluorine-doped tin oxide (FTO). Concerning the zinc oxyde, the most stable crystalline structure is the Wurtzite (Wz), characterized by an hexagonal structure and a stacking sequence such that the nanorod axis 
coincides with the [0001] crystalline direction (see \cite{Wang_2004} for details).  It follows that, the resulting nanorods have hexagonal cross section and principal axis nearly orthogonal to the base support. The overall material is, thus, characterized by a pronounced uni-axial anisotropy and the spontaneous polarization direction coincides with the axis of the nanorods. 
The ZnO nanorods, having roughly constant section sizes and heights, arrange themselves in clusters characterized in general by nearly regular distributions. It is important to emphasize that, during the synthesis process of such  material, it is possible to tune both the density and the heights of the nanorods. The electro-active polymeric matrix is subsequently added, filling the gaps between the nanorods.
Moreover, different clusters can be stacked
together along the nanorods axis direction, in order to enhance their piezoelectric performances. The device is complemented by the presence of two  electrodes located at the opposite external top and bottom  faces.
Due to the nearly regular distribution 
of the nanorods, for the sake of simplicity, it is possible to consider periodic topologies, in which the nanorods are equally spaced. 
The resulting material is a microstructured piezoelectric composite, whose global response is strongly influenced by its microstructure, i.e. by the geometry and materials properties of each constituents and by their collaborative behaviour.\\
With a view to investigating such kind of materials,  a possibility is resorting to micromechanical approaches, in which the material is described in detail, but generally they result in cumbersome analyses.  In order to overcome these drawbacks, multiscale  techniques, based on homogenization approaches, are a very valuable tool to gather both a synthetic and thorough description of the complex material behaviour. The investigation of the overall static and dynamic behaviour of periodic elastic composite materials 
has been performed resorting either to asymptotic approaches \citep{Bakhvalov1984,GambinKroner1989,
 allaire1992homogenization,Boutin1996,
 fish2001higher,andrianov2008higher,Tran2012,
 Bacigalupo2014}, or to 
variational-asymptotic approaches \citep{Smyshlyaev2000,PeerlingsFleck2004,BACIGALUPO201216,
 BacigalupoGambarotta2014}, or also to identification techniques, among which computational approaches \citep{FOREST1998,
 KouznetsovaGeers2004,
kaczmarczyk2008scale,
 BacigalupoGambarotta2010,DeBellis-Addessi11,li2011micro,
 Addessi2013,lesivcar2014second,TROVALUSCI2015396,
 Addessi2016,biswas2017micromorphic,RECCIA201839,
 TROVALUSCI2017164} and analytical approaches \citep{Bigoni2007,muhlich2012estimation,Bacca2013a,Bacca2013b,
Bacigalupo2013,bacigalupo2017identification,hutter2017homogenization}. \\
Generalized homogenization approaches have been proposed to date to handle 
multi-field problems, ranging from thermo-elastic,
thermo-diffusive, to piezoelectric  and thermo-piezoelectric problems \citep{galka1996some,PETTERMANN20005447,Aboudi2001,
BERGER200553,Kanoute2009,ZhangZhang2007,
DeraemaekerNasser2010,ZAH2013487,SALVADORI2014114,
BacigalupoMorini2016,fantoni2017multi,
de2017auxetic,FANTONI2018319}. \\
In this context, we propose a dynamic multi-field 
asymptotic homogenization approach
for the analysis of hybrid piezoelectric nanogenerators
 with periodic microstructure. 
The key point of such approach is that the microscopic displacement and the electric potential fields are asymptotically expanded and plugged into the microscopic governing equations. It follows that a series of recursive differential problems are 
defined, in terms of the sensitivities of the microscopic fields.
Such problems give rise to down-scaling relations, hierarchical cell problems, in terms of perturbation functions, and so-called average field equations of infinite order. 
The overall constitutive tensors and the overall inertial terms
are, thus, rigorously derived from the generalized macro-homogeneity condition, properly relating the macroscopic Lagrangian and the microscopic mean Lagrangian, referred to a representative portion of the material either at the macroscopic or at the microscopic scale. 
By truncating the asymptotic expansion of the microscopic mean Lagrangian at different orders, either first order or higher order equivalent homogeneous continua can be identified. 
As an alternative, first and higher order approximations can be also obtained by solving, via perturbative approaches, the average field equations of infinite order.\\
The dispersive  wave propagation in the hybrid piezoelectric nanogenerators has been, then, investigated and the frequency band structure has been determined  consistently with the Floquet-Bloch theory. Moreover, the dispersion functions in equivalent homogeneous materials are derived, able to accurately approximate the acoustic branches of the Floquet-Bloch spectrum in the long wavelength regime.\\
Applications has been devoted both to the static and dynamic analysis of periodic hybrid piezoelectric nanogenerators. 
In the framework of a first order homogenization approach,
the equivalent constitutive properties are determined, as a function of the heights and the density of the nanorods, since such geometric parameters can be easily controlled, in the synthesis of the material, in order to tune the overall piezoelectric response. Considering a benchmark test of an extensional microstrutcured nanogenerator, the reliability of the proposed homogenization model has been proved by comparing the 
analytical homogenized solutions with the corresponding solutions of the heterogeneous model. \\
In addition, three piezoelectric  microstructured nanogenerators, based on different working principles, are investigated as a set of geometrical parameters changes. More specifically, microstructured extension nanogenerators, bending nanogenerators and transversal extension nanogenerators are taken into account and their static behaviour is critically discussed, in order to provide broad guidelines to maximize their efficiency.\\
Finally, a dynamic characterization of the piezoelectric periodic nanostructured material is performed for the purpose of analysing Bloch waves propagations. Focus is on detecting detect possible partial or total band gaps, strongly characterizing the dynamic response of the composite material. A good agreement has been found between the dispersion functions analytically obtained by the homogenized model, and the acoustic branches of the Floquet-Bloch spectrum of the heterogeneous material. \\

\section{Piezoelectric material modelled at two-scales}
\begin{figure}[hbtp]
\centering
\begin{overpic}[scale=0.8]{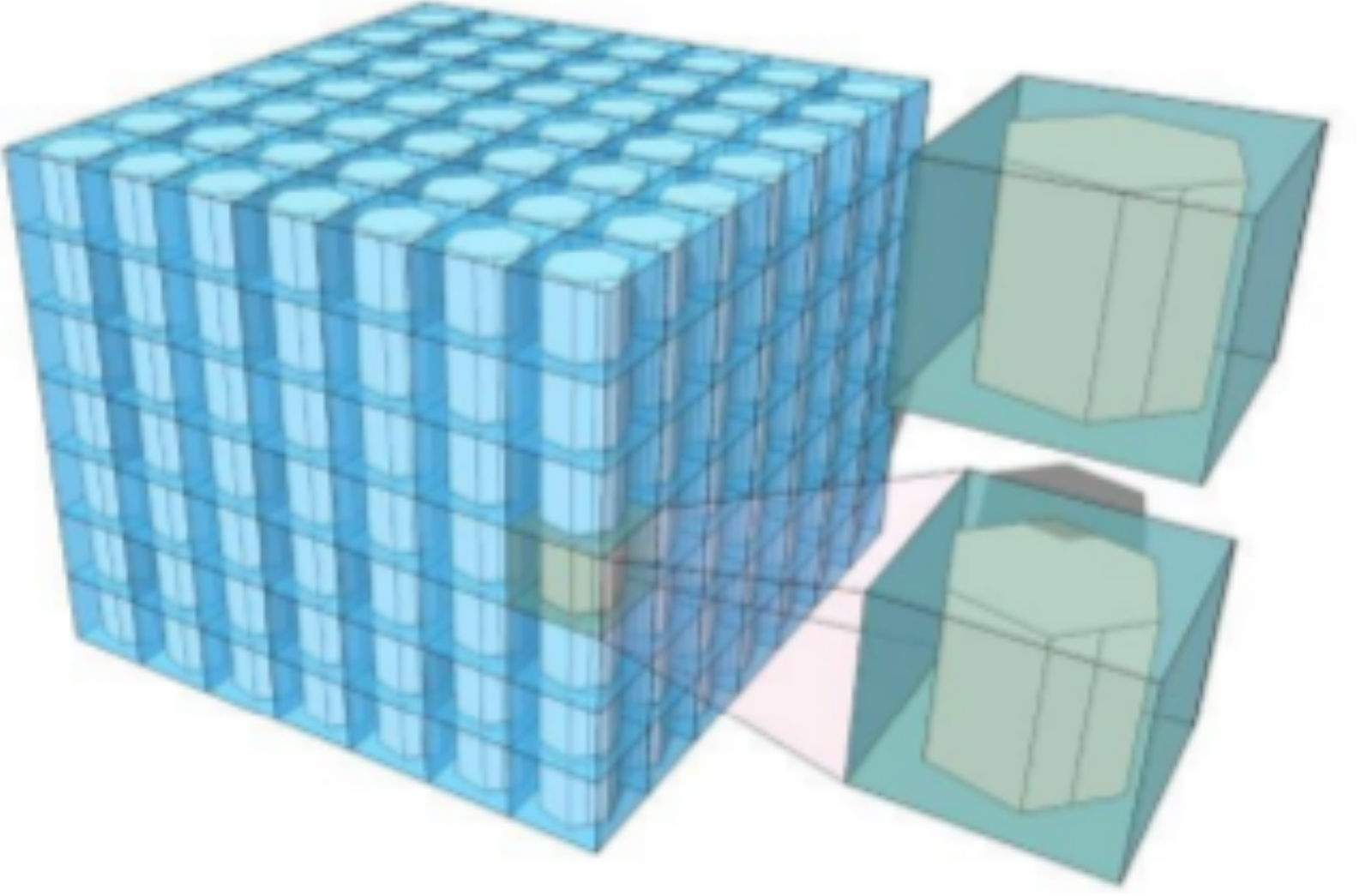}
 \put (20,8) {$L$}
  \put (-1,35) {$\eta L$}
  \put (14,62) {$L$}
 \put (68,1) {$d=\varepsilon$}
  \put (86,5) {$d=\varepsilon$}
    \put (90,16) {$h=\eta \varepsilon$}
     \put (85,60) {$1$}
  \put (94,33) {$1$}
    \put (97,45) {$\eta $}
 \put(2,50){\colorbox{white}{$\mathfrak{L}$}}
  \put(64,9){\colorbox{white}{$\mathfrak{U}$}}
   \put(68,50){\colorbox{white}{$\mathfrak{Q}$}}
\end{overpic}
\caption{Representative portion of the heterogeneous material: $\mathfrak{L}$ Cluster  of Periodic Cells, $\mathfrak{U}$ Periodic Cell, and $\mathfrak{Q}$ Unit Cell.}
\label{figurePrima}
\end{figure}
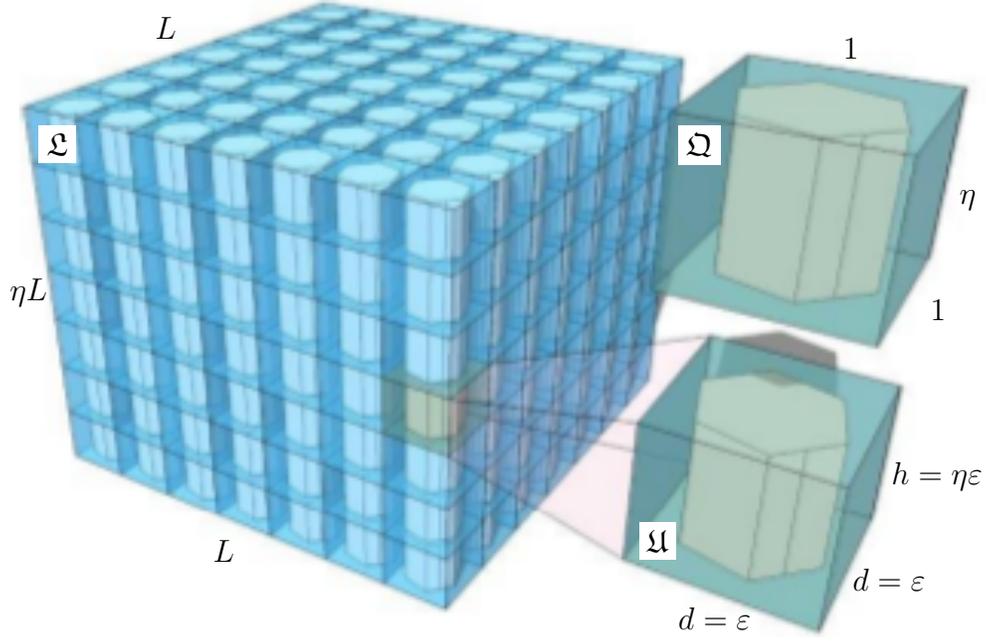

Let us consider a heterogeneous piezoelectric material, with periodic microstructure, which phases are described as a first order continuum in the framework of the linear theory of piezoelectricity. Each material point is identified by its position vector $\textbf{x} = x_1 \textbf{e}_1 + x_2 \textbf{e}_2+x_3 \textbf{e}_3$, referred to a coordinate system with origin at point $O$ and orthogonal base with the fixed set of basis vectors $\textbf{e}_1 ,\textbf{e}_2,\textbf{e}_3$, see Figure \ref{figurePrima}.
 Under the action of external sources, i.e. body forces $\textbf{b}(\textbf{x},t) = b_i(\textbf{x},t)\textbf{e}_i$ and free charge densities $\rho_e(\textbf{x},t)$, the microscopic displacement field $\textbf{u}(\textbf{x},t) = u_i(\textbf{x},t)\textbf{e}_i$ and the microscopic electric potential field $\phi(\textbf{x},t)$ are induced. The fields depends both on point $\textbf{x}$ and on time $t$.
By exploiting the periodicity of the medium, a Periodic Cell 
$\mathfrak{A}=[-d/2, d/2] \times [-d/2, d/2] \times [-h/2, h/2]$ is identified, denoted by  the three orthogonal periodicity vectors $\textbf{v}_1= d \textbf{e}_1= \varepsilon \textbf{e}_1$, $\textbf{v}_2= d \textbf{e}_2= \varepsilon \textbf{e}_2$ and $\textbf{v}_3= h \textbf{e}_3= \eta \varepsilon \textbf{e}_3$, being $\varepsilon$ the characteristic size of the
cell $\mathfrak{A}$, see Fig. 1(b). Consistently with standard asymptotic homogenization approaches, the Unit Cell
$\mathfrak{Q}=[-1/2, 1/2] \times [-1/2, 1/2] \times [-\eta/2, \eta/2]$ is obtained by applying the rescaling factor of $\varepsilon$ to the Periodic Cell  $\mathfrak{A}$. In agreement with \citep{Mindlin1974}, the partial differential equations governing the piezoelectric problem, in component form, are
\begin{align}
\begin{split}
&\frac{D}{Dx_j} \left(  C_{ijkl}^{(m,\varepsilon)} \frac{D u_k}{Dx_l} \right) +\frac{D}{Dx_j} \left(  e_{ijk}^{(m,\varepsilon)} \frac{D \phi}{Dx_k} \right)+b_i= \rho^{(m,\varepsilon)} \frac{D^2 u_i}{Dt^2}  ,\\
& \frac{D}{Dx_i} \left(  \widetilde{e}_{ikl}^{(m,\varepsilon)} \frac{D u_k}{Dx_l} \right)  - \frac{D}{Dx_i} \left(  \beta_{il}^{(m,\varepsilon)} \frac{D \phi}{Dx_l} \right)= \rho_e,
\end{split}\label{eq:1}
\end{align}
where $ C_{ijkl}^{(m,\varepsilon)}$ are the components of the fourth order micro elasticity tensor, $e_{ijk}^{(m,\varepsilon)}$ are the components of  the third order piezoelectric stress-charge coupling tensor, with the following relation between the components  $\widetilde{e}^{(m,\varepsilon)}_{ijk}={e}^{(m,\varepsilon)}_{jki}$, 
$\beta_{il}^{(m,\varepsilon)}$
are the components of the second order dielettric
permittivity tensor, and $ \rho^{(m,\varepsilon)}$ is the mass density. In Equations (\ref{eq:1}) the superscripts $m$ and $\varepsilon$ refer to microscopic fields, and to the characteristic size of the Periodic Cell, respectively. Note that the derivatives are intended in a generalized sense.\\
\noindent The constitutive tensors and the mass density are characterized by $\mathfrak{A}$-periodicity, i.e.
\begin{align}
C_{ijkl}^{(m,\varepsilon)}\left( {{\mathbf{x}} + {{\mathbf{v}}_{\alpha}}} \right) &= C_{ijkl}^{(m,\varepsilon)}\left( {\mathbf{x}} \right),\nonumber \\
e_{ijk}^{(m,\varepsilon)}\left( {{\mathbf{x}} + {{\mathbf{v}}_{\alpha}}} \right) &= e_{ijk}^{(m,\varepsilon)}\left( {\mathbf{x}} \right),\nonumber \\
\beta_{il}^{(m,\varepsilon)}\left( {{\mathbf{x}} + {{\mathbf{v}}_{\alpha}}} \right) &= \beta_{il}^{(m,\varepsilon)}\left( {\mathbf{x}} \right),\nonumber \\
\rho^{(m,\varepsilon)}\left( {{\mathbf{x}} + {{\mathbf{v}}_{\alpha}}} \right) &=\rho^{(m,\varepsilon)}\left( {\mathbf{x}} \right), \quad \alpha=1,2,3, \quad\forall \textbf{x} \in \mathfrak{A},\label{PerD}
\end{align}
therefore, they only depend on the variable 
$\boldsymbol{\xi}=\textbf{x}/\varepsilon$, so  that the $Q$-periodicity holds, as
\begin{align}
C_{ijkl}^{(m,\varepsilon)}\left( {{\mathbf{x}} } \right) &= C_{ijkl}^{(m,\varepsilon)}\left( \boldsymbol{\xi} =\frac{\textbf{x}}{\varepsilon} \right),\nonumber \\
e_{ijk}^{(m,\varepsilon)}\left({{\mathbf{x}} } \right) &= e_{ijk}^{(m,\varepsilon)}\left(\boldsymbol{\xi} =\frac{\textbf{x}}{\varepsilon} \right),\nonumber \\
\beta_{il}^{(m,\varepsilon)}\left( \mathbf{x}  \right) &= \beta_{il}^{(m,\varepsilon)}\left( \boldsymbol{\xi} =\frac{\textbf{x}}{\varepsilon} \right),\nonumber \\
\rho^{(m,\varepsilon)}\left({{\mathbf{x}} }  \right) &=\rho^{(m,\varepsilon)}\left( \boldsymbol{\xi} =\frac{\textbf{x}}{\varepsilon} \right), \quad\forall \boldsymbol{\xi} \in \mathfrak{Q}.\label{PerD1}
\end{align}
It is assumed that the body forces are $\mathfrak{L}$-periodic with period $\mathfrak{L} = [-L/2, L/2] \times [-L/2, L/2] \times [- \eta L/2, \eta L/2]$, and have zero mean values on $\mathfrak{L}$. Moreover,  $\mathfrak{L}$ can be considered as a representative portion of the whole body, under the assumption that the structural length $L$ is much greater than the microstructural size $\varepsilon$ ($L >> \varepsilon$), consistently with the scale separation principle. 
As a consequence, the microscopic fields can be expressed in the form
$u_i\left( \textbf{x},\boldsymbol{\xi}=\textbf{x}/\varepsilon,t \right)$ and $\phi\left(\textbf{x},\boldsymbol{\xi}=\textbf{x}/\varepsilon,t \right)$, where $\textbf{x}$ and 
$\boldsymbol{\xi}=\textbf{x}/\varepsilon $ play the roles of 
macroscopic variable ($slow$ variable), and microscopic variable ($fast$ variable), respectively.\\
Due to this double $\mathfrak{Q}$- and $\mathfrak{L}$-periodicity, finding the solution of partial differential equations (\ref{eq:1}) is, in general, very burdensome adopting numerical
approaches and very difficult via analytic approaches. This implies that valuable alternative approaches are found. In particular, homogenization techniques, allowing to replace the periodic medium
with an equivalent homogeneous one, provide an efficient tool 
to accurately describe the overall behaviour of the microstructured piezoelectric material, with low computational costs. We here derive a macroscopic equivalent piezoeletric continuum, which overall
constitutive tensors are analytically obtained in terms of the actual geometric and physical properties of the microsctructure.
In particular, at the macroscopic scale the variables 
$\textbf{U} \left(\textbf{x},t \right)=U_i\left( \textbf{x},t \right) \textbf{e}_i$ and $\Phi \left( \textbf{x},t \right)$ are consistently defined, depending both on the macroscopic point $\textbf{x}$ and on time $t$.

\section{Asymptotic expansion of microscopic field equations}
The microscopic fields, involved in the governing equations,  can be expressed through an asymptotic expansion in terms of the microscopic characteristic size $\varepsilon$, \citep{Bakhvalov1984} . Consistently with the principle of scale separation, the $slow$ and $fast$ variables are kept separate, and the following expressions hold
\begin{align}
\begin{split}
&{u_k}\left( {{\bf{x}}, \frac{{\bf{x}}}{\varepsilon },t} \right) =\sum\limits_{l = 0}^{ + \infty } {{\varepsilon ^l}} {u_k^{(l)}}\left( {{\bf{x}}, \frac{{\bf{x}}}{\varepsilon },t} \right)= {u_k^{(0)}}\left( {{\bf{x}}, \frac{{\bf{x}}}{\varepsilon },t} \right)+ \varepsilon {u_k^{(1)}}\left( {{\bf{x}}, \frac{{\bf{x}}}{\varepsilon },t} \right)+ \varepsilon^2 {u_k^{(2)}}\left( {{\bf{x}}, \frac{{\bf{x}}}{\varepsilon },t} \right)+\mathcal{O}\left( {\bf{\varepsilon^3}} \right),\\
&{\phi}\left( {{\bf{x}}, \frac{{\bf{x}}}{\varepsilon },t} \right) =\sum\limits_{l = 0}^{ + \infty } {{\varepsilon ^l}} {\phi^{(l)}}\left( {{\bf{x}}, \frac{{\bf{x}}}{\varepsilon },t} \right)= {\phi^{(0)}}\left( {{\bf{x}}, \frac{{\bf{x}}}{\varepsilon },t} \right)+ \varepsilon {\phi^{(1)}}\left( {{\bf{x}}, \frac{{\bf{x}}}{\varepsilon },t} \right)+ \varepsilon^2 {\phi^{(2)}}\left( {{\bf{x}}, \frac{{\bf{x}}}{\varepsilon },t} \right)+\mathcal{O}\left( {\bf{\varepsilon^3}} \right).
\end{split}
\label{eq:4}
\end{align}
\noindent Moreover, recalling the derivation rule valid for the function $f\left( \textbf{x},\boldsymbol{\xi} =\frac{\textbf{x}}{\varepsilon} \right)$, that is
\begin{align}  \frac{D}{Dx_j} f\left( \textbf{x},\boldsymbol{\xi} =\frac{\textbf{x}}{\varepsilon} \right)={\left. {\left( {\frac{{\partial f}}{{\partial {x_j}}} + \frac{1}{\varepsilon }{\frac{\partial f}{\partial \xi_j}}} \right)} \right|_{{\boldsymbol{\xi}} = \frac{{\bf{x}}}{\varepsilon }}}={\left. {\left( {\frac{{\partial f}}{{\partial {x_j}}} + \frac{1}{\varepsilon }{f_{,j}}} \right)} \right|_{{\boldsymbol{\xi}} = \frac{{\bf{x}}}{\varepsilon }}},
\label{regolaDerivata}
\end{align}
and plugging the asymptotic expansions (\ref{eq:4}), into the governing equations (\ref{eq:1}), the asymptotic expansion of miscroscopic field equations is obtained as
\begin{align}
\begin{split}
& \left.\left\{\varepsilon^{-2} \left( \left(C_{ijkl}^{m} u^{(0)}_{k,l}\right)_{,j} + \left(e_{ijk}^{m}  \phi^{(0)}_{,k} \right)_{,j} \right)+ \varepsilon^{-1}\left[ \left(C_{ijkl}^{m}  \left(\frac{\partial u^{(0)}_k}{\partial x_l}+u_{k,l}^{(1)} \right) \right)_{,j}+\frac{\partial }{\partial x_j} \left( C_{ijkl}^{m} u^{(0)}_{k,l}\right)+ \right. \right. \right.\\
& \left. \left. \left.  +\left(e_{ijk}^{m} \left(\frac{\partial \phi^{(0)}}{\partial x_l}+\phi_{,k}^{(1)} \right)\right)_{,j}+ \frac{\partial }{\partial x_j} \left( e_{ijk}^{m} \phi^{(0)}_{,k}\right)\right]+\left[ \left(C_{ijkl}^{m}  \left(\frac{\partial u^{(1)}_k}{\partial x_l}+u_{k,l}^{(2)} \right) \right)_{,j}+ \right. \right. \right. \\
& \left. \left. \left.+\frac{\partial }{\partial x_j} \left( C_{ijkl}^{m} \left(\frac{\partial u^{(0)}_k}{\partial x_l}+u_{k,l}^{(1)} \right)\right)
  +\left(e_{ijk}^{m} \left(\frac{\partial \phi^{(1)}}{\partial x_l}+\phi_{,k}^{(2)} \right)\right)_{,j}+ \frac{\partial }{\partial x_j} \left( e_{ijk}^{m} \left(\frac{\partial \phi^{(0)}}{\partial x_l}+\phi_{,k}^{(1)} \right)\right)+ \right. \right. \right. \\
& \left. \left. \left. -\rho^m  \ddot{u}_i^{(0)}+\mathcal{O}\left( {\bf{\varepsilon}} \right) \right] \right\}  \right|_{{\boldsymbol{\xi}} = \frac{{\bf{x}}}{\varepsilon }}+b_i (\textbf{x})= 0, \label{eq.12}\\
& \left.\left\{\varepsilon^{-2} \left( \left(e_{kli}^{m} u^{(0)}_{k,l}\right)_{,i} - \left(\beta_{il}^{m}  \phi^{(0)}_{,l} \right)_{,i} \right)+ \varepsilon^{-1}\left[ \left(e_{kli}^{m} \left(\frac{\partial u^{(0)}_k}{\partial x_l}+u_{k,l}^{(1)} \right) \right)_{,i}+\frac{\partial }{\partial x_i} \left( e_{kli}^{m} u^{(0)}_{k,l}\right)+ \right. \right. \right.\\
& \left. \left. \left.  -\left(\beta_{il}^{m} \left(\frac{\partial \phi^{(0)}}{\partial x_l}+\phi_{,l}^{(1)} \right)\right)_{,i}- \frac{\partial }{\partial x_i} \left( \beta_{il}^{m} \phi^{(0)}_{,l}\right)\right]+\left[ \left(e_{kli}^{m}  \left(\frac{\partial u^{(1)}_k}{\partial x_l}+u_{k,l}^{(2)} \right) \right)_{,i}+ \right. \right. \right. \\
& \left. \left. \left.+\frac{\partial }{\partial x_i} \left( e_{kli}^{m} \left(\frac{\partial u^{(0)}_k}{\partial x_l}+u_{k,l}^{(1)} \right)\right)
 -\left(\beta_{il}^{m} \left(\frac{\partial \phi^{(1)}}{\partial x_l}+\phi_{,l}^{(2)} \right)\right)_{,i}- \frac{\partial }{\partial x_i} \left( \beta_{il}^{m} \left(\frac{\partial \phi^{(0)}}{\partial x_l}+\phi_{,l}^{(1)} \right)\right)+ \right. \right. \right. \\
& \left. \left. \left. +\mathcal{O}\left( {\bf{\varepsilon}} \right) \right] \right\}  \right|_{{\boldsymbol{\xi}} = \frac{{\bf{x}}}{\varepsilon }}- \rho_e (\textbf{x})= 0.
\end{split}
\end{align}
\noindent 
After proper manipulations, by collecting the terms with
equal power $\varepsilon$,  a
hierarchical set of recursive partial differential problems, in terms of the sensitivities $u_k^{(j)}$ and $\phi^{(j)}$, is obtained.
The statement and the solution of such recursive problems are reported in Appendix A. In particular, the solution of the $\varepsilon^{-2}$-order problem, in (\ref{eq.A1}),  takes the following form, where the dependence on the microscopic variable disappears
\begin{align}
u_{k}^{(0)}\left( {{\mathbf{x}},\boldsymbol{\xi},t } \right) &= U_{k}\left( \mathbf{x},t  \right),\nonumber\\
\phi^{(0)}\left( {{\mathbf{x}},\boldsymbol{\xi},t } \right) &= \Phi\left( \mathbf{x},t  \right).\label{Phi0}
\end{align}
Analogously, the solution of the $\varepsilon^{-1}$-order problem, in (\ref{eq.A2}), results in
\begin{align}
u_{k}^{(1)}\left( {{\mathbf{x}},\boldsymbol{\xi},t } \right) &= N^{(1)}_{kpq_1} \left( \boldsymbol{\xi}\right) \frac{\partial U_p}{\partial x_{q_1}} + \widetilde{N}^{(1)}_{k q_1} \left( \boldsymbol{\xi}\right) \frac{\partial \Phi}{\partial x_{q_1}},\nonumber\\
\phi^{(1)}\left( {{\mathbf{x}},\boldsymbol{\xi},t } \right) &= W^{(1)}_{q_1} \left( \boldsymbol{\xi}\right) \frac{\partial \Phi}{\partial x_{q_1}} + \widetilde{W}^{(1)}_{p q_1} \left( \boldsymbol{\xi}\right) \frac{\partial U_p}{\partial x_{q_1}},\label{Phi1}
\end{align}
where $N^{(1)}_{kpq_1}$, $\widetilde{N}^{(1)}_{k q_1}$, $W^{(1)}_{q_1}$, $\widetilde{W}^{(1)}_{p q_1}$ are the first order perturbation functions, only depending on the geometric and physical properties of the microstructure, that will be evaluated in the following. \\
Finally, the solution of the $\varepsilon^{0}$-order problem, in (\ref{eq.A4}), is
\begin{align}
u_{k}^{(2)}\left( {{\mathbf{x}},\boldsymbol{\xi},t } \right) &= N^{(2)}_{kpq_1q_2} \left( \boldsymbol{\xi}\right) \frac{\partial^2 U_p}{\partial x_{q_1} \partial x_{q_2}} + \widetilde{N}^{(2)}_{k q_1 q_2} \left( \boldsymbol{\xi}\right) \frac{\partial^2 \Phi}{\partial x_{q_1} \partial x_{q_2}}+\widehat{N}^{(2)}_{kp}\left( \boldsymbol{\xi}\right) \ddot{U}_p,\nonumber\\
\phi^{(2)}\left( {{\mathbf{x}},\boldsymbol{\xi},t } \right) &= W^{(2)}_{q_1 q_2} \left( \boldsymbol{\xi}\right) \frac{\partial^2 \Phi}{\partial x_{q_1} \partial x_{q_2}} + \widetilde{W}^{(2)}_{p q_1 q_2} \left( \boldsymbol{\xi}\right) \frac{\partial^2 U_p}{\partial x_{q_1} \partial x_{q_2}}+ \widehat{W}^{(2)}_{q_1 q_2}\left( \boldsymbol{\xi}\right) \ddot{U}_p,\label{Phi2}
\end{align}
being $N^{(2)}_{kpq_1q_2}$, $\widetilde{N}^{(2)}_{k q_1 q_2} $,
$\widehat{N}^{(2)}_{kp}$, $W^{(2)}_{q_1 q_2}$, $\widetilde{W}^{(2)}_{p q_1 q_2}$, and $ \widehat{W}^{(2)}_{q_1 q_2}$ are the second order perturbation functions.\\
The perturbation functions are determined, in turn, by solving two sets of hierarchical differential problems referred to as cell problems. The cell problems are obtained by properly manipulating the recursive problems, reported in Appendix A, and exploiting the structure of the sensitivities in (\ref{Phi0})-(\ref{Phi2}), according with \citep{Smyshlyaev2000,fantoni2017multi}. It is remarked that
the source terms involved in such cell problems are characterized by zero mean values in the Unit Cell $\mathfrak{Q}$. As a consequence, they admit sufficiently regular and $\mathfrak{Q}$-periodic solutions. Moreover, the uniqueness of cell problems solutions is guaranteed by enforcing the normalization condition of the perturbation functions, i.e. assuring zero mean values in $\mathfrak{Q}$. \\
  The two cell problems at the order $\varepsilon^{-1}$ read
\begin{align}
\begin{split}
&   \left(C_{ijkl}^{m} N^{(1)}_{kpq_1,l}\right)_{,j} + \left(e_{ijk}^{m}  \widetilde{W}^{(1)}_{p q_1,k} \right)_{,j} + C^m_{ijpq_1,j}= 0, \label{cella11}\\
&  \left(e_{kli}^{m} N^{(1)}_{kpq_1,l}\right)_{,i} - \left(\beta_{il}^{m}  \widetilde{W}^{(1)}_{pq_1,l} \right)_{,i}+ e^m_{pq_1i,i}=0,
\end{split}
\end{align}
\begin{align}
\begin{split}
&   \left(C_{ijkl}^{m} \widetilde{N}^{(1)}_{kq_1,l}\right)_{,j} + \left(e_{ijk}^{m}  W^{(1)}_{q_1,k} \right)_{,j} + e^m_{ijq_1,j}= 0, \label{cella12}\\
&  \left(e_{kli}^{m} \widetilde{N}^{(1)}_{kq_1,l}\right)_{,i} - \left(\beta_{il}^{m}  {W}^{(1)}_{q_1,l} \right)_{,i}- \beta^m_{iq_1,i}=0,
\end{split}
\end{align}
in terms of the first order perturbation functions.\\
Finally, the three cell problems at the order $\varepsilon^{0}$ are 
\begin{align}
\begin{split}
&   \left(C_{ijkl}^{m} N^{(2)}_{kpq_1 q_2,l}\right)_{,j} + \left(e_{ijk}^{m}  \widetilde{W}^{(2)}_{p q_1 q_2,k} \right)_{,j} + \frac{1}{2} \left[ \left( C^m_{ijpq_2 }N^{(1)}_{kpq_1}\right)_{,j} + C^m_{i q_2kl}+C^m_{iq_2kl}N^{(1)}_{kpq_1,l}+ \left(e^m_{ijq_2} \widetilde{W}^{(1)}_{pq_1} \right)_{,j}+\right. \\
& \left.+ e^m_{iq_2k} \widetilde{W}^{(1)}_{pq_1,k}+\left( C^m_{ijpq_1 }N^{(1)}_{kpq_2}\right)_{,j} + C^m_{i q_1kl}+C^m_{iq_1kl}N^{(1)}_{kpq_2,l}+\left(e^m_{ijq_1} \widetilde{W}^{(1)}_{pq_2} \right)_{,j}+e^m_{iq_1k} \widetilde{W}^{(1)}_{pq_2,k} \right]= \\
& = \frac{1}{2} \langle C^m_{iq_1pq_2}+C^m_{iq_2kl} N^{(1)}_{kpq_1,l}+e^m_{iq_2k} \widetilde{W}^{(1)}_{pq_1,k}+C^m_{iq_2pq_1}+C^m_{iq_1kl} N^{(1)}_{kpq_2,l}+e^m_{iq_1k} \widetilde{W}^{(1)}_{pq_2,k} \rangle, \label{cella12}\\
&  \left(e_{kli}^{m} N^{(2)}_{kpq_1q_2,l}\right)_{,i} - \left(\beta_{il}^{m}  \widetilde{W}^{(2)}_{pq_1q_2,l} \right)_{,i}+ \frac{1}{2} \left[\left( e^m_{kq_2i} N^{(1)}_{kpq_1}\right)_{,i}+e^m_{klq_2}N^{(1)}_{kpq_1,l}
+e^m_{pq_2q_1}- \left( \beta^m_{iq_2} \widetilde{W}^{(1)}_{pq_1}\right)_{,i}+ \right.\\
& \left. -\beta^m_{q_2l} \widetilde{W}^{(1)}_{pq_1,l}+\left( e^m_{kq_1i} N^{(1)}_{kpq_2}\right)_{,i}+e^m_{klq_1}N^{(1)}_{kpq_2,l}
+e^m_{pq_1q_2}- \left( \beta^m_{iq_1} \widetilde{W}^{(1)}_{pq_2}\right)_{,i}-\beta^m_{q_1l} \widetilde{W}^{(1)}_{pq_2,l}\right]=\frac{1}{2} \left\langle e^m_{klq_2} N^{(1)}_{kpq_1,l}+\right.\\
&\left. +e^m_{pq_2q_1}-\beta^m_{q_2l} \widetilde{W}^{(1)}_{pq_1,l}+e^m_{klq_1} N^{(1)}_{kpq_2,l}+e^m_{pq_1q_2}-\beta^m_{q_1l} \widetilde{W}^{(1)}_{pq_2,l}\right\rangle,
\end{split}
\end{align}
\begin{align}
\begin{split}
&   \left(C_{ijkl}^{m} \widetilde{N}^{(2)}_{kq_1 q_2,l}\right)_{,j} + \left(e_{ijk}^{m}  {W}^{(2)}_{q_1 q_2,k} \right)_{,j} + \frac{1}{2} \left[ \left( C^m_{ijkq_2 } \widetilde{N}^{(1)}_{kq_1}\right)_{,j} + C^m_{i q_2kl} \widetilde{N}^{(1)}_{kq_1,l} + \left(e^m_{ijq_2} {W}^{(1)}_{q_1} \right)_{,j}+\right. \\
& \left.+ e^m_{iq_1q_2}+e^m_{iq_2k} {W}^{(1)}_{q_1,k}+\left( C^m_{ijkq_1 } \widetilde{N}^{(1)}_{kq_2}\right)_{,j} + C^m_{i q_1kl} \widetilde{N}^{(1)}_{kq_2,l} +\left(e^m_{ijq_1} {W}^{(1)}_{q_2} \right)_{,j} +
e^m_{iq_2q_1}
+e^m_{iq_1k} W^{(1)}_{q_2,k}\right]= \\
& = \frac{1}{2} \langle C^m_{iq_2kl} \widetilde{N}^{(1)}_{kpq_1,l}+e^m_{iq_1q_2}+e^m_{iq_2k} {W}^{(1)}_{q_1,k}+ C^m_{iq_1kl} \widetilde{N}^{(1)}_{kpq_2,l}+e^m_{iq_2q_1}+e^m_{iq_1k} {W}^{(1)}_{q_2,k}
 \rangle, \label{cella13}\\
&  \left(e_{kli}^{m} \widetilde{N}^{(2)}_{kq_1q_2,l}\right)_{,i} - \left(\beta_{il}^{m}  {W}^{(2)}_{q_1q_2,l} \right)_{,i}+ \frac{1}{2} \left[\left( e^m_{kq_2i} \widetilde{N}^{(1)}_{kq_1}\right)_{,i}+e^m_{klq_2}\widetilde{N}^{(1)}_{kq_1,l}
- \left( \beta^m_{iq_2} {W}^{(1)}_{q_1}\right)_{,i}-\beta^m_{q_1q_2} +\right.\\
& \left. -\beta^m_{q_2l} {W}^{(1)}_{q_1,l}+\left( e^m_{kq_1i} \widetilde{N}^{(1)}_{kq_2}\right)_{,i}+e^m_{klq_1}\widetilde{N}^{(1)}_{kq_2,l}
- \left( \beta^m_{iq_1} {W}^{(1)}_{q_2}\right)_{,i}-\beta^m_{q_2q_1}-\beta^m_{q_1l} {W}^{(1)}_{q_2,l}\right]=\frac{1}{2} \left\langle e^m_{klq_2} \widetilde{N}^{(1)}_{kq_1,l}+\right.\\
&\left. -\beta^m_{q_1q_2}-\beta^m_{q_2l} {W}^{(1)}_{q_1,l}+e^m_{klq_1} \widetilde{N}^{(1)}_{kq_2,l} -\beta^m_{q_2q_1}-\beta^m_{q_1l} {W}^{(1)}_{q_2,l}\right\rangle
\end{split}
\end{align}
\begin{align}
\begin{split}
&   \left(C_{ijkl}^{m} \widehat{N}^{(2)}_{kp,l}\right)_{,j} + \left(e_{ijk}^{m}  \widehat{W}^{(2)}_{p,k} \right)_{,j}-\rho^m \delta_{pi}= -\langle \rho^m \rangle \delta_{pi}, \label{cella14}\\
&  \left(e_{kli}^{m} \widehat{N}^{(2)}_{kp,l}\right)_{,i} - \left(\beta_{il}^{m}  \widehat{W}^{(2)}_{p,l} \right)_{,i}=0
\end{split}
\end{align}
in terms of the first order perturbation functions, where a proper symmetrization with respect to  indices $q_1$ and $q_2$
has been introduced. The symbol $\langle (\cdot) \rangle= 1/|\mathfrak{Q}| \int_{\mathfrak{Q}}(\cdot) d \boldsymbol{\xi}$ has been introduced, $|\mathfrak{Q}|=\eta$, and $\delta_{pi}$ is the Kronecker delta function.\\
After determining the perturbation functions, by solving the cell problems, the so-called down-scaling relations can be rigorously determined. 
In particular, by plugging the terms in (\ref{Phi0})-(\ref{Phi2})  into the asymptotic expansion (\ref{eq:4}), the following down-scaling relations are found
\begin{align}
\begin{split}
&{u_k}\left( {{\bf{x}}, \boldsymbol{\xi},t} \right) = U_k \left( {\bf{x}},t\right)+ \varepsilon \left( N^{(1)}_{kpq_1}\left(  \boldsymbol{\xi} \right)\frac{\partial U_p}{\partial x_{q_1}} +\widetilde{N}^{(1)}_{kq_1}\left(  \boldsymbol{\xi} \right)\frac{\partial \Phi}{\partial x_{q_1}}\right)+ \varepsilon^2 \left(N^{(2)}_{kpq_1q_2}\left(  \boldsymbol{\xi} \right)\frac{\partial^2 U_p}{\partial x_{q_1}\partial x_{q_2}}+\right.\\ & \left.+\widetilde{N}^{(2)}_{kq_1q_2}\left(  \boldsymbol{\xi} \right)\frac{\partial^2 \Phi}{\partial x_{q_1}\partial x_{q_2}}+\widehat{N}^{(2)}_{kp}\left(  \boldsymbol{\xi} \right) \ddot{U}_p\right)+\mathcal{O}\left( {\bf{\varepsilon^3}} \right),\\
&{\phi}\left( {{\bf{x}}, \boldsymbol{\xi},t} \right) =\Phi \left( {\bf{x}},t\right)+ \varepsilon \left( W^{(1)}_{q_1}\left(  \boldsymbol{\xi} \right)\frac{\partial \Phi}{\partial x_{q_1}} +\widetilde{W}^{(1)}_{pq_1}\left(  \boldsymbol{\xi} \right)\frac{\partial U_p}{\partial x_{q_1}}\right)+ \varepsilon^2 \left(W^{(2)}_{q_1q_2}\left(  \boldsymbol{\xi} \right)\frac{\partial^2 \Phi}{\partial x_{q_1}\partial x_{q_2}}+\right.\\ & \left.+\widetilde{W}^{(2)}_{pq_1q_2}\left(  \boldsymbol{\xi} \right)\frac{\partial^2 U_p}{\partial x_{q_1}\partial x_{q_2}}+\widehat{W}^{(2)}_{p}\left(  \boldsymbol{\xi} \right) \ddot{U}_p\right)+\mathcal{O}\left( {\bf{\varepsilon^3}} \right),
\end{split}
\label{polynExp}
\end{align}
where the microscopic fields are made dependent on the corresponding macroscopic ones.\\
The up-scaling relations are, in turn, determined as
\begin{align}
U_{k}\left( \mathbf{x},t  \right) &\buildrel\textstyle.\over= \left\langle u_k\left( {{\bf{x}}, \frac{\bf{x}}{\varepsilon}+\boldsymbol{\zeta},t} \right)\right\rangle,\nonumber\\
\Phi\left( \mathbf{x},t  \right)) &\buildrel\textstyle.\over= \left\langle \phi \left( {{\bf{x}}, \frac{\bf{x}}{\varepsilon}+\boldsymbol{\zeta},t} \right) \right\rangle,\label{Phik}
\end{align}
where 
$\boldsymbol{\zeta }  \in \mathfrak{Q}$ is a $translation$ $variable$, such that the vector $\varepsilon \, \boldsymbol{\zeta }$ $\in$ $\mathfrak{A}$ defines a translation of the heterogeneous medium with respect to $\mathfrak{L}$-periodic source terms \citep{Smyshlyaev2000,Bacigalupo2014}.  It is pointed out that the perturbation functions fulfil
the   invariance  property
satisfied by $\mathfrak{Q}$-periodic functions $g(\boldsymbol{\xi}+\boldsymbol{\zeta})|_{\boldsymbol{\xi}=\textbf{x}/\varepsilon}$, such that 
$\int\limits_{\mathfrak{Q}} {g({\textbf{x}/\varepsilon} + {\boldsymbol{\zeta }})} d{\boldsymbol{\zeta }} =\int\limits_{\mathfrak{Q}} {g({\boldsymbol{\xi}} + {\boldsymbol{\zeta }})} d{\boldsymbol{\zeta }} = \int\limits_{\mathfrak{Q}} {g({\boldsymbol{\xi}} + {\boldsymbol{\zeta }})} d{\boldsymbol{\xi}}$.
\\


\section{Identification of the homogenized constitutive tensors}
The macroscopic  piezoelectric constitutive properties and the  macroscopic inertial term are derived by exploiting a generalized macro-homogeneity condition,  establishing an energy equivalence between the macroscopic and
the microscopic scales. To this aim, the generalized microscopic mean Lagrangian $\overline{\mathcal{L}}_m$, inspired by \citep{Smyshlyaev2000,Bacigalupo2014}, and the corresponding 
generalized macroscopic Lagrangian ${\mathcal{L}}_M$ are taken into account. Note that focus is on the identification of a first order homogenized continuum.\\
More specifically, the generalized microscopic mean Lagrangian  is defined on the basis of the microscopic kinetic energy density  $t_m$ and of the microscopic electric enthalpy density $h_m$, as
\begin{align}
\begin{split}
\overline{\mathcal{L}}_m=&\int_{\mathfrak{L} } \left[ \frac{1}{|\mathfrak{Q} |} \int_{\mathfrak{Q} } \left( t_m \left(\textbf{x}, \frac{\textbf{x}}{\varepsilon}+\boldsymbol{\zeta},t \right)-h_m \left(\textbf{x}, \frac{\textbf{x}}{\varepsilon}+\boldsymbol{\zeta},t \right)\right)d\boldsymbol{\zeta} \right] d\textbf{x}=\\
=&\int_{\mathfrak{L} } \langle t_m-h_m \rangle d\textbf{x},\label{L}
\end{split}
\end{align}
where the down-scaling relations have been exploited to express the densities $t_m$ and $h_m$ as asymptotic expansions in terms of the macroscopic displacement $U_i$ and electric potential $\Phi$, respectively, i.e.
\begin{align}
\begin{split}
&t_m= \frac{1}{2} \rho^m \dot{u}_i \dot{u}_i=\frac{1}{2} \rho^m \dot{U}_i \dot{U}_i+\mathcal{O}\left( {\bf{\varepsilon}} \right), \label{tm}\\ 
& h_m=\frac{1}{2}  C^m_{ijhk} \frac{D u_i}{D x_j} \frac{D u_h}{D x_k}+ e^m_{ijh}  \frac{D \phi}{D x_h} \frac{D u_i}{D x_j}- \frac{1}{2} \beta^m_{ij} \frac{D \phi}{D x_i} \frac{D \phi}{D x_j}= \\
& = \frac{1}{2} \left[ C^m_{ijhk} \left( \delta_{ip} \delta_{jq_1}+N^{(1)}_{ipq_1,j}\right) \left( \delta_{hs} \delta_{kr_1}+N^{(1)}_{hsr_1,k}\right)+2e^m_{ijh} \left( \delta_{ip} \delta_{jq_1}+N^{(1)}_{ipq_1,j}\right)
 \widetilde{W}^{(1)}_{sr_1,h}+ \right.\\
 & \left. -\beta^m_{ij}\widetilde{W}^{(1)}_{pq_1,j} \widetilde{W}^{(1)}_{sr_1,i}  \right] \frac{\partial U_p}{\partial x_{q_1}}\frac{\partial U_s}{\partial x_{r_1}} +\left[ C^m_{ijhk} \left( \delta_{ip} \delta_{jq_1}+N^{(1)}_{ipq_1,j}\right)
\widetilde{N}^{(1)}_{hr_1,k}+e^m_{ijh}\left( \delta_{ip} \delta_{jq_1}+N^{(1)}_{ipq_1,j}\right) \right.\\
& \left. \left( \delta_{hr_1}+W^{(1)}_{r_1,h}\right)+e^m_{ijh} \widetilde{N}^{(1)}_{ir_1,j} \widetilde{W}^{(1)}_{pq_1,h} -\beta^m_{ij} \widetilde{W}^{(1)}_{pq_1,j} \left( \delta_{ir_1}+W^{(1)}_{r_1,i}\right)\right]\frac{\partial U_p}{\partial x_{q_1}}\frac{\partial \Phi}{\partial x_{r_1}}-\frac{1}{2} \left[ \beta^m_{ij} \left( \delta_{jq_1}+W^{(1)}_{q_1,j}\right) \right.\\
& \left. \left( \delta_{ir_1}+W^{(1)}_{r_1,i}\right)-C^m_{ijhk} \widetilde{N}^{(1)}_{iq_1,j} \widetilde{N}^{(1)}_{hr_1,k}-2e^m_{ijh} \widetilde{N}^{(1)}_{iq_1,j}\left( \delta_{hr_1}+\widetilde{W}^{(1)}_{r_1,h}\right)\right]\frac{\partial \Phi}{\partial x_{q_1}} \frac{\partial \Phi}{\partial x_{r_1}}+\mathcal{O}\left( {\bf{\varepsilon}} \right).
\end{split}
\end{align}
Analogously, the generalized macroscopic Lagrangian is given as
\begin{align}
\begin{split}
\mathcal{L}_	M=&\int_{\mathfrak{L}} \left( t_M\left(\textbf{x}, t \right) -h_M\left(\textbf{x}, t \right) \right) d\textbf{x},\label{LM}
\end{split}
\end{align}
in terms of the macroscopic kinetic energy density  $t_M$ and of the macroscopic electric enthalpy density $h_M$, express in the following form
\begin{align}
\begin{split}
&t_M= \frac{1}{2} \rho \, \dot{U}_i \dot{U}_i, \label{tm}\\ 
& h_M=\frac{1}{2}  C_{pq_1sr_1} \frac{\partial U_p}{\partial x_{q_1}} \frac{\partial U_s}{\partial x_{r_1}}+ e_{pq_1r_1}  \frac{\partial U_p}{\partial x_{q_1}} \frac{\partial \Phi}{\partial x_{r_1}} - \frac{1}{2} \beta_{q1r1} \frac{\partial \Phi}{\partial x_{q_1}} \frac{\partial \Phi}{\partial x_{r_1}},
\end{split}
\end{align}
involving the components of the macroscopic constitutive 
tensors and of the macroscopic inertial term of the piezoelectric first order continuum.\\
The generalized macro-homogeneity condition is, thus, defined by 
\begin{align}
\begin{split}
\overline{\mathcal{L}}_m^0  &\buildrel\textstyle.\over= \mathcal{L}_	M,\label{LM}
\end{split}
\end{align}
where $\overline{\mathcal{L}}_m^0$ identifies the generalized microscopic mean Lagrangian truncated at order 0-th, i.e. retaining only the coefficients of $\varepsilon^0$. It follows that the components of macroscopic  piezoelectric constitutive properties and the  macroscopic inertial term  result as
\begin{align}
\begin{split}
& \rho = \langle \rho^m \rangle,\label{22}\\
& C_{pq_1sr_1} = \frac{1}{2} \left \langle  C^m_{ijhk} \left( \delta_{ip} \delta_{jq_1}+N^{(1)}_{ipq_1,j}\right) \left( \delta_{hs} \delta_{kr_1}+N^{(1)}_{hsr_1,k}\right)+2e^m_{ijh} \left( \delta_{ip} \delta_{jq_1}+N^{(1)}_{ipq_1,j}\right)
 \widetilde{W}^{(1)}_{sr_1,h}+ \right.\\
 & \left. -\beta^m_{ij}\widetilde{W}^{(1)}_{pq_1,j} \widetilde{W}^{(1)}_{sr_1,i}  \right \rangle,\\
 & e_{pq_1r_1}= \left \langle  C^m_{ijhk} \left( \delta_{ip} \delta_{jq_1}+N^{(1)}_{ipq_1,j}\right)
\widetilde{N}^{(1)}_{hr_1,k}+e^m_{ijh}\left( \delta_{ip} \delta_{jq_1}+N^{(1)}_{ipq_1,j}\right)  \left( \delta_{hr_1}+W^{(1)}_{r_1,h}\right)+ \right.\\
& \left.+e^m_{ijh} \widetilde{N}^{(1)}_{ir_1,j} \widetilde{W}^{(1)}_{pq_1,h} -\beta^m_{ij} \widetilde{W}^{(1)}_{pq_1,j} \left( \delta_{ir_1}+W^{(1)}_{r_1,i}\right) \right \rangle,\\
& \beta_{q_1r_1}= \frac{1}{2} \left \langle   \beta^m_{ij} \left( \delta_{jq_1}+W^{(1)}_{q_1,j}\right) \left( \delta_{ir_1}+W^{(1)}_{r_1,i}\right)-C^m_{ijhk} \widetilde{N}^{(1)}_{iq_1,j} \widetilde{N}^{(1)}_{hr_1,k}-2e^m_{ijh} \widetilde{N}^{(1)}_{iq_1,j}\left( \delta_{hr_1}+\widetilde{W}^{(1)}_{r_1,h}\right) \right \rangle,
\end{split}
\end{align}
expressed in terms of the perturbation functions and in terms of the components of the microscopic constitutive tensors and of the microscopic inertial term.\\
A more accurate description of the response of the heterogeneous material can be consistently obtained resorting to higher order homogenization approaches , in which the generalized macro-homogeneity condition is properly modified, both taking into account higher order terms in the asymptotic expansions at the microscopic 
scale, and nonlocal constitutive tensors and inertial terms at the macroscopic level. An alternative approach is schematically reported in Appendix B. 
It consists in solving the so-called average field equations of infinite order  (\ref{eq.B1}),  via perturbation methods. An infinite hierarchical set of macroscopic partial differential problems is, thus, obtained and higher order approximations can be obtained by properly truncating the asymptotic expansion of $U_i$ and $\Phi$.\\

\section{Characterization of dispersive wave propagation in piezoelectric periodic materials}
The problem of characterizing the wave propagation in piezoelectric periodic materials can be addressed through the Floquet-Bloch theory 
able to deduce the frequency band structure of the material characterized by periodic microstructure.
Nevertheless, this approach can eventually be computationally cumbersome in the case of very complex microstructural topologies.
In this respect, a valuable alternative
is the use of homogenization techniques, adopting either local or non-local approaches.  More specifically, when first order approaches are used, an accurate description is obtained in the case of in the long wavelength regime. \\
With this in mind, in the following Section \ref{5.1}  the heterogeneous material is investigated to determine its frequency spectrum, with both acoustic and optical branches, while the Section \ref{5.2} is devoted to determine the dispersive functions approximating only 
the acoustic branches in the long wavelength regime, adopting a first order asymptotic homogenization approach.

\subsection{Frequency band structure of the heterogeneous material with periodic microstructure}\label{5.1}
In this Section, consistently with the rigorous Floquet-Bloch theory \citep{Floquet1883,Bloch1928,Brillouin1960}, a generalization to  piezoelectric materials is used to study the band structure
of the microstructured periodic material. To this aim we apply the time Fourier transform to the partial differential equations (\ref{eq:1}) at the microscopic scale, in the case of zero source terms, i.e. $b_i$=0 and $\rho_e$=0.
For the sake of completeness, we recall the time Fourier transform for a generic  $g(\mathbf{x},t)$ is defined as
\begin{align} \label{TFT}
\mathcal{F}_{t} \left[ {g}(\mathbf{x},t) \right]= \int_{-\infty}^{+\infty} {g}(\mathbf{x},t)e^{-\iota \omega t}d{\mathbf{x}}=\mathord{\buildrel{\lower3pt\hbox{$\scriptscriptstyle\frown$}} \over g} (\mathbf{x},\omega),
\end{align}
where the angular frequency $\omega \in \mathbb{R}$. 
The resulting generalized Christoffel equations are, thus, obtained as
\begin{align}
\begin{split}
&\frac{D}{Dx_j} \left(  C_{ijkl}^{(m,\varepsilon)} \frac{D \mathord{\buildrel{\lower3pt\hbox{$\scriptscriptstyle\frown$}} \over u_k} }{Dx_l} \right) +\frac{D}{Dx_j} \left(  e_{ijk}^{(m,\varepsilon)} \frac{D \mathord{\buildrel{\lower3pt\hbox{$\scriptscriptstyle\frown$}} \over \phi} }{Dx_k} \right)+\rho^{(m,\varepsilon)}  \omega^2 \mathord{\buildrel{\lower3pt\hbox{$\scriptscriptstyle\frown$}} \over u_i} =0  ,\\
& \frac{D}{Dx_i} \left(  \widetilde{e}_{ikl}^{(m,\varepsilon)} \frac{D \mathord{\buildrel{\lower3pt\hbox{$\scriptscriptstyle\frown$}} \over u_k} }{Dx_l} \right)  - \frac{D}{Dx_i} \left(  \beta_{il}^{(m,\varepsilon)} \frac{D \mathord{\buildrel{\lower3pt\hbox{$\scriptscriptstyle\frown$}} \over \phi} }{Dx_l} \right)= 0,
\end{split}
\end{align}
in which the  the well-known property $\mathcal{F}_{t}(\frac{\partial^n g(\mathbf{x},t)}{\partial t^n})=(\iota \omega)^n  \mathord{\buildrel{\lower3pt\hbox{$\scriptscriptstyle\frown$}} \over g}(\mathbf{x},\omega)$ has been exploited.\\
Due to the periodicity of the microstructured medium, only the Periodic Cell $\mathfrak{A}$ is analysed, subject to the Floquet-Bloch boundary conditions, that is 
\begin{align}
&\mathord{\buildrel{\lower3pt\hbox{$\scriptscriptstyle\frown$}} \over u_i^+}   = \mathord{\buildrel{\lower3pt\hbox{$\scriptscriptstyle\frown$}} \over u_i^-}  {e^{\iota k_j v_j^{(p)}}},\label{eq:FBdec1}\\
&\mathord{\buildrel{\lower3pt\hbox{$\scriptscriptstyle\frown$}} \over \phi^+}  = \mathord{\buildrel{\lower3pt\hbox{$\scriptscriptstyle\frown$}} \over \phi^-}  {e^{\iota k_j v_j^{(p)}}},\label{eq:FBdec2}\\
& \mathord{\buildrel{\lower3pt\hbox{$\scriptscriptstyle\frown$}} \over {\sigma}_{lr}^{+}}  =-  \mathord{\buildrel{\lower3pt\hbox{$\scriptscriptstyle\frown$}} \over {\sigma}_{lr}^{-}}  (m_r^{(p)})^- e^{\iota k_j v_j^{(p)}},\label{eq:FBdec3}\\
&   \mathord{\buildrel{\lower3pt\hbox{$\scriptscriptstyle\frown$}} \over {d}_{r}^{+} }  (m_r^{(p)})^+=- \mathord{\buildrel{\lower3pt\hbox{$\scriptscriptstyle\frown$}} \over {d}_{r}^{-} }   (m_r^{(p)})^- e^{\iota k_j v_j^{(p)}},\label{eq:FBdec3}
\end{align}
 being $v_j^{(p)}$ the components of the vector of periodicity 
$\textbf{v}_p= v_j^{(p)} \textbf{e}_j$,  and $(m_r^{(p)})^{\pm}$ the components of the outward normal $\textbf{m}_p^{\pm}= (m_j^{(p)})^{\pm} \textbf{e}_j$ to the boundary $\partial \textbf{A}$, $j,p=1,2,3$. Moreover, the apexes $^{\pm}$ 
are referred to the positive part  $\partial \textbf{A}^{+}$ (with outward normal $\textbf{m}_p^{+}$  ) and the corresponding negative parts  $\partial \textbf{A}^{-}$ (with outward normal $\textbf{m}_p^{-}$  ) of the Periodic Cell boundary. Note that for the generic function $\widehat{g}$ the following notation is used, i.e.
$\mathord{\buildrel{\lower3pt\hbox{$\scriptscriptstyle\frown$}} \over g}^{\pm} 
\buildrel\textstyle.\over=\mathord{\buildrel{\lower3pt\hbox{$\scriptscriptstyle\frown$}} \over g}(\textbf{x}^{\pm}) $
where $\textbf{x}^{\pm} \in \partial \textbf{A}^{\pm}$, and  $\textbf{x}^{+}=\textbf{x}^{-}+\textbf{v}_p$.
Finally, $k_j$ are the components of the wave vector $\textbf{k}$. In this context, the dimensionless first Brillouin zone $\mathfrak{B}=[-\pi, \pi] \times [-\pi, \pi] \times [-\pi, \pi]$ is defined in the  space of the dimensionless wave vectors (whose components are $k_1 d$, $k_2 d$ and $k_3 h$) and is associated to the Periodic Cell $\mathfrak{A}$. Such Brillouin zone is characterized by three orthogonal vectors $\pi \textbf{n}_i$, parallel to $\textbf{e}_i$, with $i$=1,2,3.\\
\subsection{Dispersion functions in the first order homogeneous material}\label{5.2}
A first order asymptotic homogenization approach is here adopted to study the dispersion functions piezoelectric periodic material. The  equations of motion for the homogenized continuum, in the absence of source terms take the following form
\begin{align}
\begin{split}
&C_{ijkl}  \frac{\partial^2 U_p}{\partial x_{q_1} \partial x_{q_2}}+e_{ijk}  \frac{\partial^2 \Phi}{\partial x_{q_1} \partial x_{q_2}} = \rho \ddot{U_i} , \label{eq.e29}\\
& \widetilde{e}_{ikl}  \frac{\partial^2 U_p}{\partial x_{q_1} \partial x_{q_2}}- \beta_{il}  \frac{\partial^2 \Phi}{\partial x_{q_1} \partial x_{q_2}}  = 0,
\end{split}
\end{align}
where $C_{ijkl} $ are the components of the fourth order macro elasticity tensor,  $e_{ijk}$ are the components of the third order macroscopic  piezoelectric stress-charge coupling tensor, with the following relation between the components  $\widetilde{e}_{ijk}={e}_{jki}$, 
$\beta_{il}$
are the components of the second order macroscopic  dielettric
permittivity tensor and $ \rho$ is the macroscopic  mass density, reported in  Equation (\ref{22}).
Equations (\ref{eq.e29}) are here properly manipulated, by first exploiting the 
time Fourier transform, see equation (\ref{TFT}),  and then the 
space Fourier transform in the macroscopic space, that is hereafter recalled
\begin{align}
\mathcal{F}_{\textbf{x}} \left[ \mathord{\buildrel{\lower3pt\hbox{$\scriptscriptstyle\frown$}} \over g} (\mathbf{x},\omega) \right]= \int_{\mathbb{R}^2}  \mathord{\buildrel{\lower3pt\hbox{$\scriptscriptstyle\frown$}} \over g} (\mathbf{x},\omega)e^{-\iota \textbf{k} \textbf{x}}d{\mathbf{x}}= \mathord{\buildrel{\lower3pt\hbox{$\scriptscriptstyle\smile$}} \over { \mathord{\buildrel{\lower3pt\hbox{$\scriptscriptstyle\frown$}} \over g}}}(\mathbf{k},\omega ),
\end{align}
where  $\textbf{k} \in \mathbb{R}^2$ is the the wave vector, and  recalling the property $\mathcal{F}_{\textbf{x}}(\frac{\partial^n \mathord{\buildrel{\lower3pt\hbox{$\scriptscriptstyle\frown$}} \over g} (\mathbf{x},\omega)}{\partial x_j^n})=(\iota k_j)^n \mathord{\buildrel{\lower3pt\hbox{$\scriptscriptstyle\smile$}} \over { \mathord{\buildrel{\lower3pt\hbox{$\scriptscriptstyle\frown$}} \over g}}}(\mathbf{k},\omega )$  the following governing equations in the
frequency and wave vector space are obtained 
\begin{align}
\begin{split}
&-C_{ijkl}  k_j k_l  \mathord{\buildrel{\lower3pt\hbox{$\scriptscriptstyle\smile$}} \over { \mathord{\buildrel{\lower3pt\hbox{$\scriptscriptstyle\frown$}} \over U_k}}}- e_{ikj}  k_j k_k \mathord{\buildrel{\lower3pt\hbox{$\scriptscriptstyle\smile$}} \over { \mathord{\buildrel{\lower3pt\hbox{$\scriptscriptstyle\frown$}} \over \Phi}}}+\rho \omega^2
 \mathord{\buildrel{\lower3pt\hbox{$\scriptscriptstyle\smile$}} \over { \mathord{\buildrel{\lower3pt\hbox{$\scriptscriptstyle\frown$}} \over U_i}}}=0, \label{eq.e31}\\
& -\widetilde{e}_{ikl}   k_l k_i \mathord{\buildrel{\lower3pt\hbox{$\scriptscriptstyle\smile$}} \over { \mathord{\buildrel{\lower3pt\hbox{$\scriptscriptstyle\frown$}} \over U_k}}}+ \beta_{il}   k_l k_i \mathord{\buildrel{\lower3pt\hbox{$\scriptscriptstyle\smile$}} \over { \mathord{\buildrel{\lower3pt\hbox{$\scriptscriptstyle\frown$}} \over \Phi}}}  = 0.
\end{split}
\end{align}
\noindent The Equation (\ref{eq.e31}) can be rewritten in an equivalent form, after simple manipulations,  in terms of the phase velocity and of the propagation direction, as
\begin{align}
\begin{split}
&-C_{ijkl}  n_j n_l  \mathord{\buildrel{\lower3pt\hbox{$\scriptscriptstyle\smile$}} \over { \mathord{\buildrel{\lower3pt\hbox{$\scriptscriptstyle\frown$}} \over U_k}}}- e_{ikj}  n_j n_k \mathord{\buildrel{\lower3pt\hbox{$\scriptscriptstyle\smile$}} \over { \mathord{\buildrel{\lower3pt\hbox{$\scriptscriptstyle\frown$}} \over \Phi}}}+\rho c^2
 \mathord{\buildrel{\lower3pt\hbox{$\scriptscriptstyle\smile$}} \over { \mathord{\buildrel{\lower3pt\hbox{$\scriptscriptstyle\frown$}} \over U_i}}}=0, \label{eq.42}\\
& -\widetilde{e}_{ikl}   n_l n_i \mathord{\buildrel{\lower3pt\hbox{$\scriptscriptstyle\smile$}} \over { \mathord{\buildrel{\lower3pt\hbox{$\scriptscriptstyle\frown$}} \over U_k}}}+ \beta_{il}   n_l n_i \mathord{\buildrel{\lower3pt\hbox{$\scriptscriptstyle\smile$}} \over { \mathord{\buildrel{\lower3pt\hbox{$\scriptscriptstyle\frown$}} \over \Phi}}}  = 0,
\end{split}
\end{align}
where the phase velocity is $c=\omega/k$, the wave number is $k=||\textbf{k}||_2$ and the unit vector of propagation is $\textbf{n}=\textbf{k}/k$ with components $n_j$.\\
A static condensation is, at this stage, performed 
so that from the second 
equation in (\ref{eq.42}) we obtain $\mathord{\buildrel{\lower3pt\hbox{$\scriptscriptstyle\smile$}} \over { \mathord{\buildrel{\lower3pt\hbox{$\scriptscriptstyle\frown$}} \over \Phi}}}= \widetilde{e}_{ikl}   n_l n_i \mathord{\buildrel{\lower3pt\hbox{$\scriptscriptstyle\smile$}} \over { \mathord{\buildrel{\lower3pt\hbox{$\scriptscriptstyle\frown$}} \over U_k}}}/(\beta_{il}   n_l n_i)$ and, after substituting in the first equation in (\ref{eq.42}), the eigenproblem governing the Bloch-wave propagation, expressed in terms of phase velocity and the transformed displacement components, is obtained as
\begin{align}
\begin{split}
& \left(C_{ijkl}  n_j n_l  + e_{ijq}  n_j n_q  \frac{\widetilde{e}_{pkl}   n_l n_p}{\beta_{rs}   n_s n_r}
-\rho c^2 \delta_{ki}
\right) \mathord{\buildrel{\lower3pt\hbox{$\scriptscriptstyle\smile$}} \over { \mathord{\buildrel{\lower3pt\hbox{$\scriptscriptstyle\frown$}} \over U_k}}}=0, \,\,\,\,\,i=1,2,3, \label{eq.43}
\end{split}
\end{align}
where $\widetilde{e}_{pkl}=e_{klp}$ is exploited. Note that for any unit vector of propagation $\textbf{n}$, the eigenvalues $c^2$
are the square of the wave velocity in the first order homogenized
continuum. The corresponding eigenvectors, whose components
$\mathord{\buildrel{\lower3pt\hbox{$\scriptscriptstyle\smile$}} \over { \mathord{\buildrel{\lower3pt\hbox{$\scriptscriptstyle\frown$}} \over U_k}}}$ correspond to the time and space Fourier transform of the components of the macroscopic
displacement in the first order homogenized continuum, identify
the components of the polarization vector.

\section{Illustrative applications}
\noindent In this Section some illustrative examples are shown. We refer to the realistic assumption 
of  materials with periodic cells characterized by orthogonal periodic vectors, as schematically shown in Figure \ref{figurePrima}. First, the components of the homogenized constitutive tensors, characterizing the composite material with periodic nano-structure, are shown for different geometric parameters, i.e. the height of the nanorods and their volumetric density. Then, both the results obtained with a micromechanical model and with a first order piezoelectric homogenized model  are compared with each other, in the case of an extensional nanoscopic generator. 
Moreover, the influence of the volumetric density on the overall efficiency of such extensional nanostructured devices has been investigated in order to provide broad guidelines to maximize their efficiency. \\
Both analytic and numerical solutions are considered. 
Concerning the latter ones,
finite elements analyses have been performed 
adopting fully coupled tetrahedral second order elements with displacements and electric potential  
independent degrees of freedom.

\subsection{Homogenized piezoelectric properties}
\begin{figure}[hbtp]
\centering
\begin{overpic}[scale=0.85]{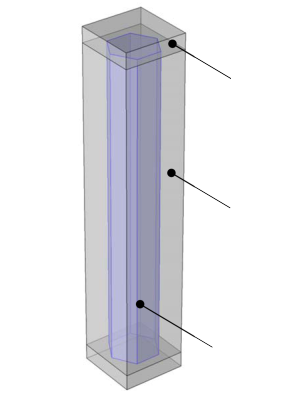}
 \put (72,90) {$(a)$}
\put (53,18) {$nanorod$}
 \put (53,12) {$ZnO$}
  \put (57,50) {$Polymeric$}
 \put (57,43) {$matrix$}
   \put (57,83) {$Polymeric$}
  \put (57,76) {$layer$}
\end{overpic}\qquad \qquad
\begin{overpic}[scale=0.6]{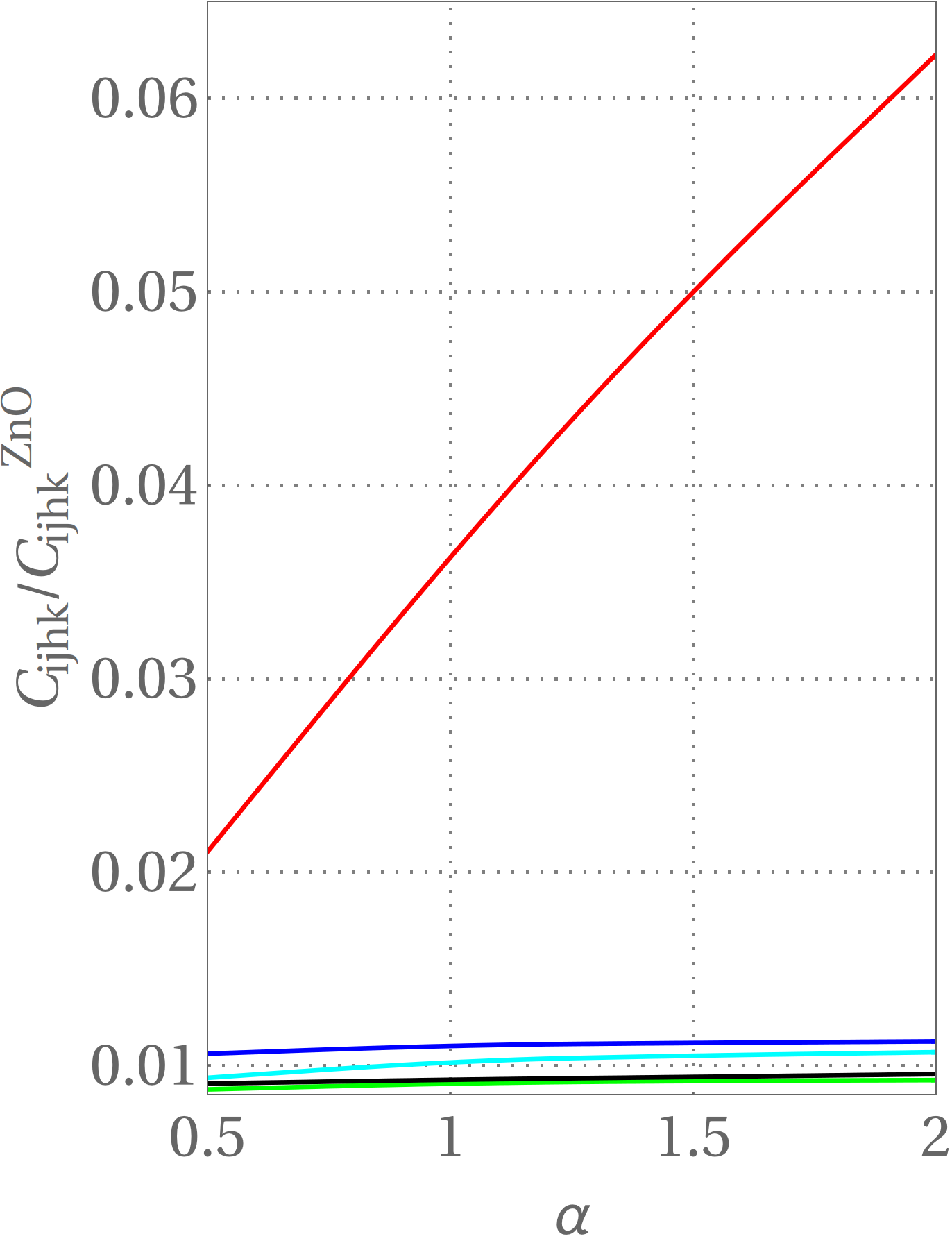}
\centering
\llap{\shortstack{%
        \includegraphics[scale=.25]{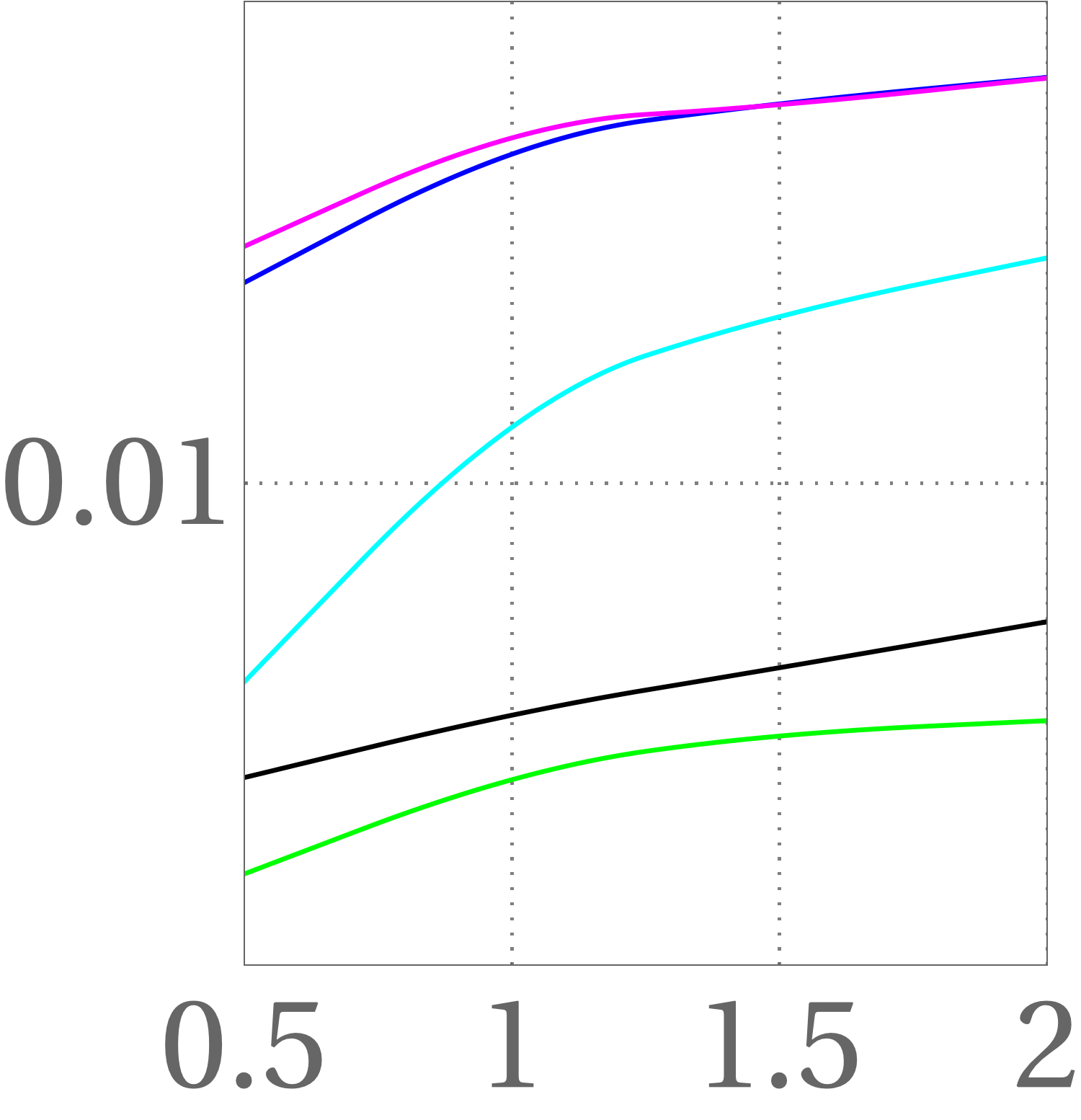}\\
        \rule{0ex}{0.6in}%
      }
  \rule{-2.5in}{0ex}} 
 \put (66,84) {$(b)$}
\end{overpic}
 \qquad
 \begin{overpic}[scale=0.6]{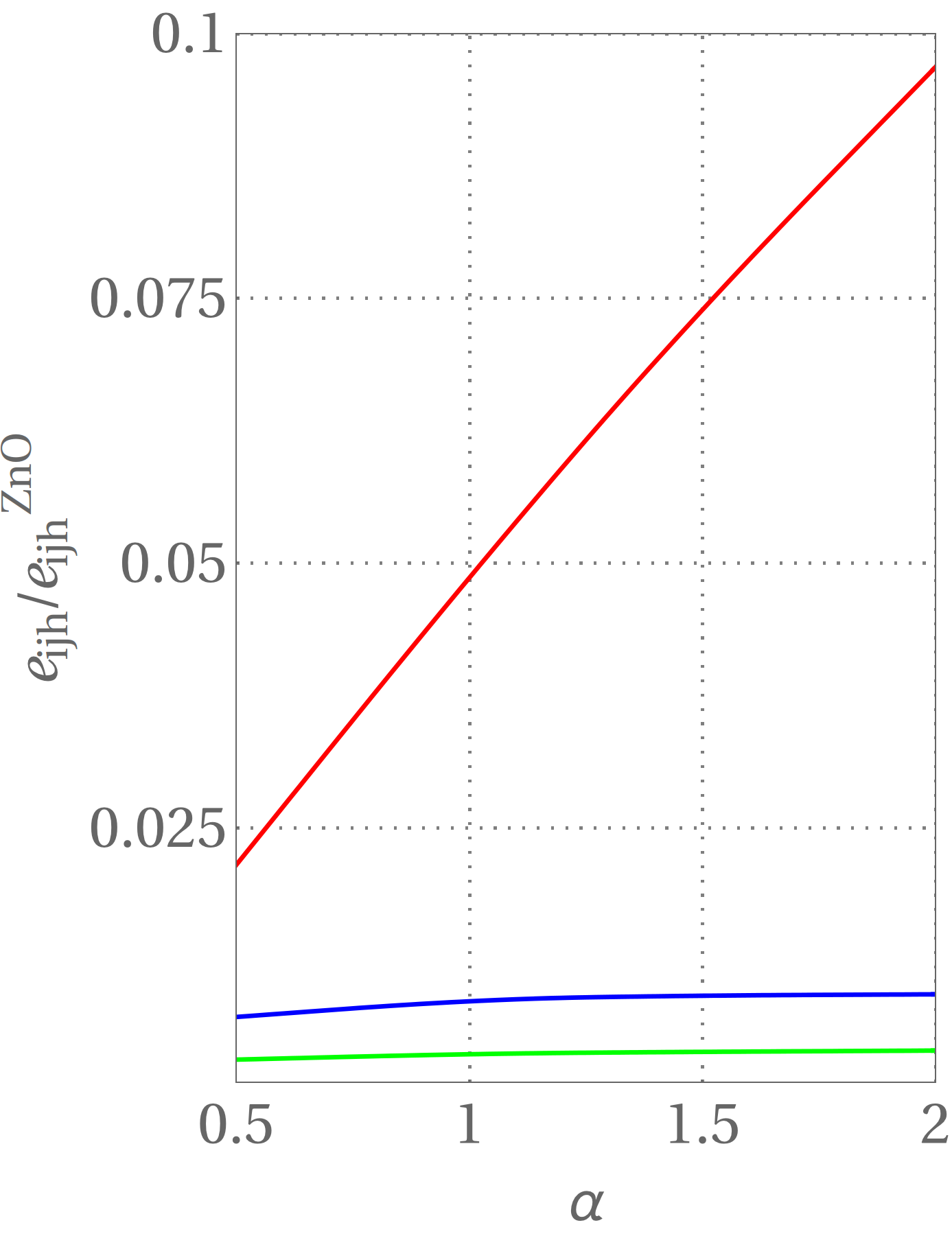}
 \centering
\llap{\shortstack{%
        \includegraphics[scale=.23]{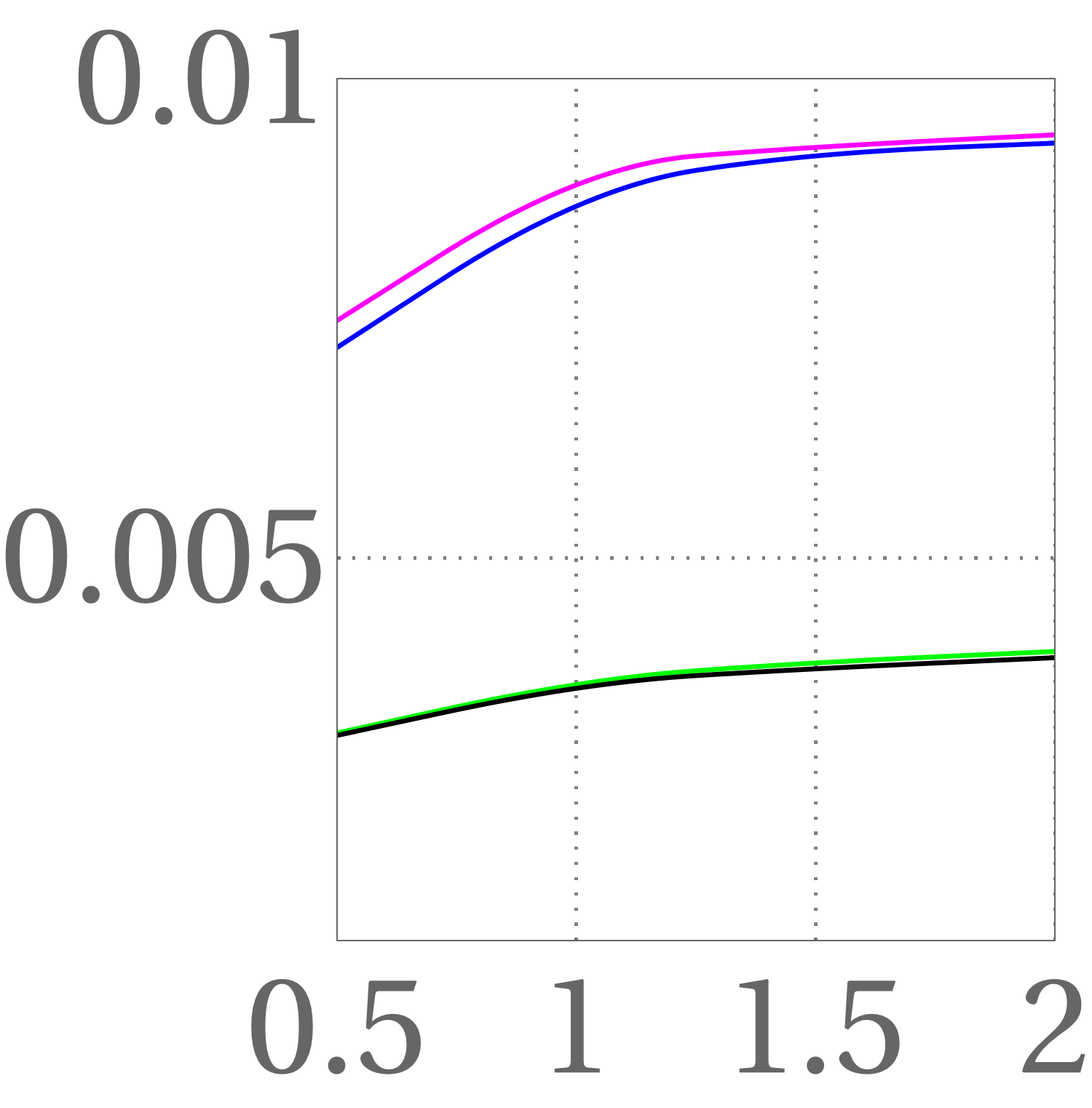}\\
        \rule{0ex}{0.7in}%
      }
  \rule{-2.5in}{0ex}}
 \put (66,82) {$(c)$}
\end{overpic}\qquad
\qquad 
\begin{overpic}[scale=0.6]{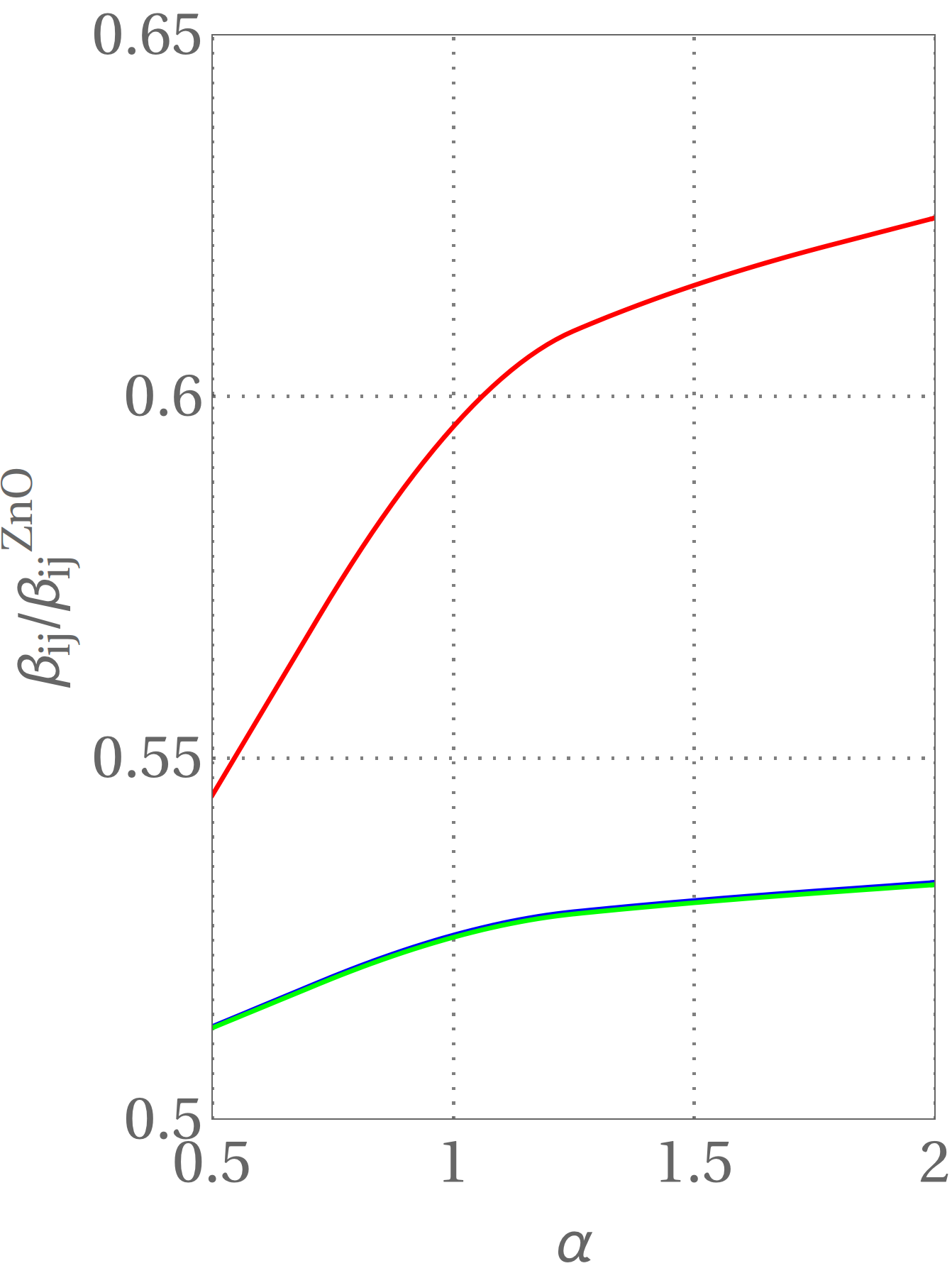}
 \centering
\llap{\shortstack{%
        \includegraphics[scale=.23]{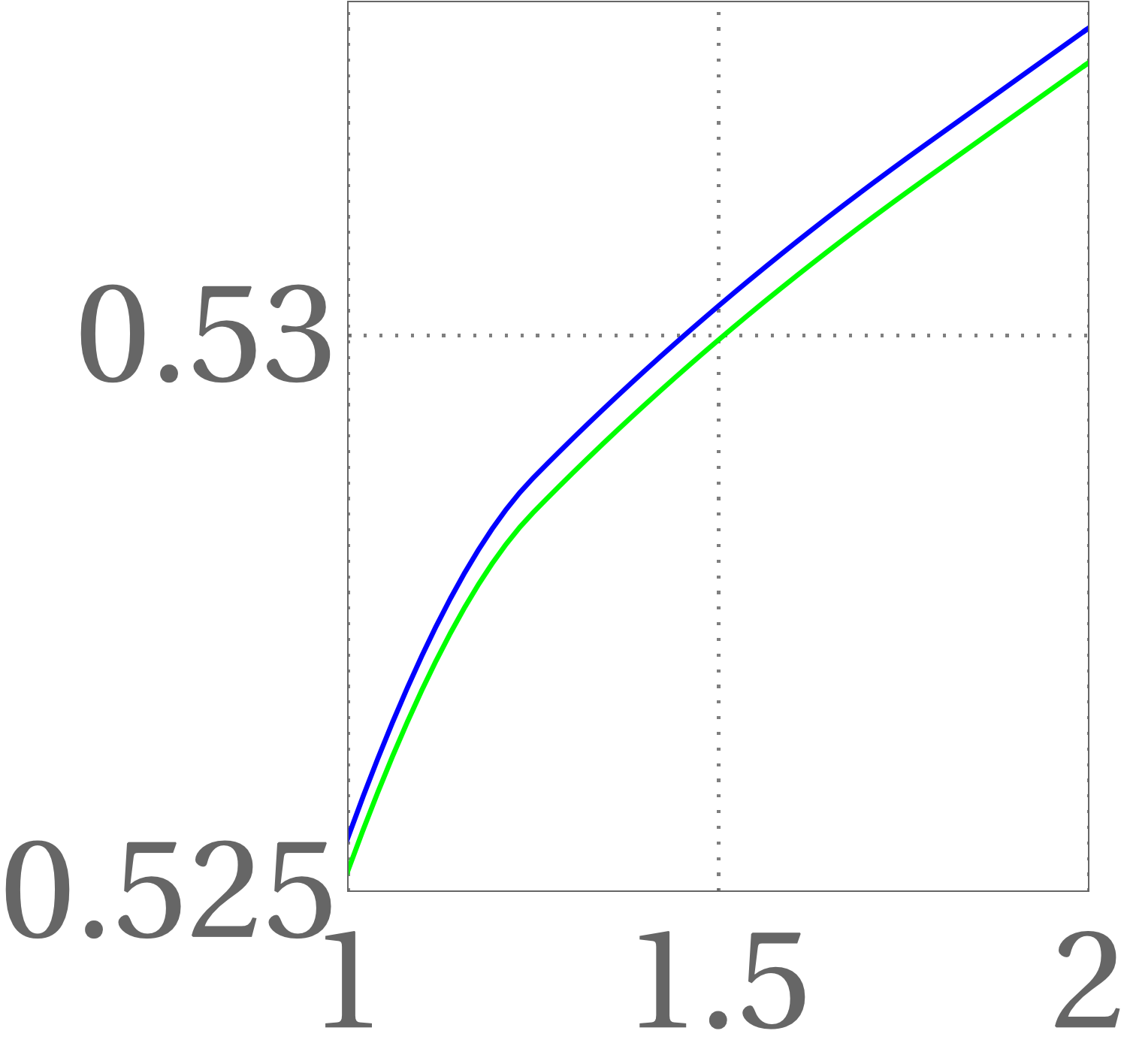}\\
        \rule{0ex}{1.3in}%
      }
  \rule{-2.5in}{0ex}}
 \put (64,85) {$(d)$}
\end{overpic}
\caption{ (a) Schematic of the Periodic Cell. Components of the homogenized constitutive tensors  versus $\alpha=h_{nr}/h^*_{nr}$, with $h^*_{nr}$=1000 nm: (b) components of the elastic tensor (the blue curve is $C_{1111}$, the magenta curve is $C_{2222}$, the red curve is $C_{3333}$, the green curve is $C_{1122}$, the black curve is $C_{1133}$ (nearly indistinguishable from $C_{2233}$), the cyan curve is $C_{1212}$ almost perfectly overlapping $C_{1313}$ and $C_{2323}$); (c) components of the coupling tensor  (the blue curve is $e_{113}$, the magenta curve is $e_{223}$, the red curve is $C_{333}$, the green curve is $e_{131}$ and the back curve is $e_{232}$); (d) components of the dielectric permittivity tensor  (the blue curve is $\beta_{11}$, the green curve is $\beta_{22}$ and the red curve is $\beta_{33}$).}
\label{figure1}
\end{figure}
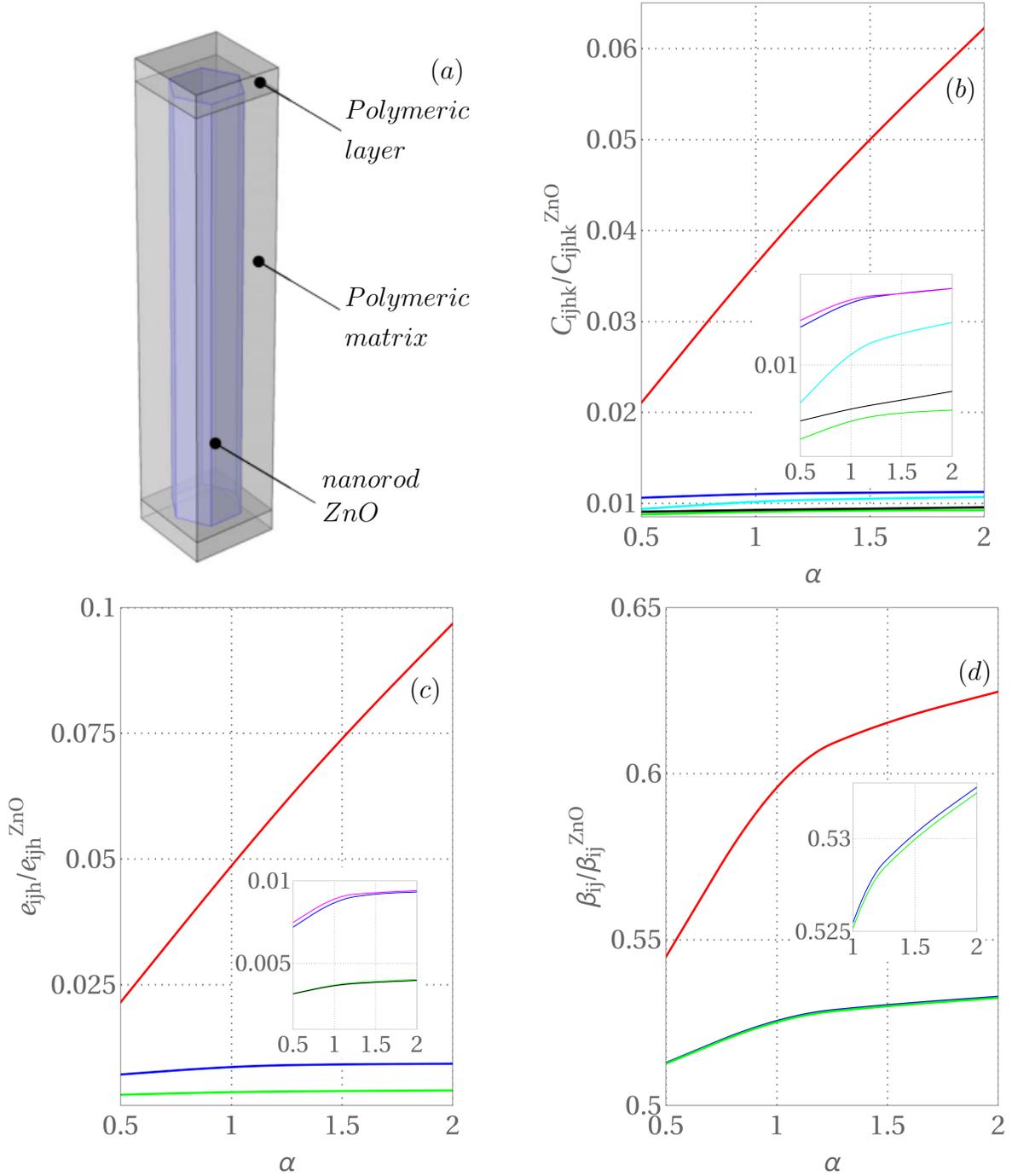

\begin{figure}[hbtp]
\centering
\begin{overpic}[scale=0.3]{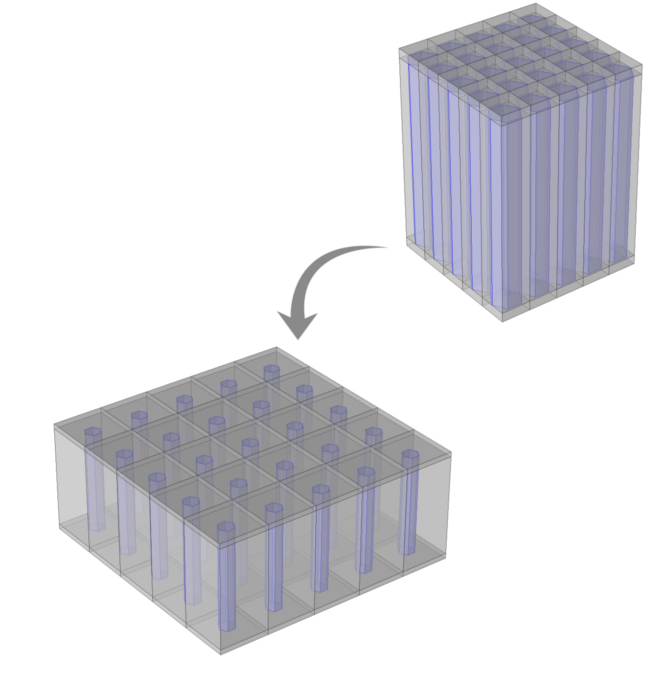}
 \put (92,96) {$(a)$}
  \put (17,80) {increasing}
 \put (17,74) {nanorods   }
  \put (17,67) {density $\delta$}
   \put (45,7) {$L=\sqrt{\delta V_{nr}/h_{nr}}$} 
     \put (67,27) {$h$} 
      \put (93,75) {$h$} 
\end{overpic}\qquad \qquad
\begin{overpic}[scale=0.6]{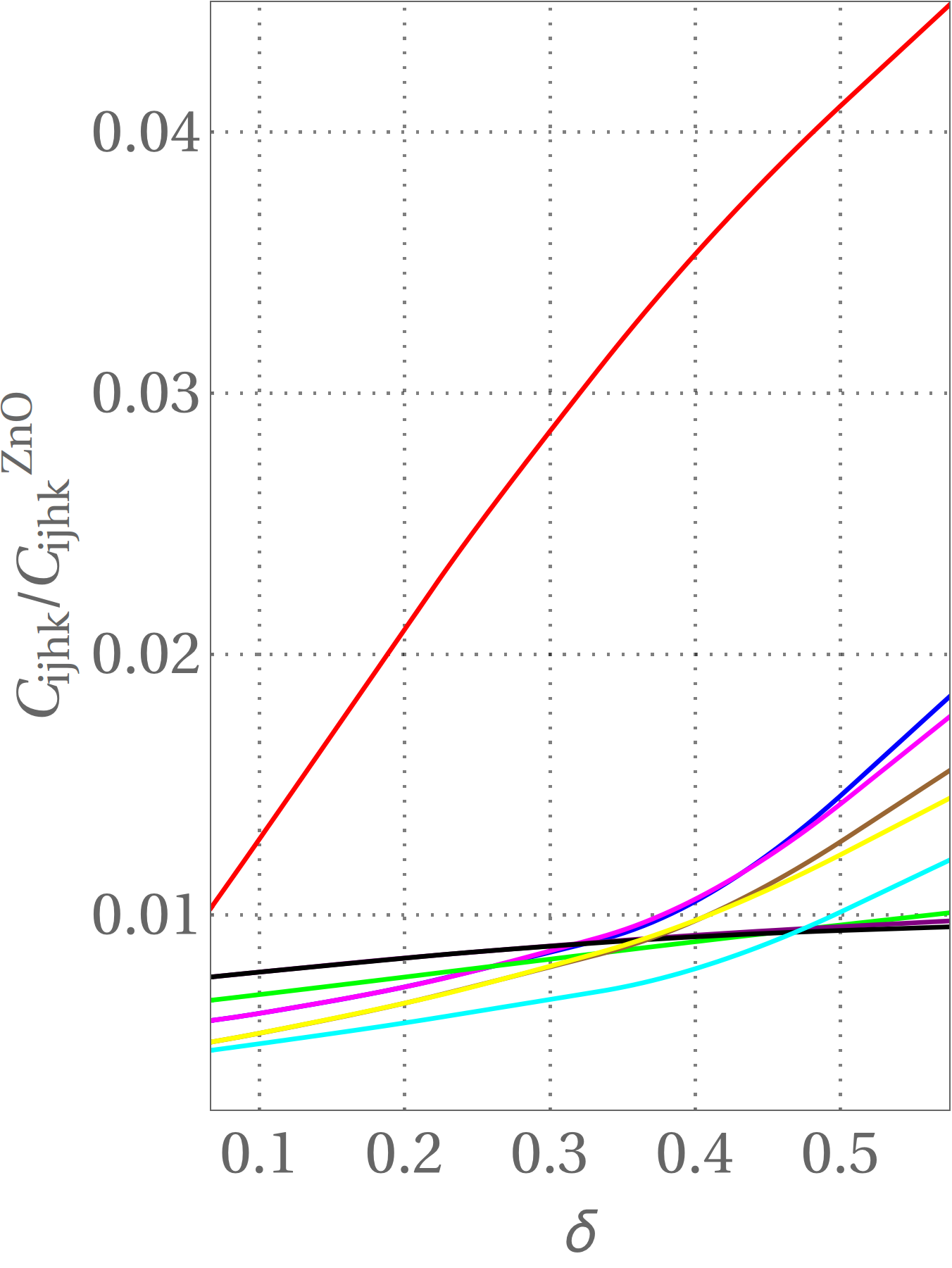}
 \put (68,86) {$(b)$}
\end{overpic}\qquad
\begin{overpic}[scale=0.6]{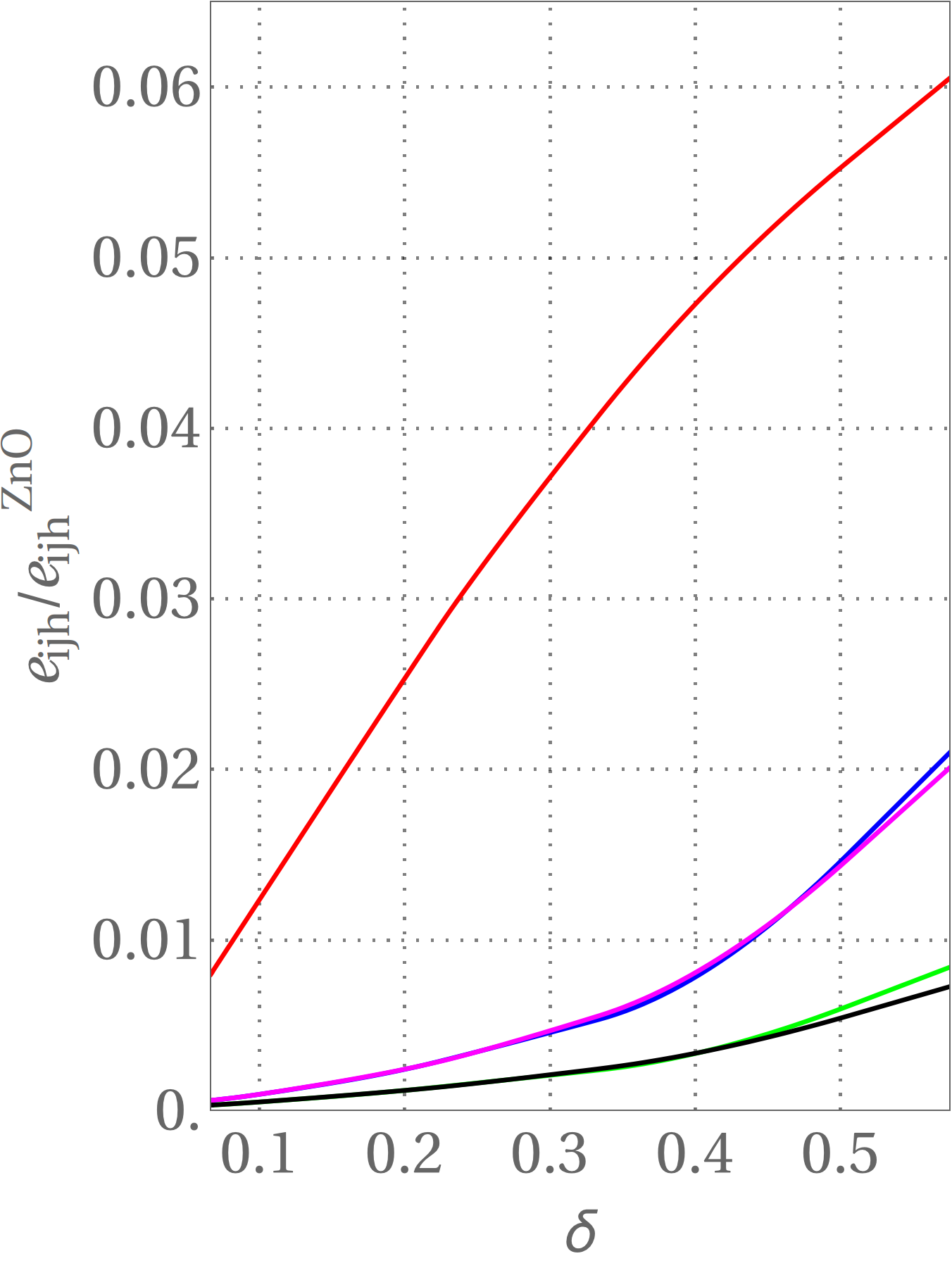}
 \put (66,80) {$(c)$}
\end{overpic}\qquad \qquad
\begin{overpic}[scale=0.6]{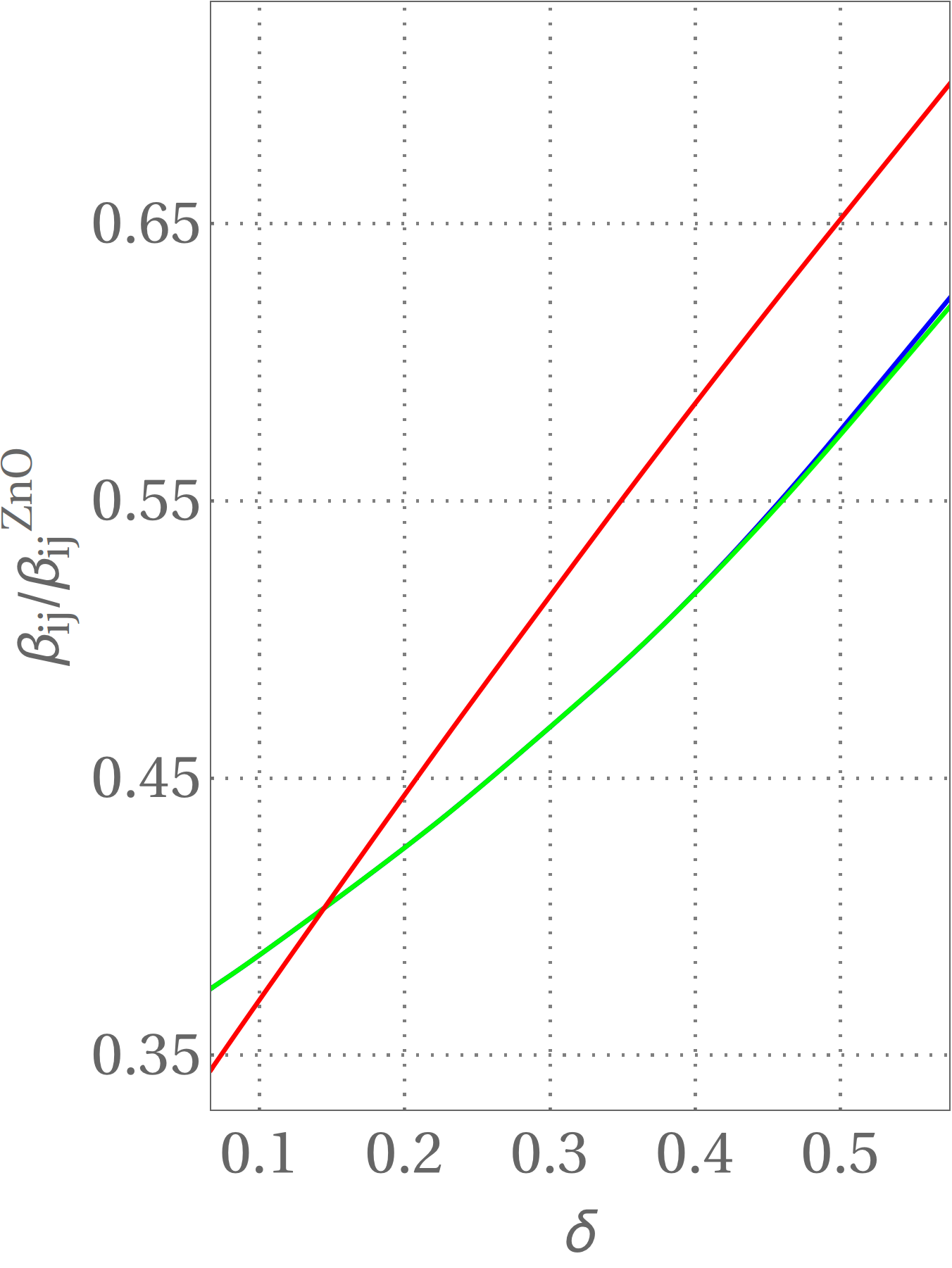}
 \put (65,90) {$(d)$}
\end{overpic}
\caption{ (a) Schematic of the 5$\times$5 cluster of nanorods for different values of density $\delta$. Components of the homogenized constitutive tensors versus $\delta$: (b) components of the elastic tensor  (the blue curve is $C_{1111}$, the magenta curve is $C_{2222}$, the red curve is $C_{3333}$, the green curve is $C_{1122}$, the black curve is $C_{1133}$ , the purple curve is  $C_{2233}$, the cyan curve is $C_{1212}$ , the yellow curve is  $C_{1313}$ and the brown curve is $C_{2323}$); (c) components of the coupling tensor  (the blue curve is $e_{113}$, the magenta curve is $e_{223}$, the red curve is $C_{333}$, the green curve is $e_{131}$ and the back curve is $e_{232}$); (d) components of the dielectric permittivity tensor  (the blue curve is $\beta_{11}$, the green curve is $\beta_{22}$ and the red curve is $\beta_{33}$).}
\label{figure2}
\end{figure}

\noindent We consider a hybrid piezoelectric nanogenerator made of equispaced ZnO-nanorods
embedded in a polymeric matrix and sandwiched 
in two homogeneous polymeric layers. The homogenized constitutive properties of the piezoelectric material are investigated with reference to the  Periodic Cell, shown in Figure \ref{figure1}(a).
The Periodic Cell $\mathfrak{A}=[-d/2, d/2] \times [-d/2, d/2] \times [-h/2, h/2]$ is characterized by the three orthogonal periodicity vectors $\textbf{v}_1$, $\textbf{v}_2$ and $\textbf{v}_3$.
 We assume  $d$=200 $nm$, the nanorod has hexagonal  section 
with edge  80 $nm$, while the height in the polarization direction is  $h$= 1100 $nm$, and the thickness of both top and bottom layers is 50 $nm$.
Such average geometric values are representative of  actual ZnO-rods based nanostructures, see \cite{Wang_2004}. \\
The nanorod with hexagonal section is made of 
Zinc oxide with polarization along the vertical axis  $\textbf{e}_3$.\\
Considering the ZnO material, see \citep{yang2004introduction}, the non-vanishing components of the elasticity tensor are:  $C^m_{1111}$ = 2.097 $\cdot 10^{11}$ Pa, 
$C^m_{2222}$ = 2.097 $\cdot 10^{11}$ Pa,
$C^m_{3333}$ = 2.111 $\cdot 10^{11}$ Pa,
$C^m_{1122}$ = 1.211$\cdot 10^{10}$ Pa,
$C^m_{1133}$ = 1.053$\cdot 10^{10}$ Pa,
$C^m_{2233}$ = 1.053$\cdot 10^{10}$ Pa,
$C^m_{1212}$= 4.237$\cdot 10^{10}$ Pa,
$C^m_{1313}$= 4.237$\cdot 10^{10}$ Pa,
$C^m_{2323}$= 4.424$\cdot 10^{10}$ Pa.
Moreover, the non-vanishing components of the coupling tensor in the stress-charge form are:
 $\widetilde{e}^m_{311}$=$\widetilde{e}^m_{322}$ =-0.567 C/$\text{m}^2$, $\widetilde{e}^m_{333}$=1.3204 C/$\text{m}^2$,
 $\widetilde{e}^m_{113}$=$\widetilde{e}^m_{223}$=-0.4805 C/$\text{m}^2$. Finally, the non-vanishing components of the dielectric permittivity tensor are  
 $\beta^m_{11} / \varepsilon_0$=  $\beta^m_{22} / \varepsilon_0$=8.5446, 
$\beta^m_{33} / \varepsilon_0$=10.204, where $\varepsilon_0$=8.854 $\cdot 10^{-12}$ C/(Vm) is the vacuum permittivity. 
Both the matrix and the top and bottom layers are made out of an isotropic
polymeric material doped with a highly conductive polymer (PANI), see \citep{huang2004,wang2005,eftekhari2011,almadhoun2014,wang2015}. The Young modulus is $E$= 535 MPA and the Poisson's coefficient is  $\nu$=0.4.
We assume the dimensionless dielectric constant  $\varepsilon_r^{P/PANI}=\beta$/$\varepsilon_0$=5. \\
A first numerical investigation concerns the influence of the heights of ZnO-nanorods on the overall piezoelectric constitutive response.
 The values of the actual height of the nanorods  $h_{nr}$ are supposed to vary between  0.5$\times h_{nr}^*$ and  2$\times h_{nr}^*$, where $ h_{nr}^*$=1100 nm is the reference initial height. In Figure \ref{figure1}(b) the non-vanishing components of the homogenized elasticity tensor,  normalized with respect to the corresponding components of bulk ZnO material, are plotted against  $\alpha= h_{nr}/h_{nr}^*$. A monotonic increasing variation is found for all the considered components. 
As expected, $C_{3333}$ (red curve) is the component most affected by the variation of $\alpha$, with a maximum value about three times the initial one. The remaining components exhibit variations significantly lower, see the zoomed plot in Figure \ref{figure1}(b). Similar considerations apply to the components of the coupling tensor, shown in Figure \ref{figure1}(c). Only the component $e_{333}$ (red curve) exhibits remarkable variations. Moreover, also the maximum variations of the components  $\beta_{ij}/\beta_{ij}^{ZnO}$  are referred to the 
component $\beta_{33}$ (red curve), see Figure \ref{figure1}(d).\\
 The second numerical study is devoted to analyse the influence of the volumetric density of ZnO-nanorods on the overall constitutive response of the Periodic Cell.
\noindent The density is defined as $\delta=V_{NR}/V$, i.e. the ratio between the volume occupied by the nanorods and the  volume of a Periodic Cell, neglecting the contribution of the top and bottom layers. We let the density $\delta$ varying between a minimum value of about $\delta$=0.066, corresponding to the case where  $d$=500 nm in the Periodic Cell $\mathfrak{A}$, and a maximum value of about $\delta$=0.65, for which $d$=160 nm, see the sketch in  Figure \ref{figure2}(a).\\
In Figure \ref{figure2}(b) the dimensionless non-vanishing components of the homogenized elasticity tensor $C_{ijhk}$ $/C_{ijhk}^{ZnO}$  are plotted versus the density $\delta$. We observe that the all the components monotonically increase, and the $C_{3333}$ (red curve) shows the maximum variation between the extremal $\delta$ values. Moreover, due to the geometrical and material symmetries of the Periodic Cell,  the curves related to the components $C_{3333}$, $C_{1133}$ and $C_{2233}$ 
are concave upward, while all the others have opposite concavity. The components $e_{ijh}/e_{ijh}^{ZnO}$ are plotted in Figure 
\ref{figure2}(c). Again, monotonic variations are observed with maximum values reached for 
$e_{333}$. Finally, in Figure 
\ref{figure2}(d) the dimensionless non-vanishing components $\beta_{ij}/\beta_{ij}^{ZnO}$ are plotted, confirming  monotonic variations 
that take maximum values for  $\beta_{33}$. Note that the maximum variations are  significantly lower than those shown by the components 
$C_{3333}$ and $e_{333}$.

\subsection{ Benchmark test:  two-phase piezoelectric hybrid composite } \label{BT}
\begin{figure}[hbtp]
\centering
\begin{overpic}[scale=1]{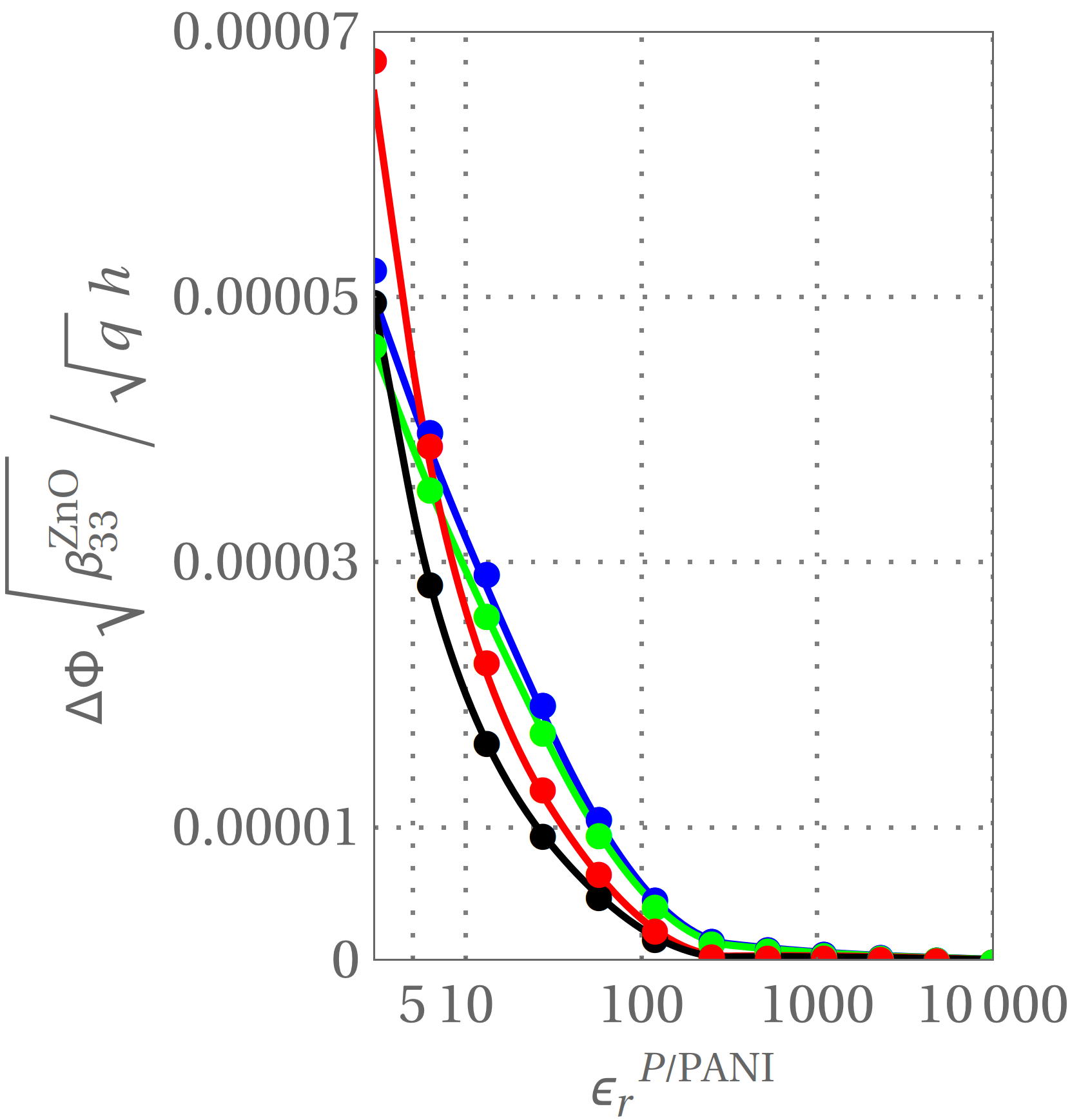}
 \put (80,92) {$(a)$}
\end{overpic}
\begin{overpic}[scale=1]{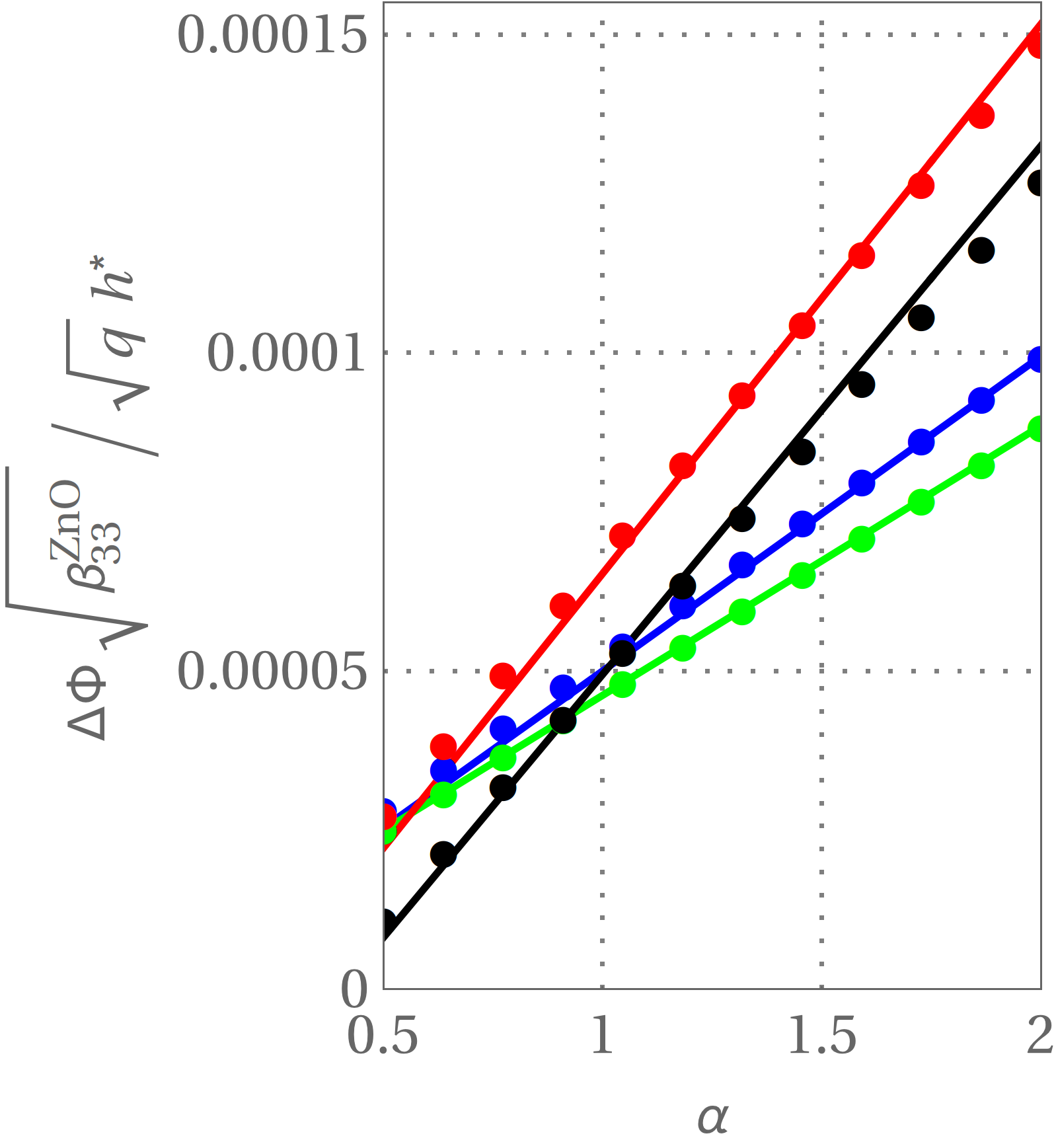}
 \put (75,92) {$(b)$}
\end{overpic}
\caption{Comparison of the macroscopic dimensionless potential differences obtained from the heterogeneous model via up-scaling relations (solid plots) and from the equivalent homogenized model (dotted plots).  (a)  $\Delta \Phi\sqrt{\beta_{33}^{ZnO}}/ (\sqrt{q} h)$ versus $\varepsilon_r^{P/PANI}$ for $\delta$=0.415 (the blue curves is for the $Case$ $1$, the green curves is for the $Case$ $2$), and for $\delta$=0.104 (the red curves is for the $Case$ $1$, the black curves is for the $Case$ $2$); (b) $\Delta \Phi\sqrt{\beta_{33}^{ZnO}}/ (\sqrt{q} h^*)$ , with $h^*$=1100 nm, versus $\alpha$ for $\delta$=0.415 (the blue curves is for the $Case$ $1$, the green curves is for the $Case$ $2$), and for $\delta$=0.104 (the red curves is for the $Case$ $1$, the black curves is for the $Case$ $2$)}
\label{figure3}
\end{figure}

\begin{figure}[t]
\centering
\begin{overpic}[scale=0.78]{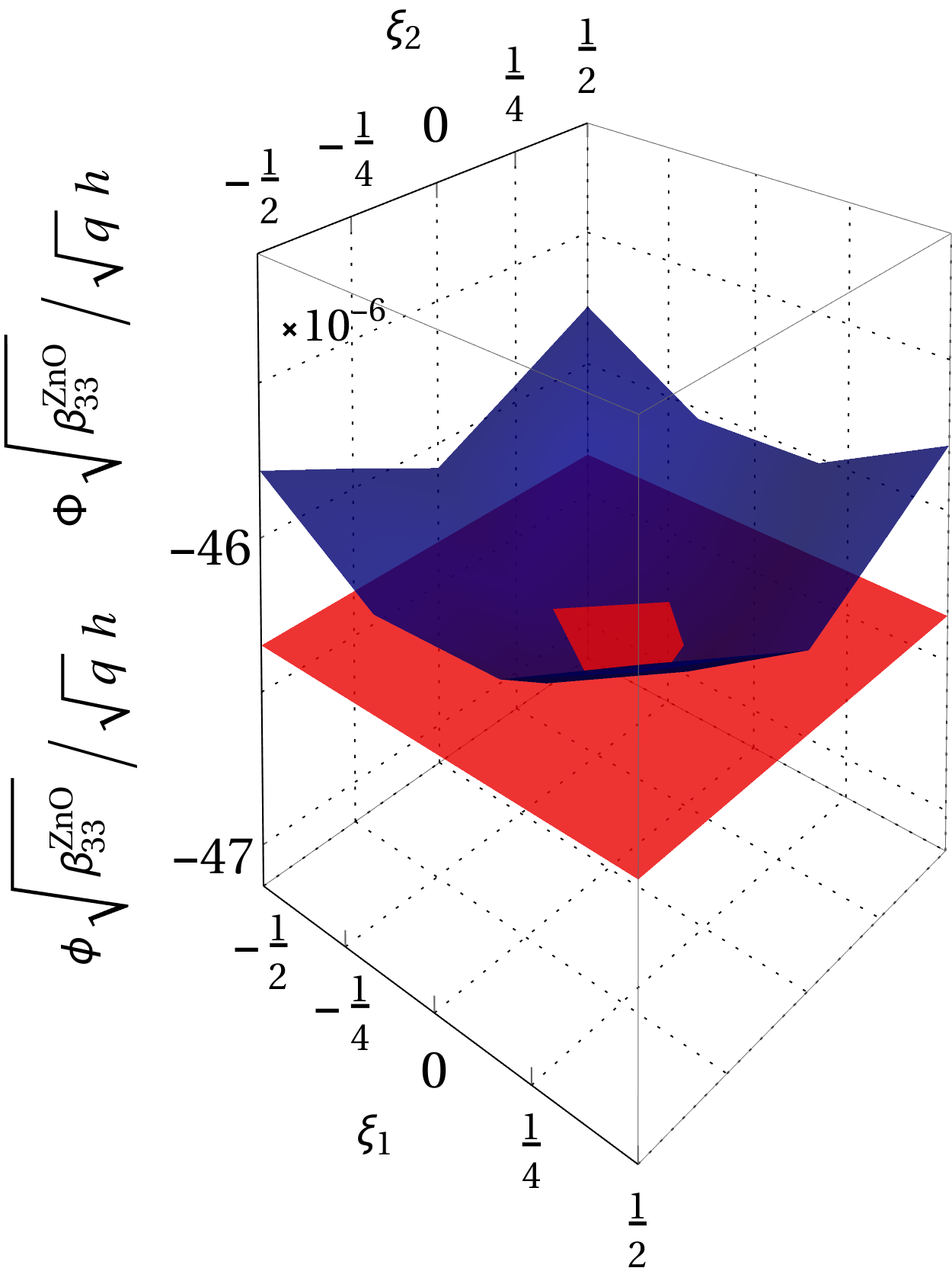}
 \put (70,92) {$(a)$}
\end{overpic}\qquad 
\begin{overpic}[scale=0.78]{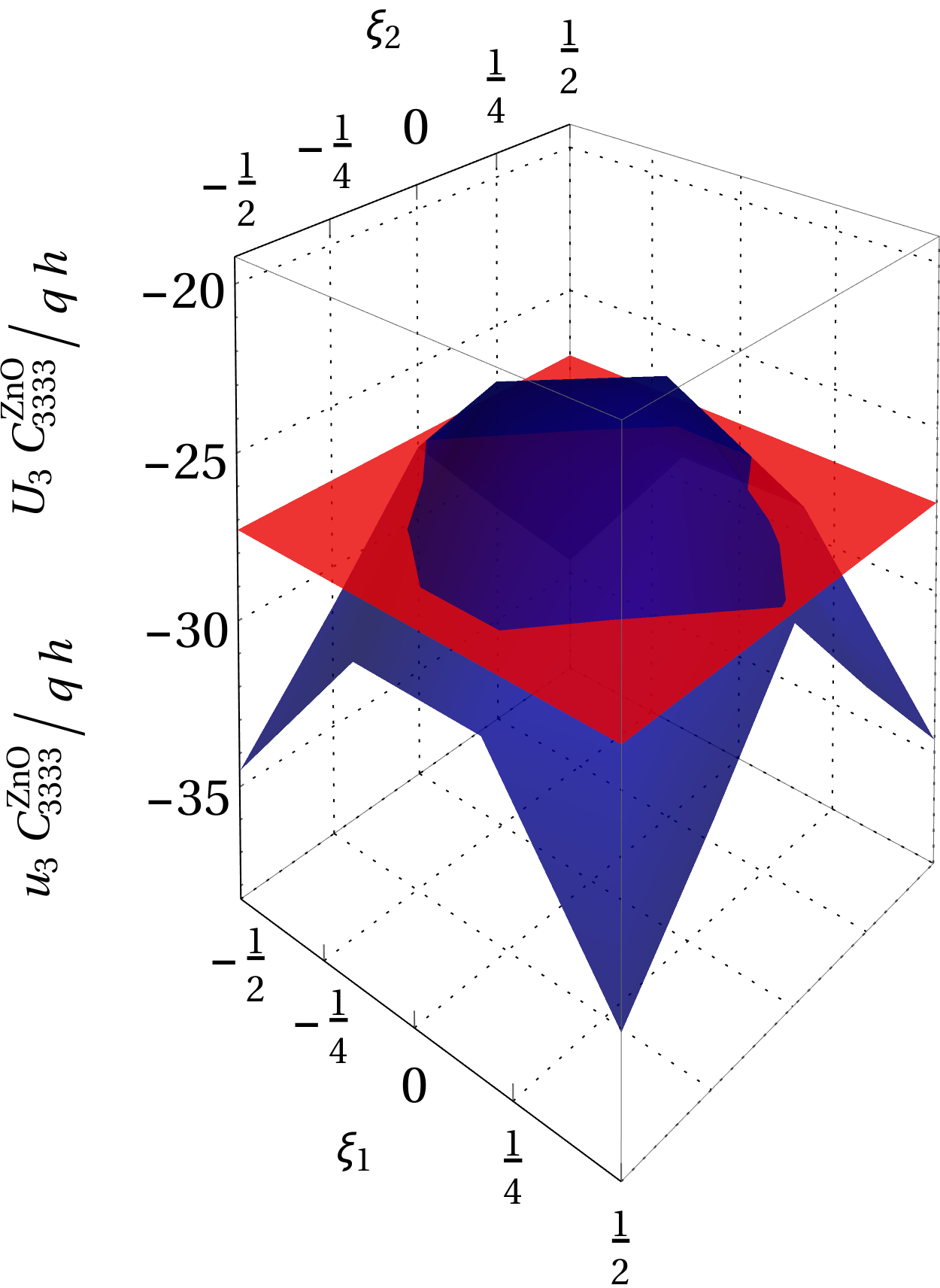}
 \put (68,92) {$(b)$}
\end{overpic}
\caption{Comparison between the heterogeneous model (blue surfaces) and the equivalent homogenized model (red surfaces). (a) Dimensionless  potential of the heterogeneous model  $ \phi\sqrt{\beta_{33}^{ZnO}}/ (\sqrt{q} h)$ and dimensionless potential of the homogenized model  $ \Phi\sqrt{\beta_{33}^{ZnO}}/ (\sqrt{q} h)$ along the top surface of the central Periodic Cell; (b) Dimensionless  displacement component of the heterogeneous model $u_3 C_{3333}^{ZnO}/ (q h)$ and dimensionless displacement component of the homogenized model $U_3 C_{3333}^{ZnO}/ (q h)$ along the top surface of the central Periodic Cell.}
\label{figure4}
\end{figure}

We study the microstructured nanogenerators under compressive loads along the nanorods  axis, considering a cluster of $n$ $\times$ $n$  cells (with base dimension $L$ and height $h$), obtained by repeating the Periodic Cell along the periodicity vectors $\textbf{v}_1$ and $\textbf{v}_2$,  see Figure \ref{figure2}(a) where a sketch is shown.   A uniform surface compressive load $q$ is applied on the top horizontal face of the specimen, while the  displacements of the points on the bottom face are restricted in $\textbf{e}_3$ direction. The  electrodes
are located along the top and bottom faces. Two alternative cases are considered, depending on the boundary conditions applied on the lateral faces. The $Case$ $1$ is characterized by free lateral faces, while in the $Case$ $2$ the displacements are restricted along the outward normal to the lateral surface. \\
In order to reduce the computational costs involved in micromechanical analyses of such  piezoelectric nanogenerators, their  macroscopic behaviour   can be satisfactorily  described via a first order equivalent homogeneous material. \\ 
The macroscopic field equations of the first order homogenized continuum, specialized to the considered material symmetries, and with zero source terms, result as
\begin{align}
&\frac{\partial}{\partial x_1} \left(C_{1111} \frac{\partial U_1}{\partial x_1}+C_{1122} \frac{\partial U_2}{\partial x_2}+C_{1133} \frac{\partial U_3}{\partial x_3} \right)+\frac{\partial}{\partial x_3} \left(C_{1313} \left(\frac{\partial U_3}{\partial x_1}+\frac{\partial U_1}{\partial x_3} \right) \right)+ \nonumber \\
&+\frac{\partial}{\partial x_2} \left(C_{1212} \left(\frac{\partial U_2}{\partial x_1}+\frac{\partial U_1}{\partial x_2} \right) \right)+ \frac{\partial}{\partial x_3} \left(e_{131} \frac{\partial \Phi}{\partial x_1}\right)+\frac{\partial}{\partial x_1} \left(e_{113} \frac{\partial \Phi}{\partial x_3}\right)=0,\label{13}\\ 
&\frac{\partial}{\partial x_2} \left(C_{2222} \frac{\partial U_2}{\partial x_2}+C_{2233} \frac{\partial U_3}{\partial x_3}+C_{1122} \frac{\partial U_2}{\partial x_2} \right)+ \frac{\partial}{\partial x_3} \left(C_{3232} \left(\frac{\partial U_3}{\partial x_2}+\frac{\partial U_2}{\partial x_3}\right)\right)+ \nonumber\\
&+\frac{\partial}{\partial x_1} \left(C_{1212} \frac{\partial U_1}{\partial x_2} \right)+\frac{\partial}{\partial x_2} \left(e_{223} \frac{\partial \Phi}{\partial x_3} \right)+\frac{\partial}{\partial x_3} \left(e_{322} \frac{\partial \Phi}{\partial x_2} \right)=0,\label{14}\\
&\frac{\partial}{\partial x_3} \left(C_{3333} \frac{\partial U_3}{\partial x_3}+C_{1133} \frac{\partial U_1}{\partial x_1}+C_{2233} \frac{\partial U_2}{\partial x_2}\right)+ \frac{\partial}{\partial x_2} \left(C_{3232} \left(\frac{\partial U_3}{\partial x_2}+\frac{\partial U_2}{\partial x_3}\right) \right)+\nonumber\\
&+\frac{\partial}{\partial x_1} \left(C_{1313} \left(\frac{\partial U_3}{\partial x_1}+\frac{\partial U_1}{\partial x_3}\right)\right)+\frac{\partial}{\partial x_3} \left(e_{333} \frac{\partial \Phi}{\partial x_3}\right)+\frac{\partial}{\partial x_2} \left(e_{322} \frac{\partial \Phi}{\partial x_2}\right)+\frac{\partial}{\partial x_1} \left(e_{131} \frac{\partial \Phi}{\partial x_1}\right)=0,   \label{15}\\ 
&\frac{\partial}{\partial x_3} \left(e_{113} \frac{\partial U_1}{\partial x_1}+e_{223} \frac{\partial U_2}{\partial x_2}\right)+\frac{\partial}{\partial x_2} \left(e_{322} \left(\frac{\partial U_2}{\partial x_3}+\frac{\partial U_3}{\partial x_2}\right)\right)+\frac{\partial}{\partial x_1}\left(e_{131} \left(\frac{\partial U_1}{\partial x_3}+\frac{\partial U_3}{\partial x_1}\right) \right)-\nonumber\\
&-\frac{\partial}{\partial x_1} \left(\beta_{11} \frac{\partial \Phi}{\partial x_1}\right)-\frac{\partial}{\partial x_2} \left(\beta_{22} \frac{\partial \Phi}{\partial x_2} \right)-\frac{\partial}{\partial x_3}\left(\beta_{33} \frac{\partial \Phi}{\partial x_3} \right)=0. \label{16}
\end{align}
Considering the $Case$ $1$, the boundary conditions are $U_1(x_1=- L/2)=0$, $U_2(x_2= -L/2)=0$, $U_3(x_3= -h/2)=0$, $\Phi(x_3=-h/2)=0$, $\Sigma_{11}(x_1=L/2)=0$, $\Sigma_{22}(x_2=L/2)=0$, $\Sigma_{33}(x_3=h/2)=-q$, $d_3(x_3=h/2)=0$.\\
The solution of this ODE problem takes the following form
\begin{align}
&U_1(x_1)=-q \left( 
2\,x_{{1}}+L \right) \Psi^{-1}\left[C_{1133}{e_{223}}^{2}-\left(C_{1122}e_{333}-C_{2233}e_{113} \right)e_{{223}}-C_{{1122}}C_{{2233}}\beta_{33}+\right. \nonumber\\ 
&\left.+C_{2222} \left( C_{1133}\beta_{33}+e_{113}e_{333} \right)  \right],\nonumber\\
&U_2(x_2)= q \left( 2
\,x_{{2}}+L \right) \Psi^{-1} \left[ -C_{{2233}}{e_{{113}}}^{2}+ \left( C_{{1,1112}}e
_{{333}}+C_{{1133}}e_{{223}} \right) e_{{113}}+C_{{1122}}C
_{{1133}}\beta_{{33}}- \right.\nonumber\\
& \left.+C_{{1111}} \left( C_{{2233}}\beta_{{33}}+e_{{223}}e_{{333}} \right)  \right],  \nonumber\\
&U_3(x_3)=q \left( 2\,x_{{3}}+h \right)\Psi^{-1}\,   \left[ -{C_{{1122}}}^{2}\beta_{{33}}-2\,C_{{1122}}e_{{113}}e_{{223}}+C_{{2222}}{e_{{113}}}^{2}+C_{{1111}} \left( C_{{2222}}\beta_{{33}}+{e_{{223}}}^{2} \right)  \right], \nonumber\\
&\Phi(x_3)=q \left( 
2\,x_{{3}}+h \right) \Psi^{-1}\left[ {C_{{1122}}}^{2}e_{{333}}+ \left( -C_{{1133}}
e_{{223}}-C_{{2233}}e_{{113}} \right) C_{{1122}}+C_{{1133
}}C_{{2222}}e_{{113}}- \right. \nonumber\\
& \left.C_{{1111}} \left( C_{{2222}}e_{{333
}}-C_{{2233}}e_{{223}} \right)  \right], 
\end{align}
where the constant $\Psi$ takes the form
\begin{align}
&\Psi=-\left[ 2\,C_{3333}\beta_{33}+2\,{e_{333}}^{2} \right){C_{1122}}^{2}+\left( \left( 4\,C_{2233}\beta_{33}+4\,e_{223}e_{333} \right) C_{1133}+ \right. \nonumber\\
& \left.-4\,e_{113} \left( C_{2233}e_{333}-C_{3333}e_{223} \right) \right) C_{1122}-2\left( \,C_{2222}\beta_{33}+\,{e_{223}}^{2} \right) {C_{1133}}^{2}+ \nonumber\\
&-4\,e_{113} \left( C_{2222}e_{333}-C_{2233}e_{223} \right) C_{1133}-2\left( \,C_{1111}\beta_{33}+\,{e_{113}}^{2} \right) {C_{2233}}^{2 }- \nonumber\\
&+4\,C_{1111}C_{2233}e_{223}e_{333}+2\,C_{2222} C_{3333}{e_{113}}^{2}+ \nonumber \\
&+2\, \left( C_{3333}{e_{223}}^{ 2}+C_{2222} \left( C_{3333}\beta_{33}+{e_{333}}^{2} \right) \right] C_{{1111}}.
\end{align}
\noindent The strain field and the electric field are consistently derived.  The electric displacement field is identically zero, while the only non-vanishing component of the resulting stress field is $\Sigma_{33}=-q$.\\
\noindent Concerning, instead, the $Case$ $2$, the new set of boundary conditions is 
$U_1(x_1=\pm L/2)=0$, $U_2(x_2= \pm L/2)=0$, $U_3(x_3= -h/2)=0$,   $\Phi(x_3=-h/2)=0$, $\Sigma_{33}(x_3=h/2)=-q$, $d_3(x_3=h/2)=0$.
The corresponding solution of the ODE problem is
\begin{align}
&U_1(x_1)=0,\nonumber\\
&U_2(x_2)= 0,  \nonumber\\
&U_3(x_3)=-\frac{q}{2} \left( 2\,x_{{3}}+h \right) \Xi^{-1} \beta_{33}, \,   \nonumber\\
&\Phi(x_3)=-\frac{q}{2}  \left( 2\,x_{{3}}+h \right) \Xi^{-1} e_{333}, 
\end{align}
with the constant $\Xi$ expressed in the form
\begin{align}
&\Xi=C_{3333}\beta_{33}+e^2_{333}.
\end{align}
The components of the electric displacement field are equal to zero, and the non-vanishing components of the stress field are
\begin{align}
&\Sigma_{11}(x_1)=-q \Xi^{-1} \left( C_{1133}\, \beta_{33}+e_{113}\, e_{333}\right),\nonumber\\
&\Sigma_{22}(x_2)= -q \Xi^{-1} \left( C_{2233}\, \beta_{33}+e_{223}\, e_{333}\right),  \nonumber\\
&\Sigma_{33}(x_3)=-q. 
\end{align}
\noindent The macroscopic fields, analytical determined, are compared against numerical results obtained with a micro-mechanical 
model. To this aim, a convergence study is performed considering, at the microscopic scale, clusters of increasing dimensions, in order to detect the minimum size of the cluster such that the results related to the central cell remain almost unchanged. We observe that already for a cluster of 5 $\times$ 5 satisfactory results are obtained, so that the numerical results presented below are referred to this minimum size cluster geometry.\\
A first investigation is referred to the behaviour of the nanogenerators as the volume fraction of PANI nanoscopic particles varies within the matrix and the top and bottom layers, i.e. for dielectric permittivity $\varepsilon_r^{P/PANI}$ ranging from 5 to 10000.
We, indeed, assume that the matrix and the layers are characterized by very high values of dielectric constants, varying with the volume fraction of PANI nanoscopic particles dispersed in the polymeric matrix. In Figure \ref{figure3}(a) the macroscopic dimensionless potential differences $\Delta \Phi\sqrt{\beta_{33}^{ZnO}}/ (\sqrt{q} h)$ obtained from the heterogeneous model via up-scaling relations (solid plots) and from the equivalent homogenized model (dotted plots) are shown. Two densities are considered for both $Case$ $1$ and $Case$ $2$. In particular,  blue and green curves refer to $Case$ $1$ and $Case$ $2$, respectively, considering a density $\delta$=0.415, while red and black curves
refer to $Case$ $1$ and $Case$ $2$, respectively, considering a density $\delta$=0.104. We observe that the macroscopic analytical solutions  are in very good agreement with the macroscopic numeric results obtained via up-scaling from the heterogeneous model. A non-linear trend is observed, with potential differences increasing as the the volume fraction of PANI decreases. Better performances are obtained for low values of 
$\varepsilon_r^{P/PANI}$. Note that the maximum values of potential differences are observed for the $Case$ $1$ with lower density. For both the densities,  the $Case$ $1$ exhibits better performances.\\
 A further investigation 
is devoted to study the influence of the height of the nanorods, considering the parameter $\alpha=h/h^*$, as the height $h$ varies between 550 nm and 2200 nm, with $h^*$=1100 nm.
In Figure \ref{figure3}(b) the macroscopic dimensionless potential differences $\Delta \Phi\sqrt{\beta_{33}^{ZnO}}/ (\sqrt{q} h^*)$ are, thus, plotted versus $\alpha$. Also in this plot, 
blue and green curves refer to $Case$ $1$ and $Case$ $2$, respectively, considering a density $\delta$=0.415, while red and black curves
refer to $Case$ $1$ and $Case$ $2$, respectively, considering a density $\delta$=0.104.
The solid lines refer to the solution of the micromechanical problems, while the dotted lines are the analytical solutions. Again, a nearly perfect match is observed. In this case, nearly linear variations are found as $\alpha$ varies. As expected, higher values of nanorods' heights are associated with better performances. Also in this case, the $Case$ $1$ is more effective than  $Case$ $2$ irrespective of the considered density.\\
\noindent A more comprehensive investigation concerns the comparison between dimensionless potentials and dimensionless displacement components, evaluated along the top surface of the central Periodic Cell in the Cluster, adopting the heterogeneous model and the corresponding quantities derived from the first order homogenized model.
 In particular, in Figure \ref{figure4}(a) the dimensionless macroscopic  and microscopic electric potentials, $ \Phi\sqrt{\beta_{33}^{ZnO}}/ (\sqrt{q} h)$ and $ \phi\sqrt{\beta_{33}^{ZnO}}/ (\sqrt{q} h)$, respectively, evaluated along the top surface of the central Periodic Cell, are  plotted, confirming the effectiveness of the homogenized model in predicting the averaged behaviour of the heterogeneous model.  Finally, in Figure \ref{figure4}(b), both the dimensionless macroscopic  and the microscopic  displacement components, $U_3 C_{3333}^{ZnO}/ (q h)$  and $u_3 C_{3333}^{ZnO}/ (q h)$, respectively, are compared with each other, again in good agreement. \\
 
\section{Applications of the asymptotic piezoelectric homogenization to microstructured nanogenerators}
In this Section, the behaviour of three piezoelectric  microstructured nanogenerators, with different working principles, is investigated as a set of geometrical parameters changes.  Both extension and bending nanogenerators are taken into account, considering either extension along the nanorods axis, or orthogonally to it. The influence of either the height of nanorods and their density on the overall behaviour of such devices is analysed.
These geometric parameters  can be, indeed,
properly modified during the synthesis process of the ZnO-nanorods cluster,  in order to tune the overall piezoelectric response and obtain improved performances, to possibly orient a final user towards optimal design choices.
\\
In Subsection \ref{Pmen},   microstructured extension nanogenerators are investigated. In this case, mechanical loads act along the nanorods axis and the related potential differences are evaluated. Moreover, in Subsection \ref{Pmbn}, 
 microstructured  bending nanogenerators
are analysed. The bending behaviour is induced by properly defined boundary conditions. Finally, in Subsection \ref{Pmten},  microstructured transversal extension nanogerators are taken into account. The extension/contraction behaviour  is due the adoption of appropriate boundary conditions. 
In all the considered cases, the homogenized equivalent response of the device at hand is examined. Finite elements analyses have been performed 
adopting fully coupled tetrahedral second order elements with displacements and electric potential  
independent degrees of freedom. Note that we are particularly interested in determining the potential difference $\Delta \Phi $ measured between the top and bottom faces of the nanogenerator.\\

\subsection{Microstructured extension nanogenerators} \label{Pmen}
We consider a microstructured device made of a cluster of 5$\times$5 Periodic Cells, with base dimension $L$ and height $h$. Electrodes are present on the upper and lower surfaces the device. The heterogeneous material is ideally replaced by a homogeneous equivalent piezoelectric first order continuum, obtained by exploiting the proposed asymptotic homogenization approach, see Figure \ref{GEOMMulti5} where a sketch is shown.
\begin{figure}[hbtp]
  \centering
  \begin{overpic}[scale=0.6]{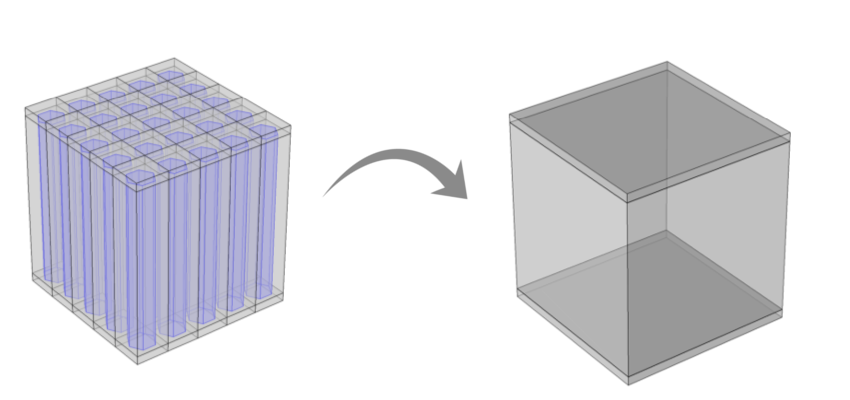}
 \put (7,46) {Microstructured}
  \put (5,42) {extension generator}
   \put (63,46) {First order equivalent}
  \put (72,42) {material}
    \put (5,7) {$L$}
   \put (86,3) {$L$}
       \put (0,22) {$h$}
   \put (94,18) {$h$}
    \put (36,31) {$homogenization$}
\end{overpic}
      \caption{ Two-scales description of the extensional nanogenerator made by a cluster of 5$\times$5 Periodic Cells.}
    \label{GEOMMulti5}
\end{figure}
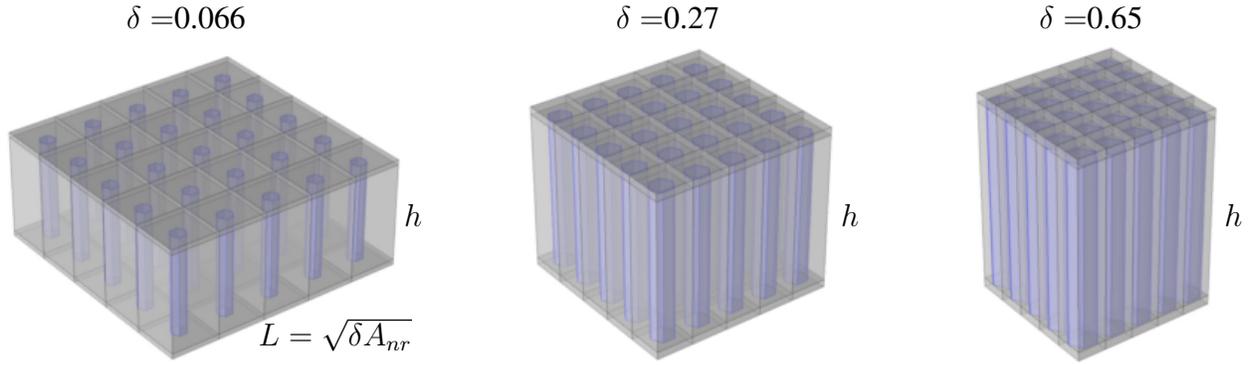
\begin{figure}[hbtp]
  \centering
  \begin{overpic}[width=1\textwidth]{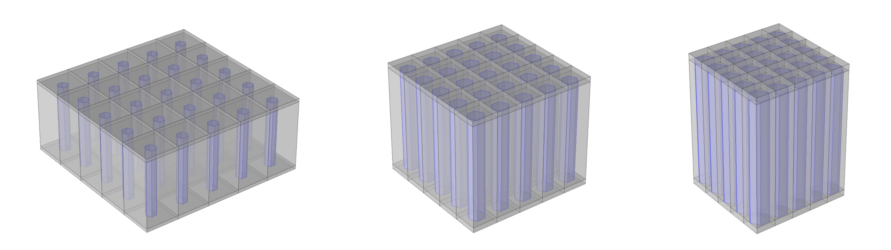}
       \put (34,12) {$h$}
     \put (67,12) {$h$}
      \put (96,12) {$h$}
     \put (23,3) {$L=\sqrt{\delta A_{nr}}$}      
     \put (13,27) {$\delta=$0.066} 
    \put (50,27) {$\delta=$0.27} 
     \put (82,27) {$\delta=$0.65} 
\end{overpic}
\caption{Schematic of the extensional nanogenerator as the density $\delta$ varies, while the height of the ZnO nanorods is kept constant. }
    \label{31}
\end{figure}
\noindent Two loading conditions are alternatively taken into account, i.e. a uniform surface load, and a load with fixed resultant, both applied on the top horizontal face. Moreover,  the multiscale analyses are performed assuming either free ($Case$ $1$) or restrained ($Case$ $2$) lateral vertical faces of the cluster, according to what proposed in the previous Section \ref{BT}.  \\
The initial investigation relates the influence of the nanorods density $\delta$ on the overall piezoelectric response of the devices. 
 In Figure \ref{31}, a set of clusters made of 5 $\times$ 5 Periodic Cells, corresponding to various densities is reported. \\
\begin{figure}[hbtp]
\centering
\begin{overpic}[scale=1]{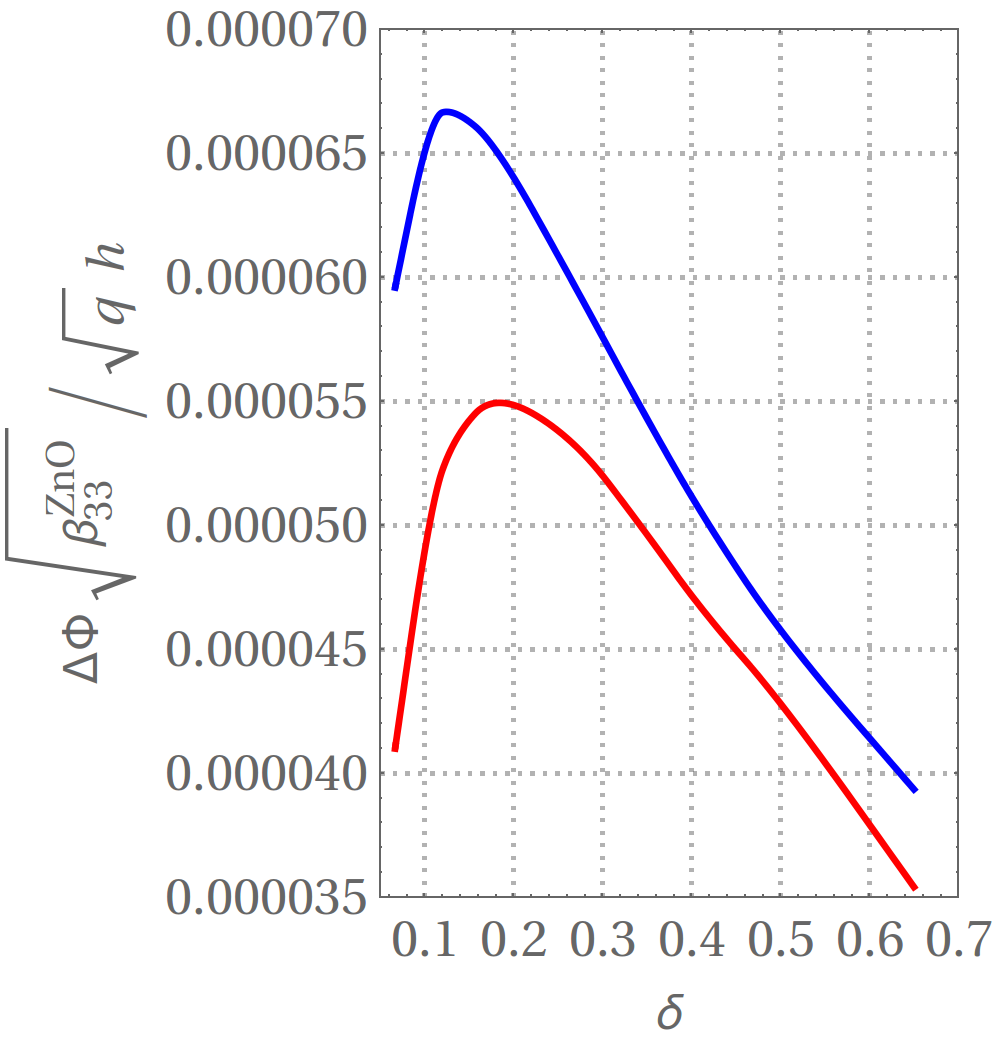}
 \put (80,90) {$(a)$}
\end{overpic}\qquad
\begin{overpic}[scale=1]{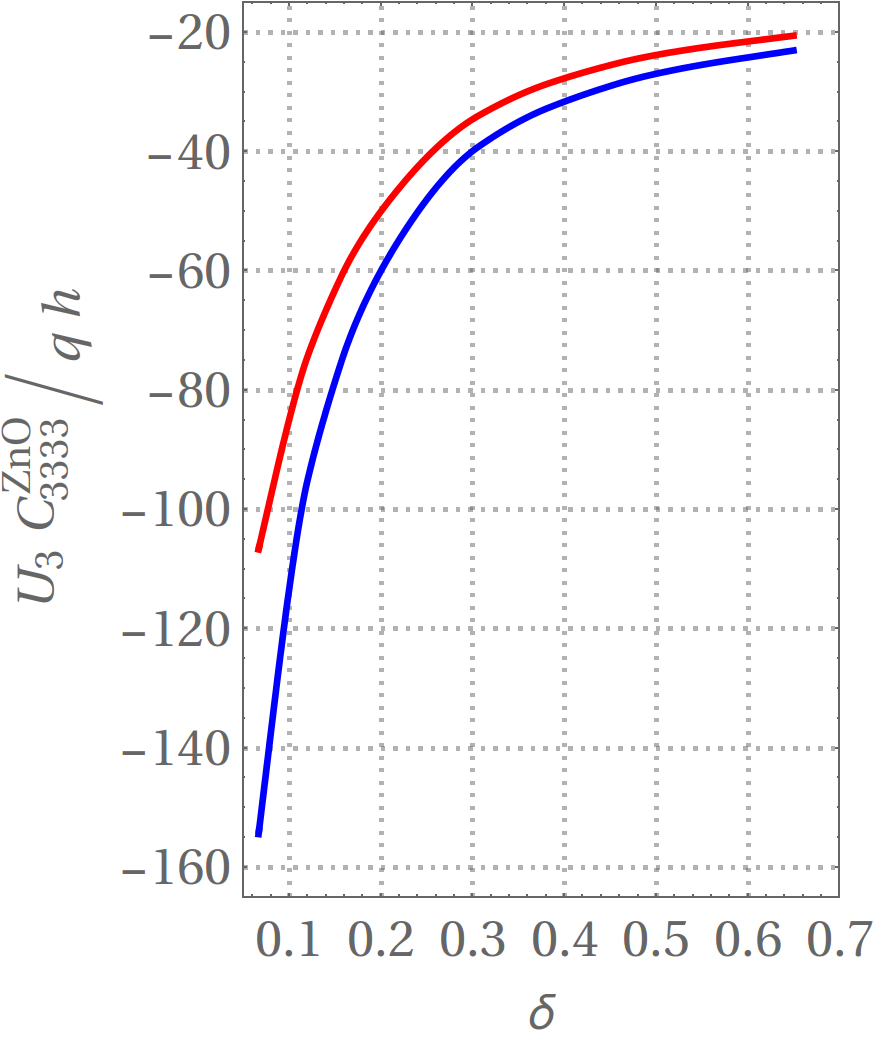}
 \put (70,90) {$(b)$}
\end{overpic}
\caption{Uniform surface load: (a) dimensionless potential difference $\Delta \Phi\sqrt{\beta_{33}^{ZnO}}/ (\sqrt{q} h)$ versus $\delta$; (b) dimensionless  displacement component $U_3 C_{3333}^{ZnO}/ (q h)$ versus $\delta$. Blue curves refer to the case with free lateral faces, while red curves to the case with fixed lateral faces.}
\label{DvULoad}
\end{figure}
\begin{figure}[hbtp]
\centering
\begin{overpic}[scale=1]{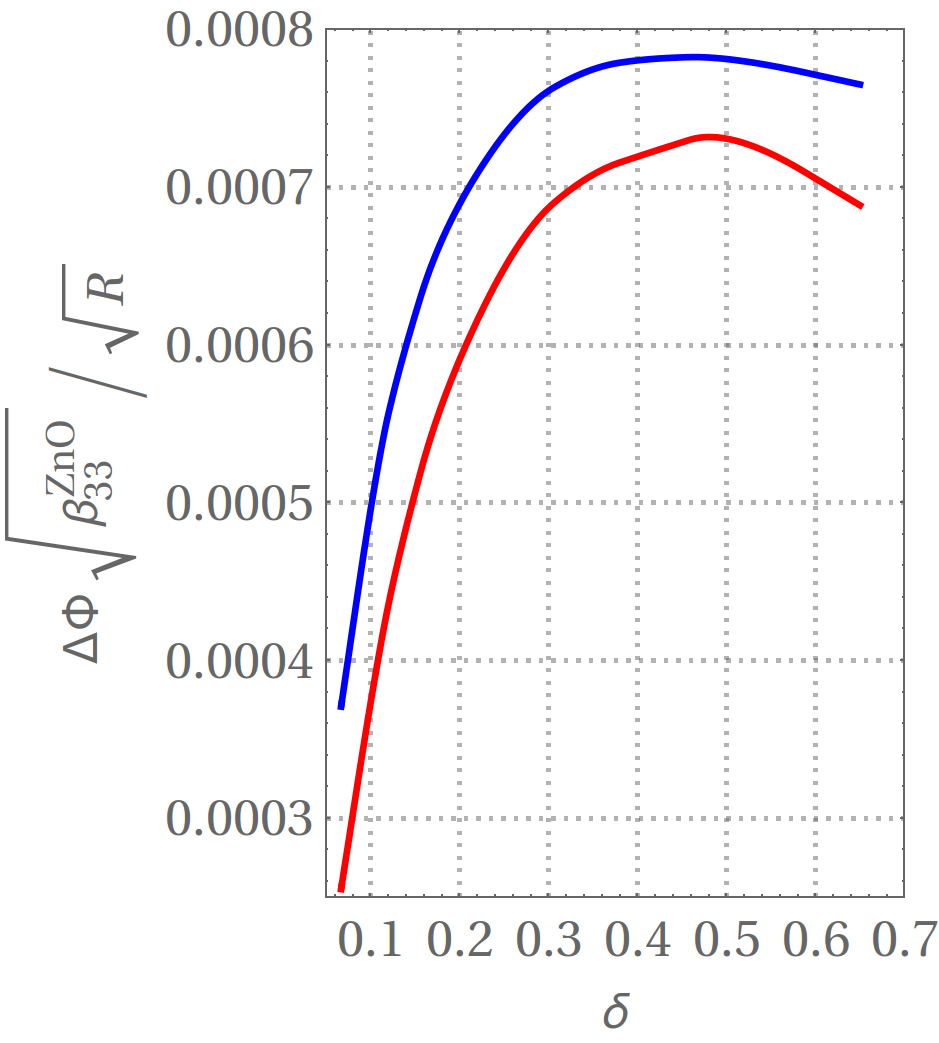}
 \put (75,87) {$(a)$}
\end{overpic}\qquad
\begin{overpic}[scale=1]{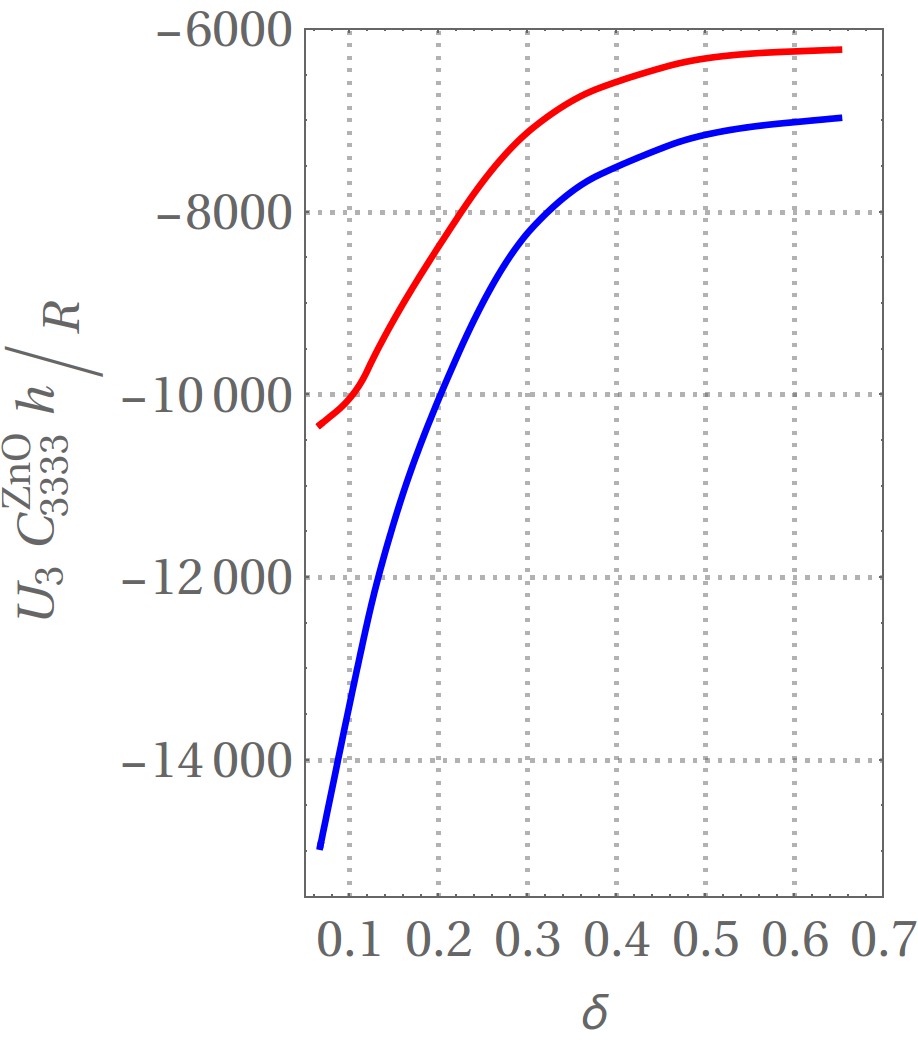}
 \put (75,90) {$(b)$}
\end{overpic}
\caption{Load with fixed resultant: (a) dimensionless potential difference $\Delta \Phi\sqrt{\beta_{33}^{ZnO}}/ \sqrt{R} $ versus $\delta$; (b) dimensionless  displacement component $U_3 C_{3333}^{ZnO}h/ R$ versus $\delta$. Blue curves refer to the case with free lateral faces, while red curves to the case with fixed lateral faces.}
\label{DvURES}
\end{figure}
\noindent We first consider the uniform surface load. In Figure \ref{DvULoad}, the overall response of the nanogenerators is reported in terms of averaged values, referred to the central Periodic Cell of the cluster, of the dimensionless potential difference and  dimensionless vertical displacements versus $\delta$ are reported. In particular, in Figure \ref{DvULoad}(a) the dimensionless values of $\Delta \Phi\sqrt{\beta_{33}^{ZnO}}/ (\sqrt{q} h)$ where $\beta_{33}$ is the $33$ component of the permittivity matrix of the bulk Zinc Oxyde material, are shown versus $\delta$. The blue curve refers to $Case$ $1$, while the red curve to $Case$ $2$. It is interesting to notice that the optimal behaviour, that maximizes the potential difference of the device, is achieved for different values of $\delta$ considering either $Case$ $1$ or $Case$ $2$. The maximum values of the curves correspond to small $\delta$ in the range between 0.1 and 0.2. Moreover, in Figure \ref{DvULoad}(b) the dimensionless values of $U_3C_{3333}^{ZnO}/ (q h)$, are shown as $\delta$ varies. As expected, the vertical displacements of the central cell increase as the density decreases. The sensor with free faces ($Case$ $1$) is more flexible than the other. Similarly, the case where the resultant load is kept constant on clusters of variable densities is considered. In Figure \ref{DvURES}, the overall response of the sensors in terms of averaged values, referred to the central cell, of $\Delta \Phi$ and   $U_3$ versus $\delta$ are plotted. 
In this case the normalization is performed considering the load resultant $R$. Referring to Figure \ref{DvURES}(a), it is confirmed  that the sensor with free faces (blue curve) is characterized by higher sensitivity (better performances). Nevertheless, the maximum of the curves is achieved for values of $\delta$ in the range between 0.4 and 0.5, that is shifted to the right with respect to the previous case. Also considering the vertical displacements in Figure \ref{DvURES}(b), again the blue curve (sensor with free lateral faces) is always below the red one.\\
\begin{figure}[hbtp]
\centering
\begin{overpic}[scale=1]{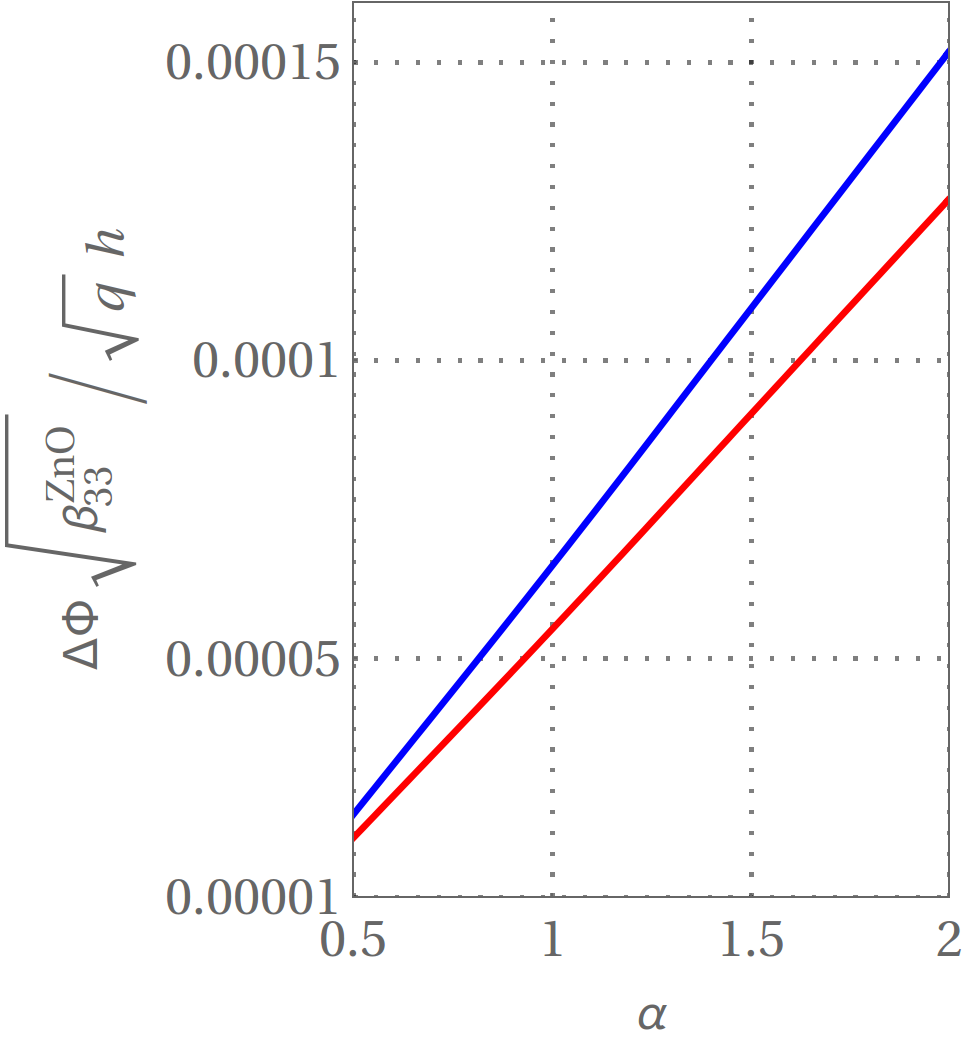}
 \put (75,92) {$(a)$}
\end{overpic}\qquad
\begin{overpic}[scale=1]{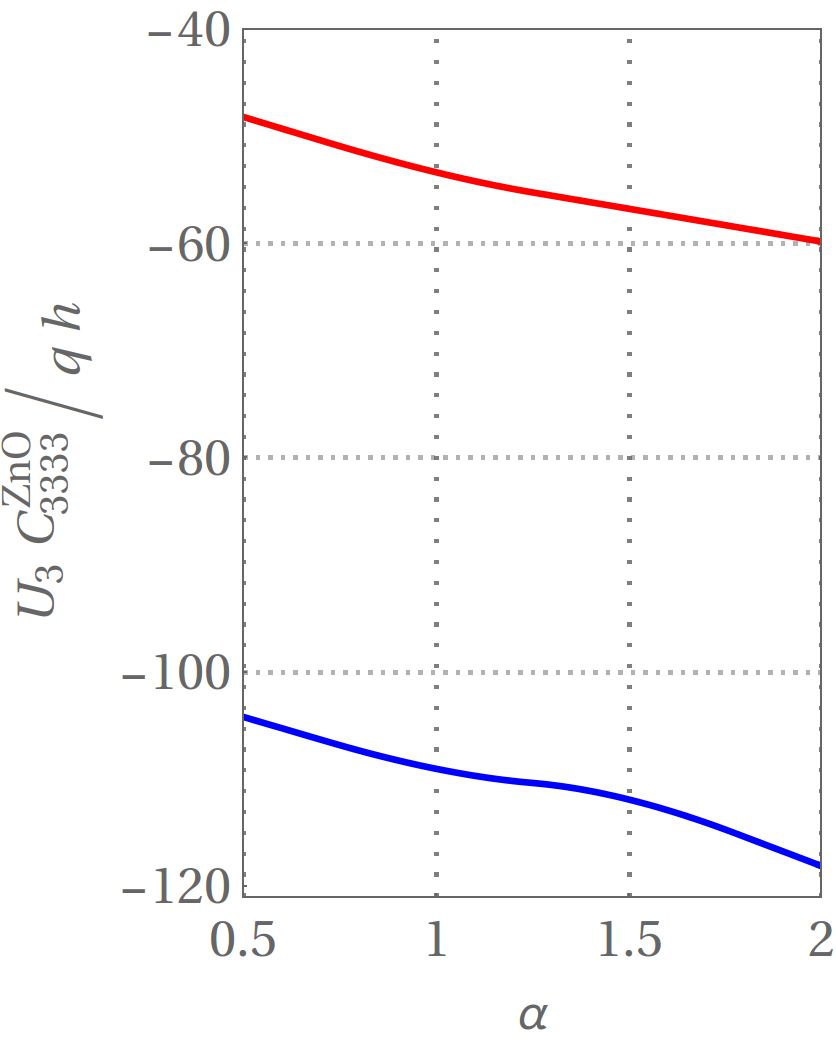}
 \put (70,92) {$(b)$}
\end{overpic}
\caption{Uniform surface load: (a) dimensionless potential difference $\Delta \Phi\sqrt{\beta_{33}}/ (\sqrt{q} L)$ versus $\alpha$; (b) dimensionless  displacement component $U_3 C_{3333}/ (q L)$ versus $\delta$. Blue curves refer to the case with free lateral faces, while red curves to the case with fixed lateral faces.}
\label{Loadalfa}
\end{figure}
\begin{figure}[hbtp]
\centering
\begin{overpic}[scale=1]{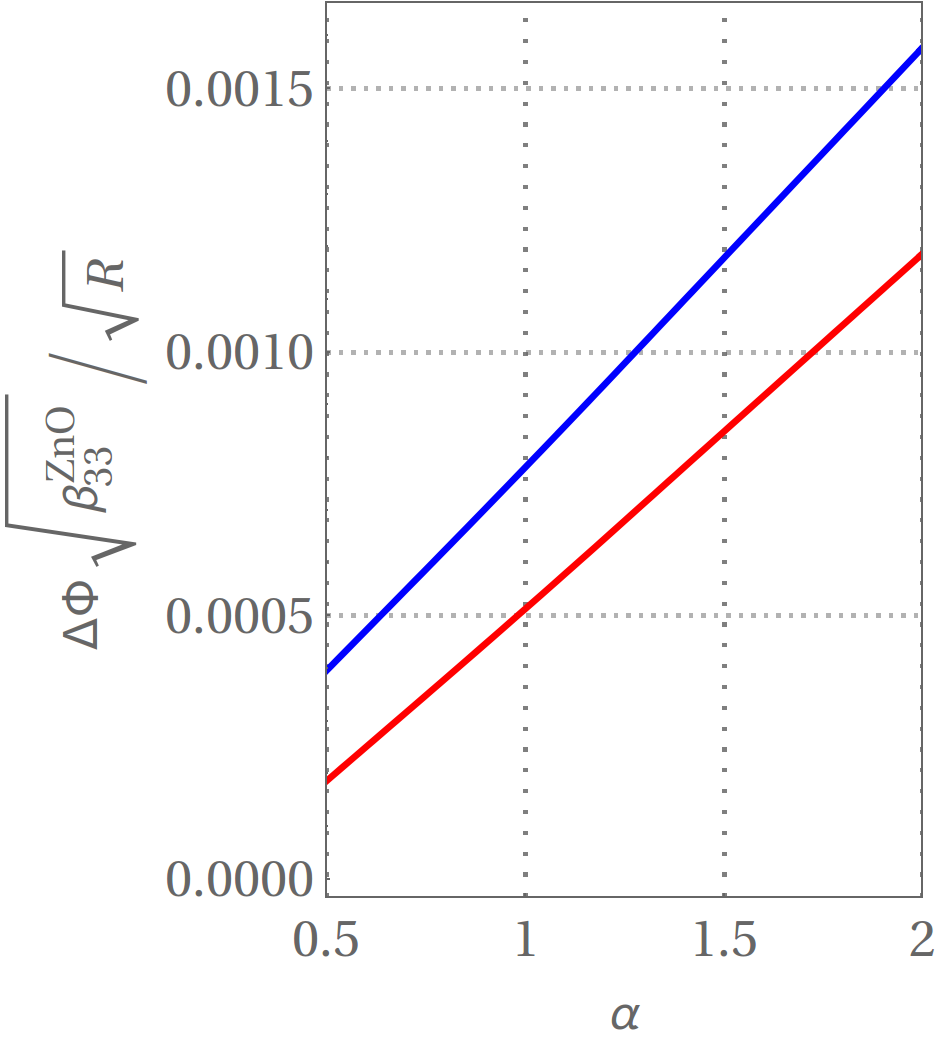}
 \put (75,92) {$(a)$}
\end{overpic}\qquad
\begin{overpic}[scale=1]{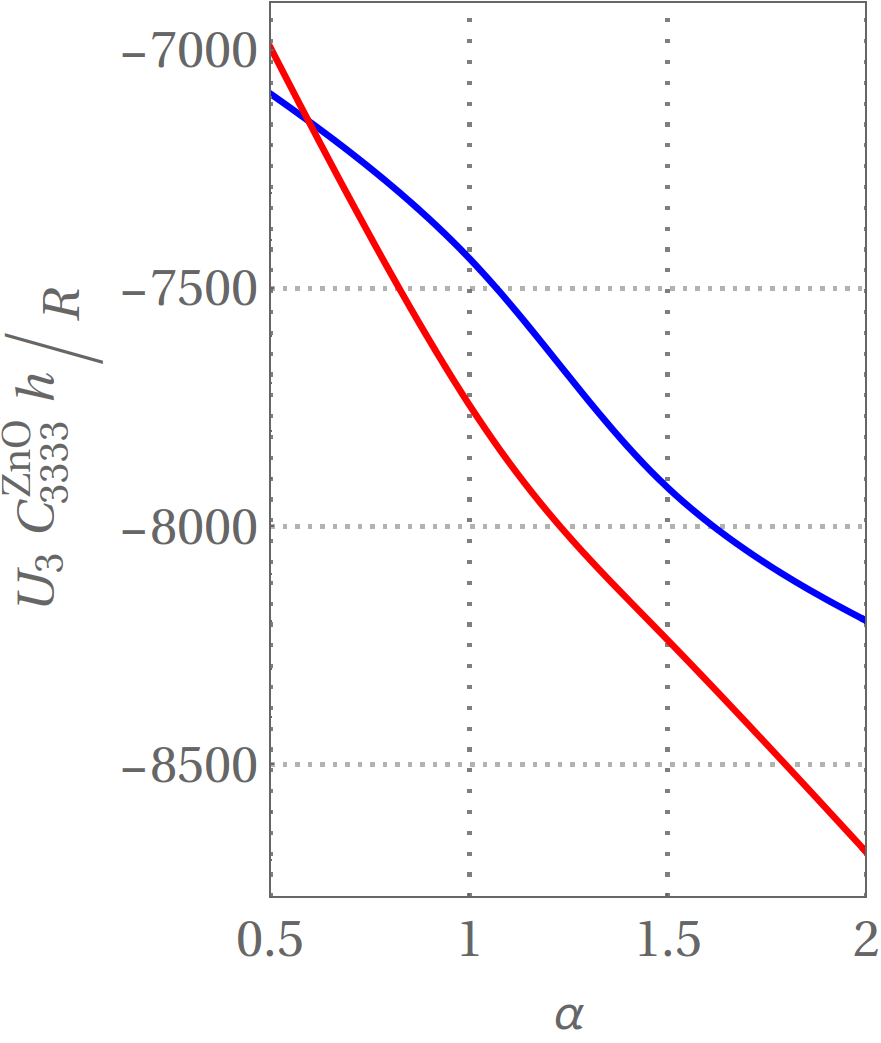}
 \put (75,92) {$(b)$}
\end{overpic}
\caption{Load with fixed resultant: (a) dimensionless potential difference $\Delta \Phi\sqrt{\beta_{33}^{ZnO}}/ \sqrt{R} $ versus $\alpha$; (b) dimensionless displacement component $U_3 C_{3333}/ R$ versus $\delta$. Blue curves refer to the case with free lateral faces, while red curves to the case with fixed lateral faces.}
\label{Resalfa}
\end{figure}
\noindent Furthermore, the influence of the height $h$, directly related to the the nanorods length, is investigated. Considering again both the cases with uniform surface load, and load with fixed resultant, the response of the sensors for $Case$ $1$ and $Case$ $2$ is investigated. For each of the four analysed cases, 
the density $\delta$ that maximize the corresponding responses in Figures \ref{DvULoad}(a)  and \ref{DvURES}(a) is taken into account. Starting with uniform surface load, in Figure \ref{Loadalfa} the response in terms of averaged values of $\Delta \Phi$ and $U_3$ versus $\alpha$ are reported. The dimensionless value $\alpha$ is let varying between 0.5 and 2 and  is equal to the ratio between the actual height of the device $h$ and a reference value of $h^*$=1100 nm. 
As expected, a nearly linear variation of $\Delta \Phi\sqrt{\beta_{33}}/ (\sqrt{q} h)$, as $\alpha$ varies, is found, see Figure \ref{Loadalfa} (a). In the case with free faces, blue curves, higher values of both potential difference and vertical displacements are observed. Finally,  the results for the load with fixed resultant are shown in Figure \ref{Resalfa}. Again a linear variation of the dimensionless potential difference versus $\alpha$ is found. On the other hand, concerning the vertical displacements, a nearly linear variation is observed. It is confirmed that the microstructured extension nanogenerators with free faces provide the better response in terms of potential difference, and are more flexible.

\subsection{ Microstructured  bending nanogenerators}\label{Pmbn}

\noindent A key point of the hybrid flexible nanogenerators is the possibility to exploit flexural mechanisms, related to the bending of the base support, to ensure energy harvesting. 
\begin{figure}[t]
  \centering
      \begin{overpic}[width=0.7\textwidth]{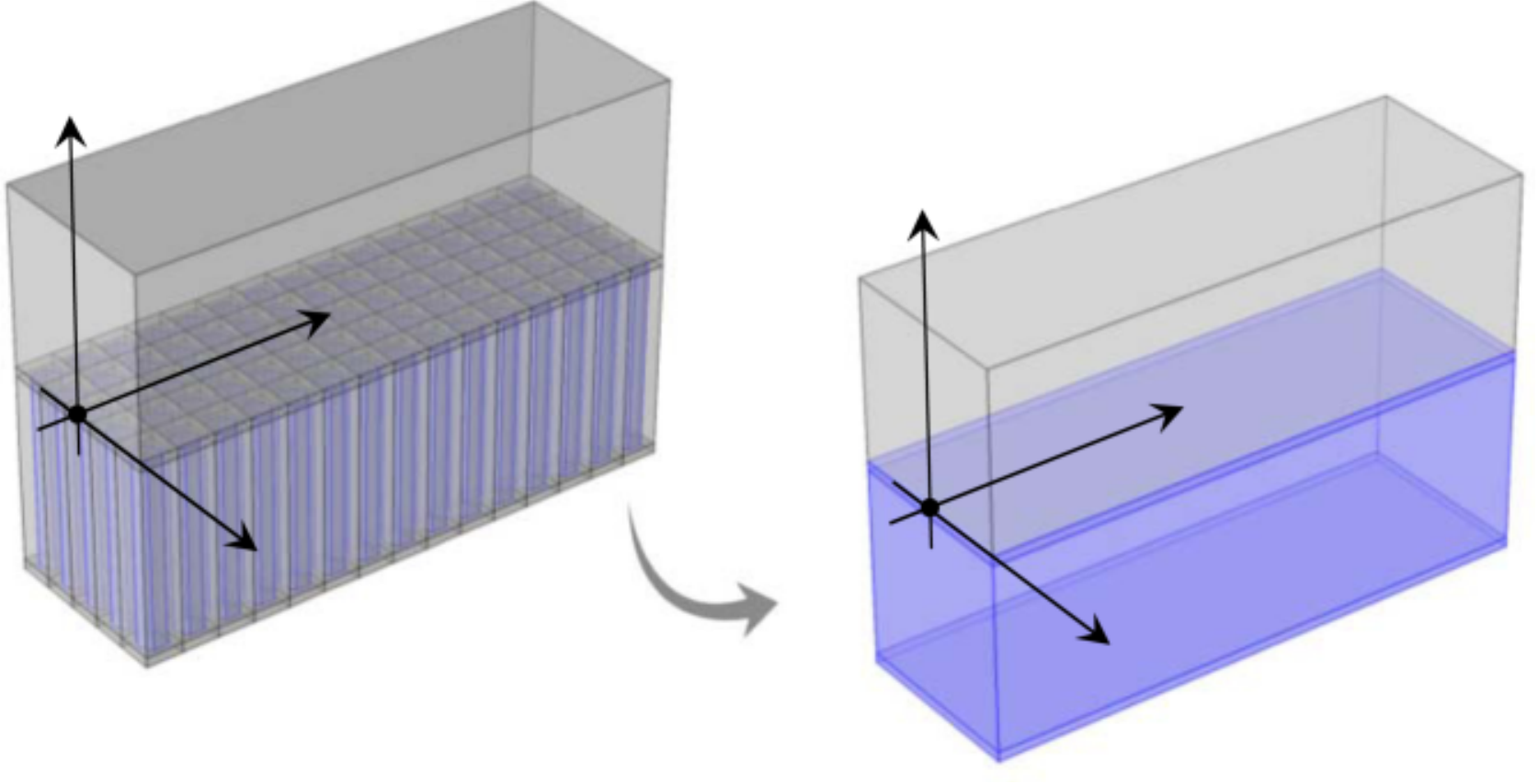}
 \put (98,54) {$(a)$}
  \put (1,56) {Microstructured}
  \put (1,52) { nanogenerator}
   \put (58,50) {Two-phase composite}
  \put (60,46) {equivalent material}
    \put (2,8) {$L$}
      \put (26,11) {3$L$}
   \put (83,5) {3$L$}
       \put (-2,19) {$h$}
       \put (-3,31) {$h_t$}
       \put (99,20) {$h$}
       \put (99,32) {$h_t$}
    \put (30,6) {$homogenization$}
       \put (12,13) {$x_1$}
       \put (20,27) {$x_2$}
        \put (0,42) {$x_3$}
       \put (68,7) {$x_1$}
       \put (75,21) {$x_2$}
        \put (55,35) {$x_3$}
\end{overpic}
      \begin{overpic}[width=0.4\textwidth]{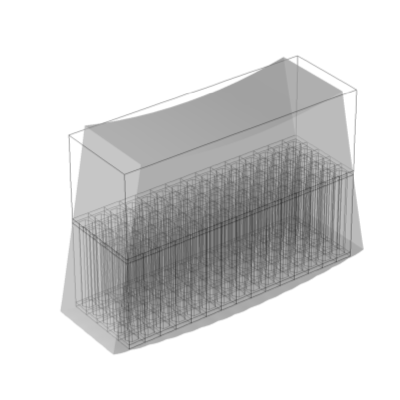}\
 \put (3,8) {$(b)$}
\end{overpic},\,\,\,\,\,\,\,
      \begin{overpic}[width=0.4\textwidth]{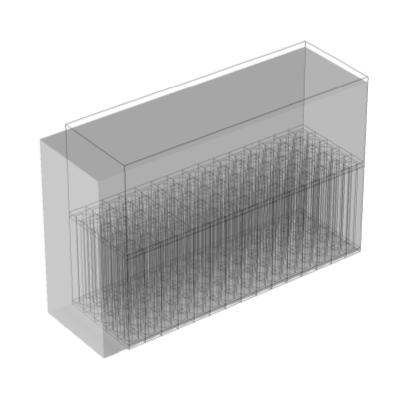}
 \put (0,8) {$(c)$}
\end{overpic}
      \caption{(a) Two-scales description of the  microstructured nanogenerator made by a cluster of 15$\times$5 Periodic Cells with a top polymeric layer of thickness $h_t$; (b) qualitative deformed shape of the microstructured bender nanogenerator; (c) qualitative deformed shape of the microstructured transverse extensional nanogenerator.}
    \label{GEOMMultiBB}
\end{figure}
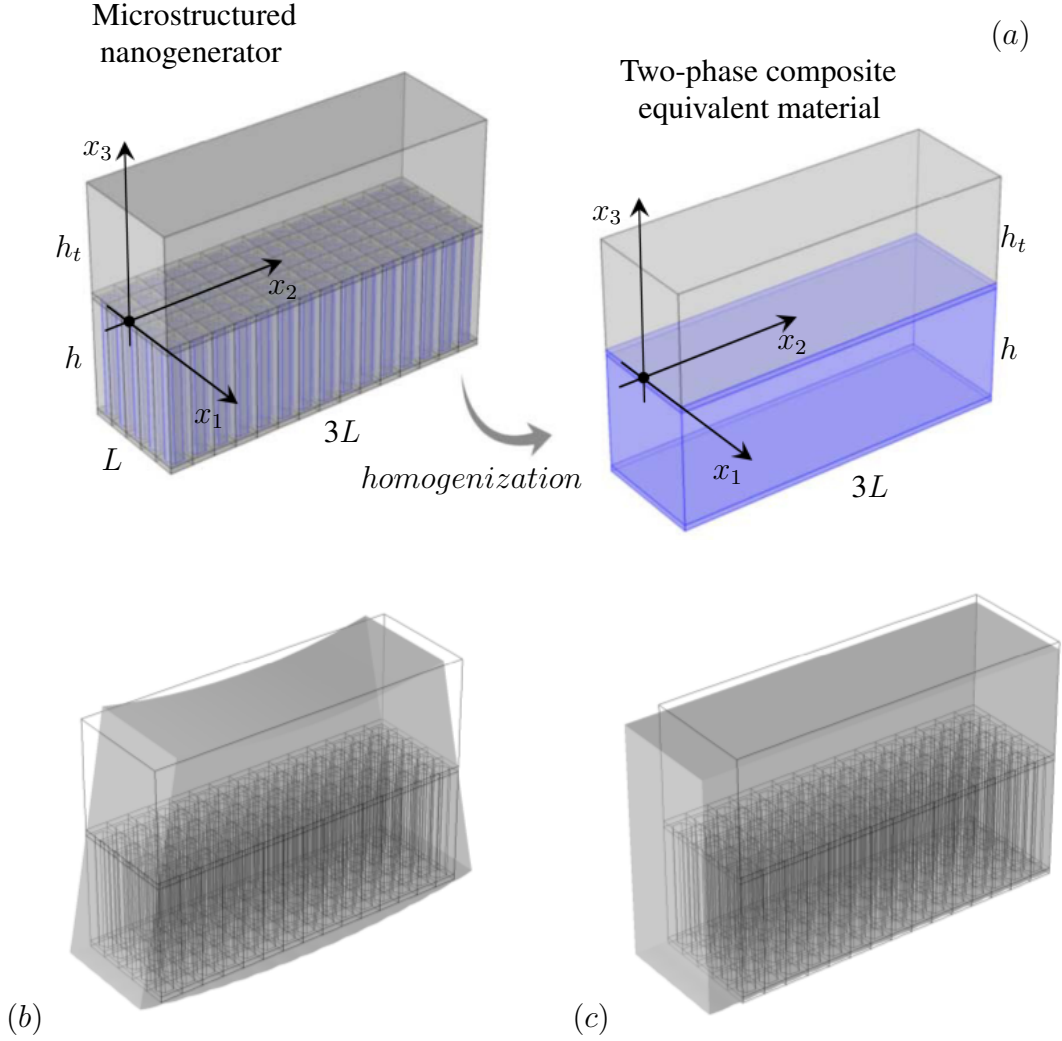
This peculiar behaviour is particularly emphasized in such innovative flexible devices, differently from what happen with standard ZnO-nanorods  nanogenerator, provided with rigid supports, in which this behaviour is inhibited.
Therefore, in this Section, the bending behaviour of such  hybrid flexible nanogenerators  is investigated and critically commented.\\
We consider devices made of a cluster of periodic cells topped by a further  layer of electroactive polymeric material, see Figure \ref{GEOMMultiBB}(a) where a schematic of the microstructured nanogenerator is depicted.
Adopting this particular geometry, we are interested 
in investigating how the thickness of the top layer influences the overall behaviour of the bending nanogenerator,  inspired by the work by Opoku and co-authors,  \citep{OPOKU2015858}. Two electrodes are located at the top and bottom faces of the microstructured device.\\
This problem is solved via a multi-scale approach, suitable  to reduce high computational costs related to the adoption of micromechanical models. In particular, the cluster of periodic cells is preliminary homogenized into an equivalent piezoelectric first order continuum by exploiting 
the asymptotic homogenization approach.  The resulting two-phase composite equivalent material is depicted in Figure \ref{GEOMMultiBB}(a), where the bottom layer is a homogeneous material characterized by overall piezoelectric constitutive tensors equivalent to periodic cells, and the top layer is a homogeneous electroactive polymer.
The two-phase device is bent by applying opposite rigid rotations $\Theta$ to the lateral faces (with normal parallel to $x_2$), around the axis $x_1$,  that falls at half the height of the device, see Figure \ref{GEOMMultiBB}(a). The qualitative deformed shape of this device is reported in  Figure \ref{GEOMMultiBB} (b). Again, two set of boundary conditions are considered, characterized either by free longitudinal faces (with normal parallel to $x_1$) or by longitudinal faces with restricted displacements in the $x_1$ direction. 
It results that in this framework, the potential $\Phi$ is only dependent on the $x_3$ axis, i.e. is uniform in $x_1$ and $x_2$, so that the potential difference, measured between two opposite points laying on the top and bottom faces (orthogonal to the $x_3$ axis), is constant.\\
A set of numerical analyses is proposed hereafter, where 
we consider the geometry of the device associated with a cluster of 15 $\times$ 5 periodic cells and adopting for the top layer the same polymeric material of the matrix, i.e. with Young modulus $E$= 535 MPa and Poisson's coefficient equal to $\nu$=0.4. The relative permittivity is $\beta$=3.\\
In a first investigation we let the thickness of the top layer $h_t$ vary between 0 nm and 1100 nm (so that the maximum total height is $h$+$h_t$= 2200 nm). Considering  a reference heterogeneous  material with nanorods density  $\delta$=0.415,  a rigid rotation $\Theta=1/100$ radiants is applied.
\begin{figure}[hbtp]
  \centering
    \begin{overpic}[scale=1]{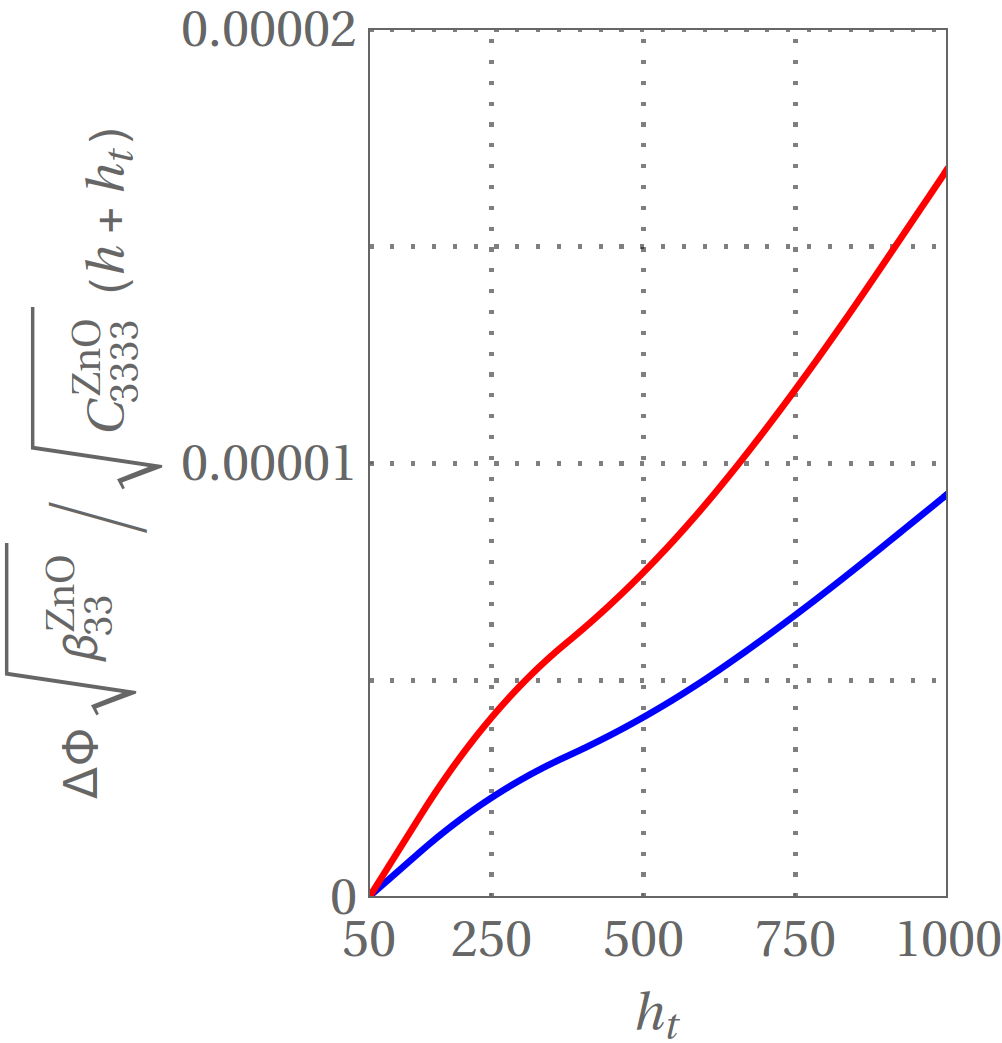}
     \put (80,90) {$(a)$}
    \end{overpic}
       \begin{overpic}[scale=1]{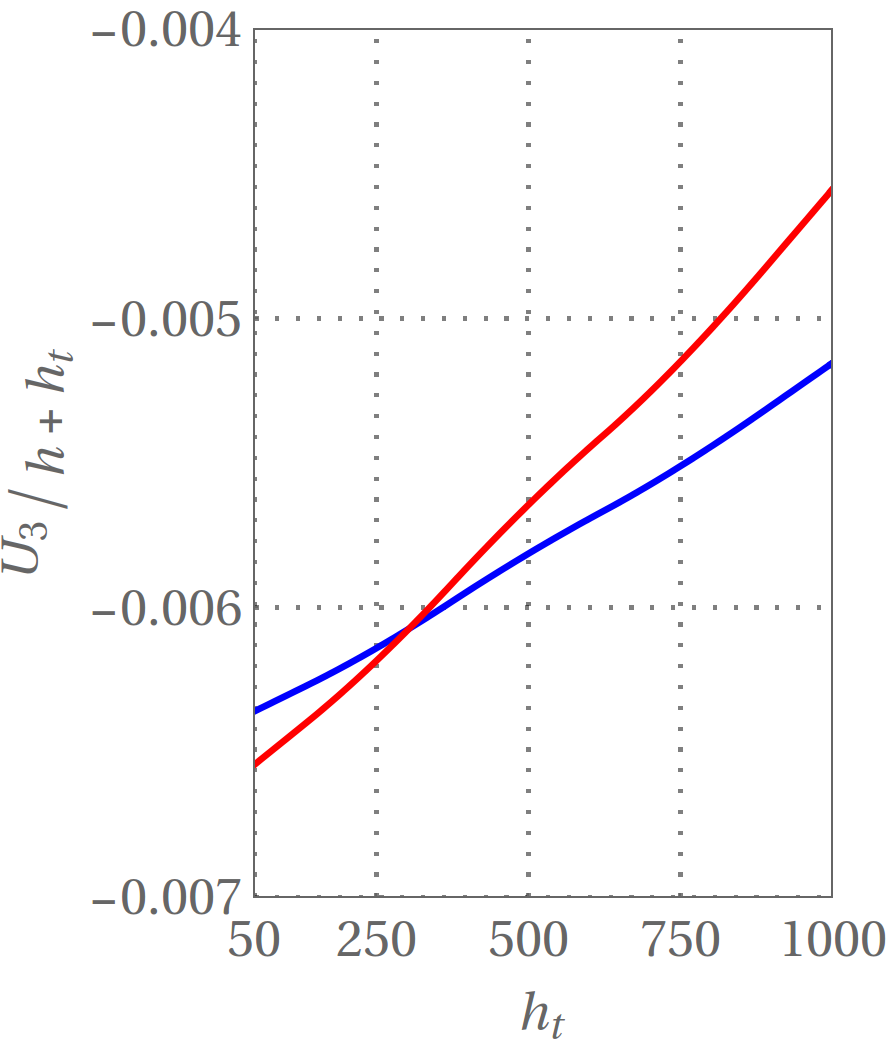}
        \put (70,90) {$(b)$}
    \end{overpic}
\caption{Bending nanogenerators obtained by homogenizing   a cluster of 15 $\times$ 5 nanorods  with $\delta$=0.415.(a) Dimensionless potential difference $\Delta \Phi\sqrt{\beta_{33}^{ZnO}}/( \sqrt{C_{3333}^{ZnO}} (h+h_t))$ versus $h_t$; (b) dimensionless  displacement component $U_3/(h+h_t)$ versus $h_t$.}
    \label{bendingdelta2}
\end{figure}
In Figure \ref{bendingdelta2} the overall response of the nanogerators in terms of the potential difference and of  displacement component $U_3$ (measured in the geometric center of the top face) versus $h_t$ are reported.
In Figure \ref{bendingdelta2}(a), $\Delta \Phi\sqrt{\beta_{33}^{ZnO}}/ (\sqrt{C^{ZnO}_{3333}} (h+h_t))$ is plotted versus $h_t$ for both the cases with free (blue curve) and prescribed (red curve) lateral displacements. It clearly emerges that when $h_t$= 0 nm, no potential difference is measured, while as soon as the thickness $h_t$ of the top layer is greater than zero,  increasingly higher values are found. Better performances are experienced by the device with prescribed displacements.
Concerning the dimensionless  displacement component $U_3/ (h+h_t)$ shown in Figure \ref{bendingdelta2}(b) versus $h_t$, nearly linear trends are observed. The red and blue curves intersect each other around $h_t$=300 nm, so that below this value the case with fixed longitudinal faces is more flexible than the other.  Note that similar trends of both dimensionless $\Delta \Phi$ and $u_3$ are found for different values of $\delta$ referred to the corresponding heterogeneous material.\\
A principal outcome of this first study is that the nanogerator exhibits the best performances with maximum values of $h_t$. It follows that the second investigation concerns the analysis of bending nanogenerators with $h_t$=1100 nm as the density $\delta$ of the corresponding heterogeneous material varies.
\begin{figure}[hbtp]
  \centering
    \begin{overpic}[scale=1]{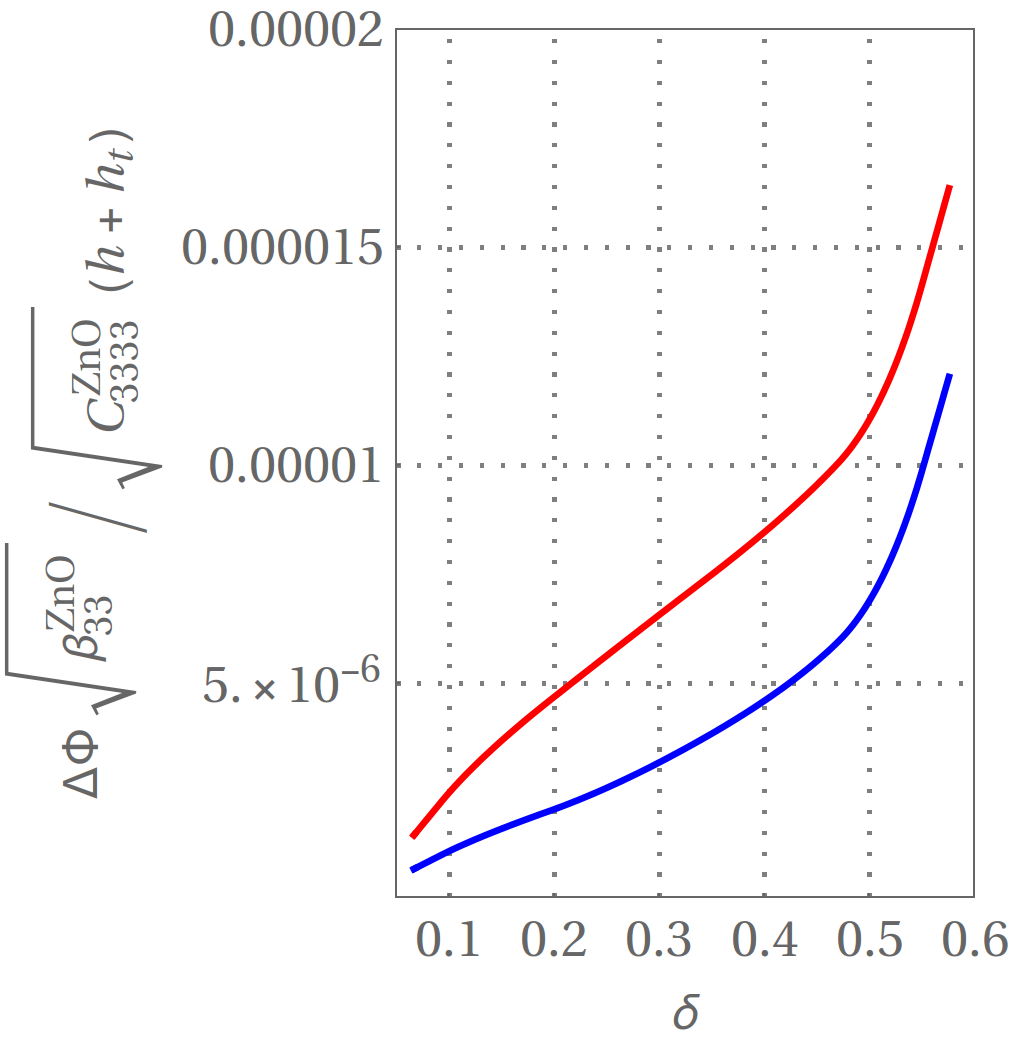}
     \put (80,90) {$(a)$}
    \end{overpic}\,\,
       \begin{overpic}[scale=1]{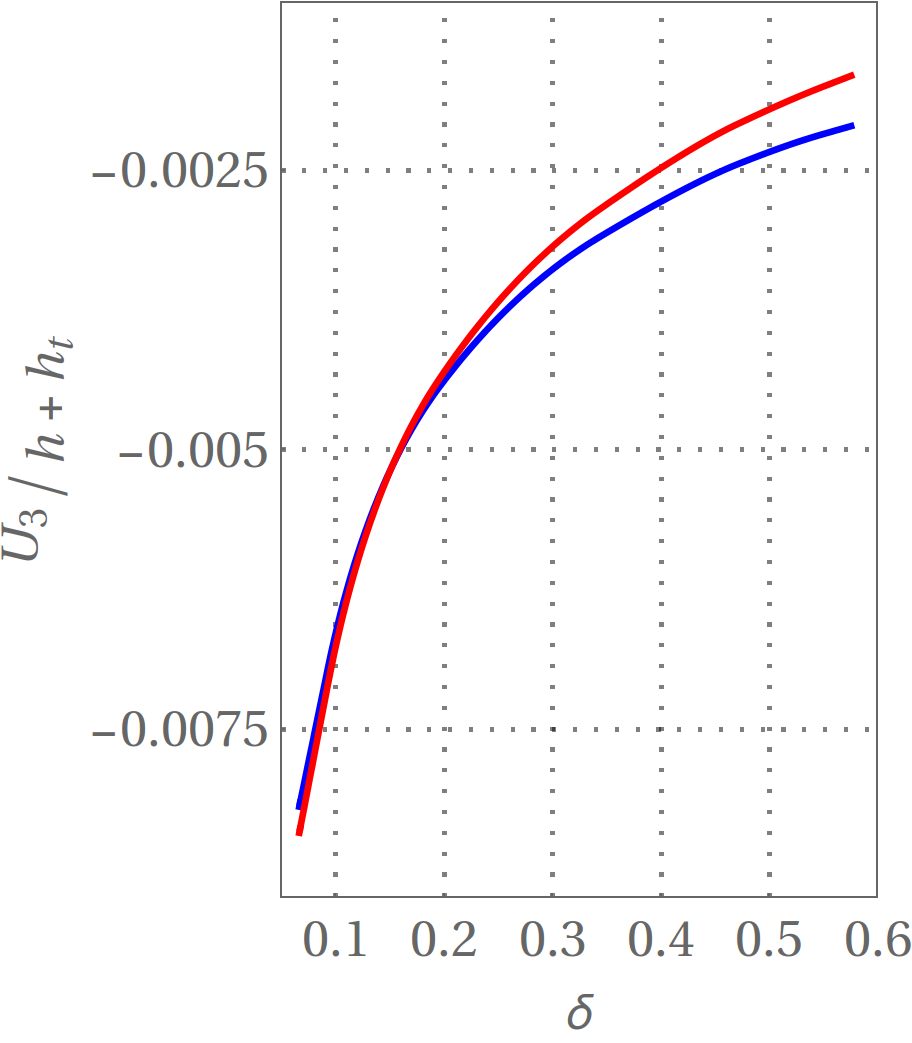}
            \put (70,95) {$(b)$}
    \end{overpic}
\caption{Bending nanogenerators obtained by homogenizing   a cluster of 15 $\times$ 5 nanorods  with $h_t$=1100 nm.(a) dimensionless potential difference $\Delta \Phi\sqrt{\beta_{33}^{ZnO}}/ (\sqrt{C_{3333}^{ZnO}} (h+h_t))$ versus $\delta$; (b) dimensionless vertical displacement $U_3/(h+h_t)$ versus $\delta$.}
    \label{bendingdelta1}
\end{figure} 
\noindent In Figure \ref{bendingdelta1}(a), the dimensionless potential difference $\Delta \Phi\sqrt{\beta^{ZnO}_{33}}/ (\sqrt{C^{ZnO}_{3333}} (h+h_t))$ is plotted versus the density $\delta$. As before, blue curves refer to the case with free longitudinal faces, while and red curves refer to the case with displacements fixed. An interesting nonlinear behaviour is found, with the potential difference that increases with $\delta$. Also in this case, better performances are provided by the device with prevented lateral displacements (red curve). A nonlinear response is also confirmed in the dimensionless displacement component $U_3/(h+h_t)$. The curves are nearly indistinguishable for low values of $\delta$, while as 
$\delta>$ 0.2 are increasingly far apart.
Finally, it can be remarked that the bending nanogenerators show optimal behaviours when designed with very thick top layers, fixed lateral faces and high nanorods densities.

\subsection{ Microstructured  transversal extension nanogenerators} \label{Pmten}
\noindent The same devices analysed in Section \ref{Pmbn}, is here investigated adopting a  different working principle related to transversal extension mechanism. Similarly to the previous case, the considered devices are made of  a cluster of periodic cells topped with a polymeric layer of variable thickness. Again the multi-scale approach is adopted to obtain a two-phase composite equivalent material, see in Figure \ref{GEOMMultiBB}(a). 
The same geometries and materials as in Section \ref{Pmbn} are taken into account. The two-phase device undergoes uniform opposite displacements along $x_2$ direction applied to the lateral faces (with normal parallel to $x_2$) inducing an overall extension $\Delta U_2$= 10 nm of the nanogenerator. Moreover, the  displacements of the points on the bottom face are restricted in the $x_3$ direction.  The qualitative deformed shape of this device is reported in  Figure \ref{GEOMMultiBB} (c).
We choose $\Delta U_2$ so that it is comparable with the maximum  displacement in the direction $x_2$ induced by the bending (when $h_t$=1100 nm) in the previous Section. Again, the displacement components along the $x_1$ direction of the longitudinal faces are alternatively free or set equal to zero.\\
\begin{figure}[hbtp]
  \centering
    \begin{overpic}[scale=1]{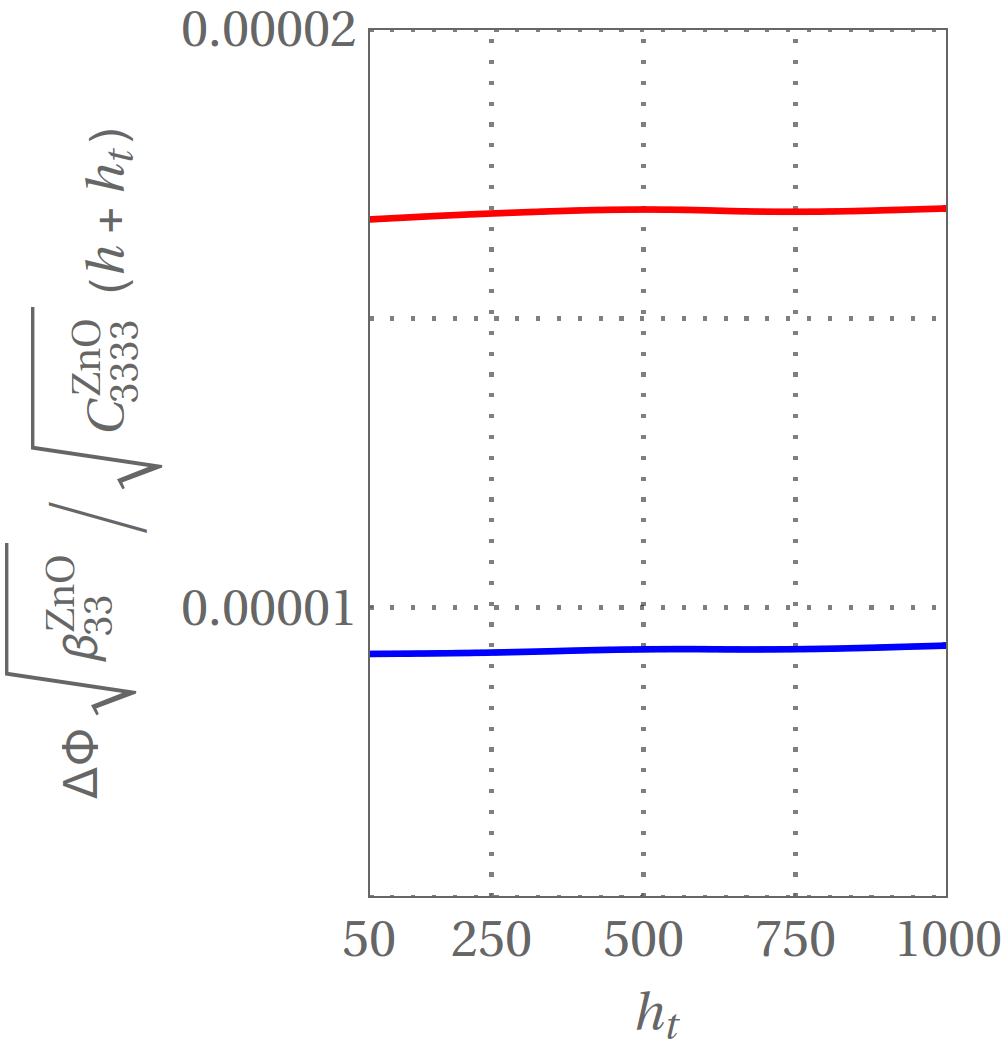}
         \put (80,90) {$(a)$}
    \end{overpic}\,\,
       \begin{overpic}[scale=1]{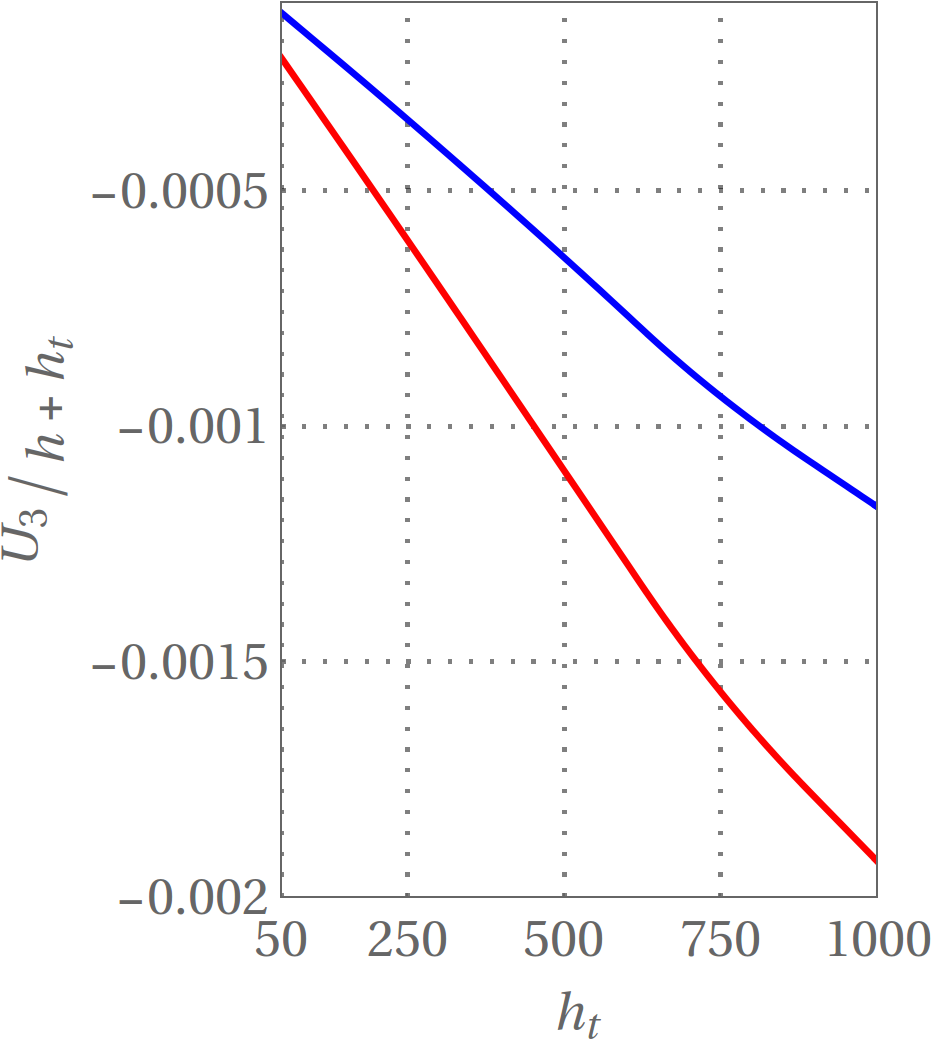}
            \put (75,90) {$(b)$}     
    \end{overpic}
\caption{Transversal extension nanogenerators  obtained by homogenizing   a cluster of  15 $\times$ 5 nanorods  with $\delta$=0.415.(a) Dimensionless potential difference $\Delta \Phi\sqrt{\beta_{33}^{ZnO}}/ (\sqrt{C_{3333}^{ZnO}} (h+h_t))$ versus $h_t$; (b) dimensionless vertical displacement $U_3/(h+h_t)$ versus $h_t$.}
    \label{bendingdelta3}
\end{figure}
Also in this case,  first the influence of the thickness $h_t$ of the top layer (for a fixed corresponding density $\delta$=0.415) is investigated, and then the response as the corresponding density $\delta$ varies (for $h_t$=1100 nm) is studied.
\noindent In Figure \ref{bendingdelta3}(a), the dimensionless potential difference $\Delta \Phi\sqrt{\beta_{33}^{ZnO}}/ (\sqrt{C_{3333}^{ZnO}} (h+h_t))$ versus $h_t$ is shown for both the cases with free longitudinal faces (blue curve) and restrained faces (red curve). As expected, nearly constant values of potential difference are found, irrespective of the thickness $h_t$. Moreover, the case with fixed longitudinal faces provides higher potential differences. On the other hand, concerning the vertical displacement $U_3$, pseudo-linear variations are observed as $h_t$ varies. \\
\noindent Finally, for the case with $h_t$=1100 nm, the dimensionless potential difference $\Delta \Phi\sqrt{\beta_{33}^{ZnO}}/ (\sqrt{C_{3333}^{ZnO}} (h+h_t))$ is shown in Figure \ref{bendingdelta4}(a), while 
 the dimensionless displacement components $U_3/(h+h_t)$, of the points laying on the top face, is plotted in  Figure \ref{bendingdelta4}(b)) versus the corresponding density $\delta$ of the heterogeneous  ZnO based hybrid nanogenerator. The plots of the potential difference with $\delta$ are almost overlapping with those in Figure \ref{bendingdelta1}(a). It is, nevertheless, noted that a qualitative different distribution of the macroscopic electric potential $\Phi$ along the height ($h+h_t$) of the specimen is observed in the case of bending nanogenerators and transversal extension nanogenerators. In particular, a nonlinear and a linear monotonic trends are observed, along the height of the bottom homogenized layer, in the former and the latter cases, respectively. Moreover, in both cases a nearly constant value is maintained within the polymeric top layer. Concerning the $U_3$ displacement components, as expected, significantly lower values are found (one order of magnitude lower) with respect to the Figure \ref{bendingdelta1}(b).
\begin{figure}[hbtp]
  \centering
    \begin{overpic}[scale=1]{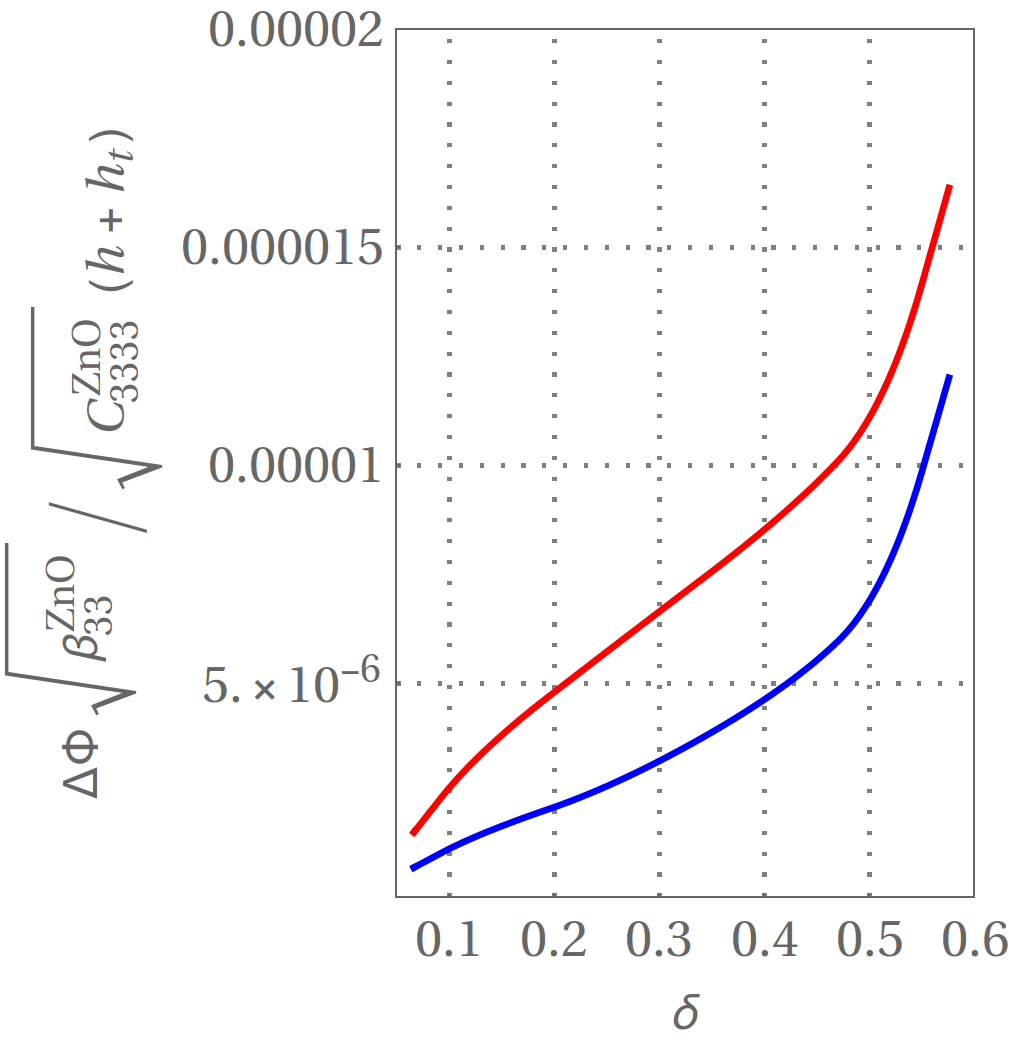}
         \put (80,90) {$(a)$}
    \end{overpic}\,\,
       \begin{overpic}[scale=1]{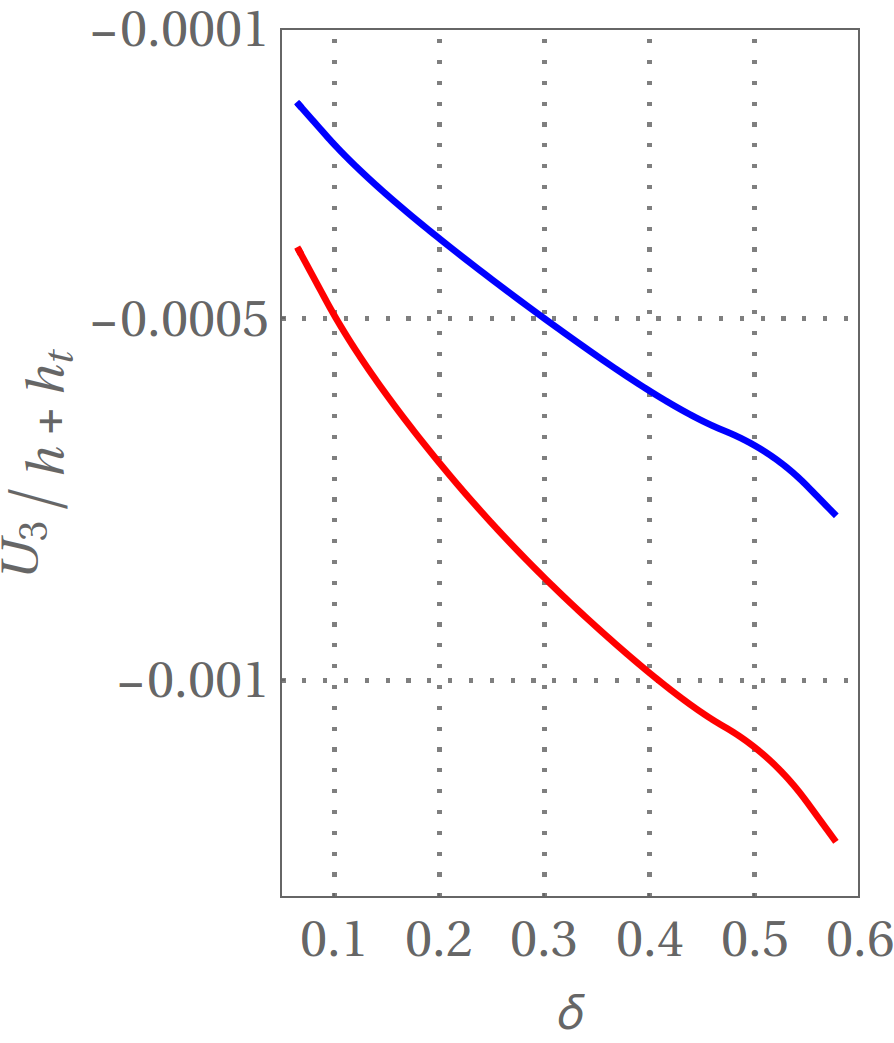}
            \put (70,90) {$(b)$}
    \end{overpic}
\caption{Uniform traction of an cluster of 15 $\times$ 5 nanorods  with $h_t$=1050 nm.(a) dimensionless potential difference $\Delta \Phi\sqrt{\beta_{33}^{ZnO}}/ (\sqrt{C_{3333}^{ZnO}} h)$ versus $\delta$; (b) dimensionless vertical displacement $U_3/h$ versus $\delta$.}
    \label{bendingdelta4}
\end{figure}
\noindent It is noteworthy that in the case of combined effects of bending and transversal extension,  optimal performances of the device are obtained by combining positive rigid rotations and extension, as qualitatively shown in Figures \ref{GEOMMultiBB}(b) and \ref{GEOMMultiBB}(c).

\section{Bloch waves in piezoelectric nanogenerators }\label{Dynamic}
This Section is devoted to investigate the dynamic characterization of the piezoelectric periodic nanostructured material for the purpose of analysing the Bloch waves propagation. In particular, the frequency spectrum of the heterogeneous material is obtained for different unit vectors of propagation. Moreover the acoustic branches of such frequency spectrum are compared against the dispersion functions obtained by the homogenized model.\\
Different Periodic Cells $\mathfrak{A}=[-d/2, d/2] \times [-d/2, d/2] \times [-h/2, h/2]$ of the microstructured nanogenerators are considered (see Figure \ref{figure1}(a)),  
characterized by various nanorods densities $\delta$.  The ZnO-nanorod has hexagonal  section 
with edge  80 nm, while the height in the polarization direction is  $h$= 1100 nm. The ZnO nanorod is embedded within a polymeric matrix ($E$=535 MPa, $\nu$=0.4 and $\beta$=3). The same polymeric material is adopted for the top and bottom layers. The mass density of the nanorod is $\rho^{ZnO}$= 5680 kg/m$^3$, while the mass density of the polymeric material is set equal to $\rho$= 1500 kg/m$^3$. 
In this framework, the Periodic Cell $\mathfrak{A}$ is associated with the dimensionless first Brillouin zone $\mathfrak{B}=[-\pi, \pi] \times [-\pi, \pi] \times [-\pi, \pi]$, defined in the  space of the dimensionless wave vectors (whose components are $k_1 d$, $k_2 d$ and $k_3 h$). Such Brillouin zone is characterized by three orthogonal vectors $\pi \textbf{n}_i$, parallel to $\textbf{e}_i$, with $i$=1,2,3.\\
The influence of the nanorods density $\delta$ on the Floquet-Bloch spectrum, as the unit vector of propagation $\textbf{n}$ varies, is thus analysed.  In particular, the wave vector  Main attention is devoted to identifying the existence, the position
and the frequency range of full or partial band gaps in the microstructured piezoelectric material. Note that in this case of nanogenerators, the Bloch spectrum analysis is particularly relevant in view of providing design guidelines. Such devices are, indeed, intended to operate under dynamical loadings that can trigger the electrical energy production. A detailed investigation concerning the band gaps is very useful to identify possible frequency ranges in which the device behaves as a filter. It is, thus, clear that a dynamical loading is supposed to be characterized by frequencies falling outside frequency band gaps, in order to be effective.\\
\begin{figure}[hbtp]
  \centering
    \begin{overpic}[width=0.32\textwidth]{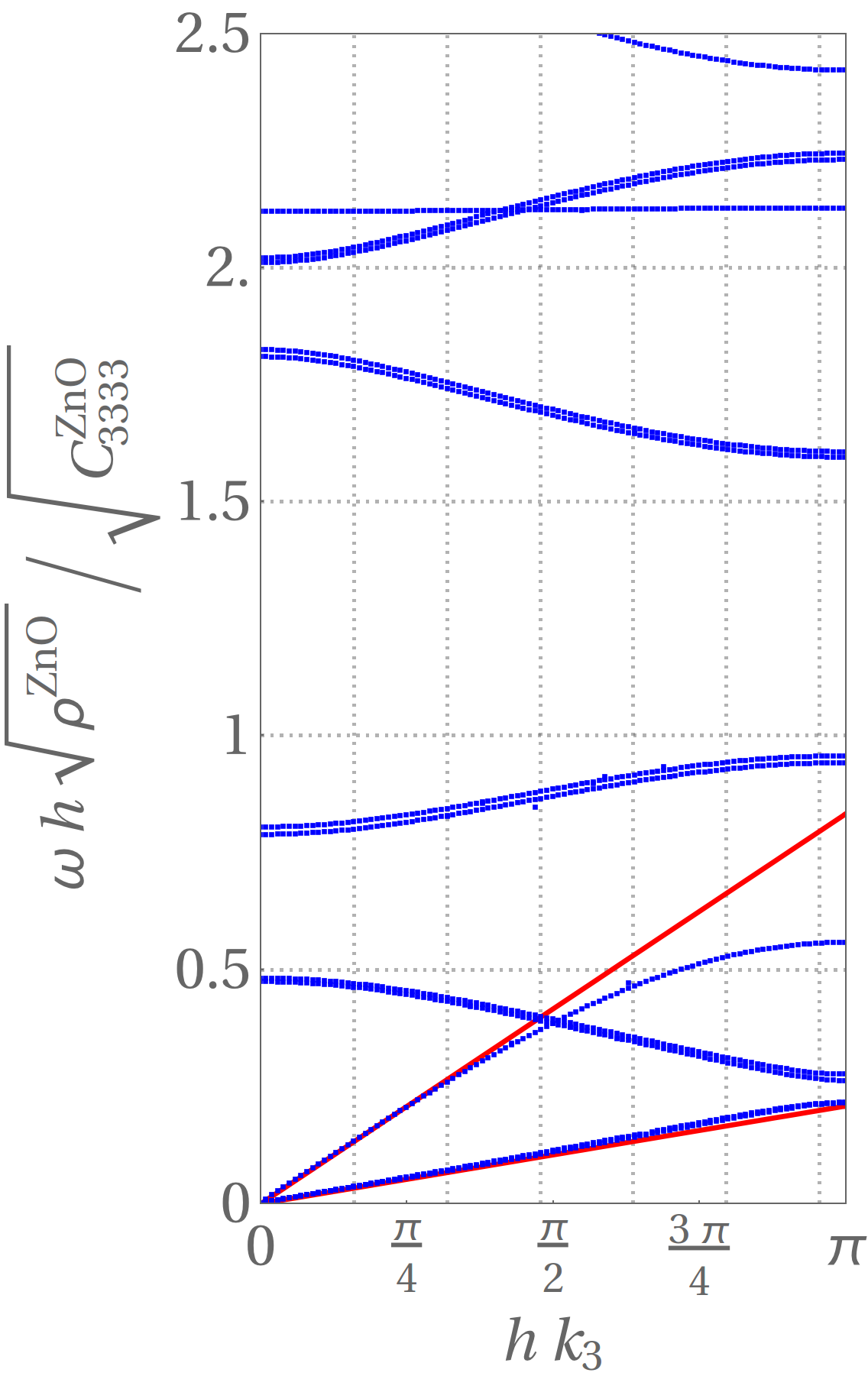}
         \put (20,90) {$(a)$}
    \end{overpic}\,\,\,
       \begin{overpic}[width=0.32\textwidth]{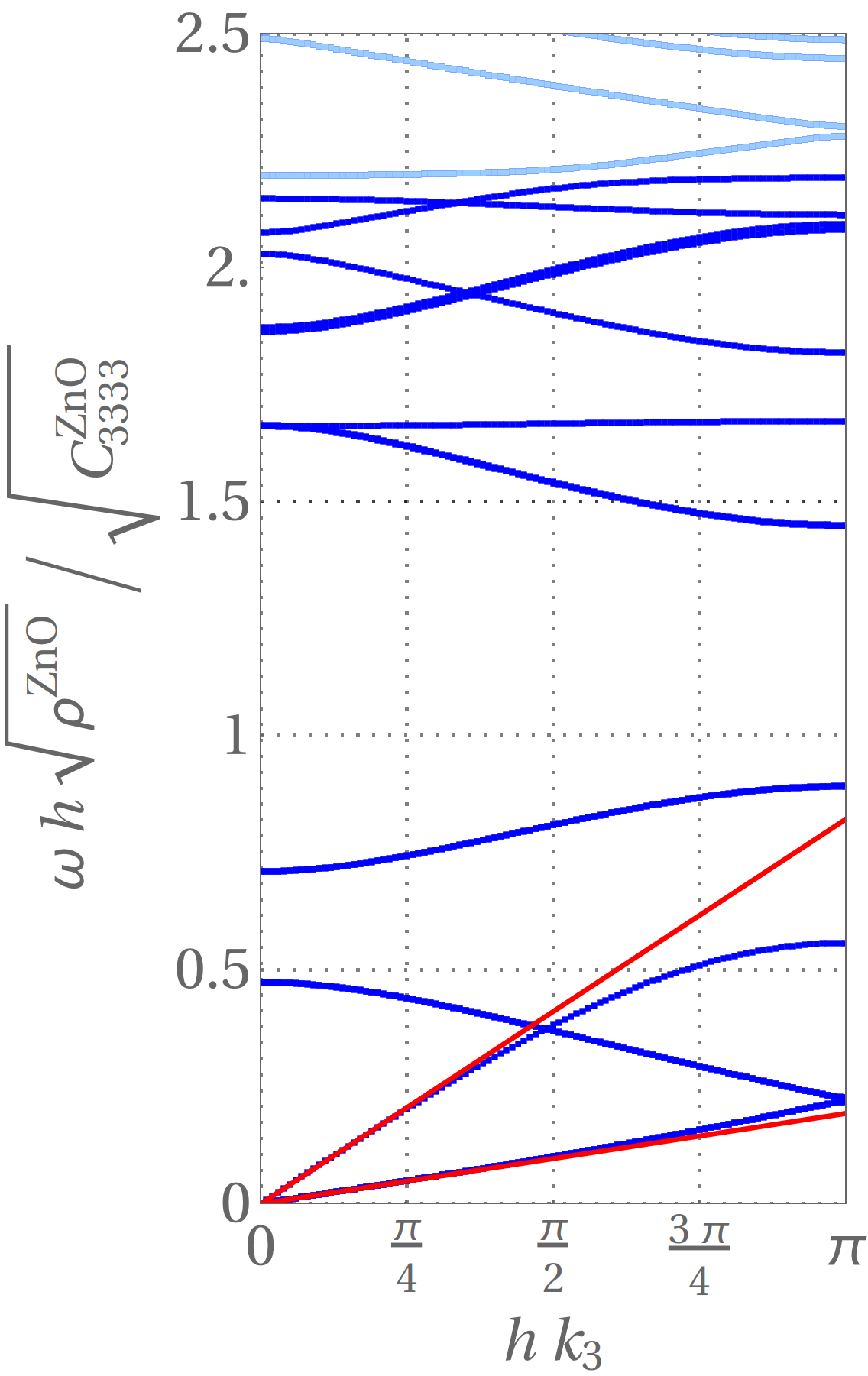}
                \put (20,90) {$(b)$}
    \end{overpic}\,\,\,
           \begin{overpic}[width=0.32\textwidth]{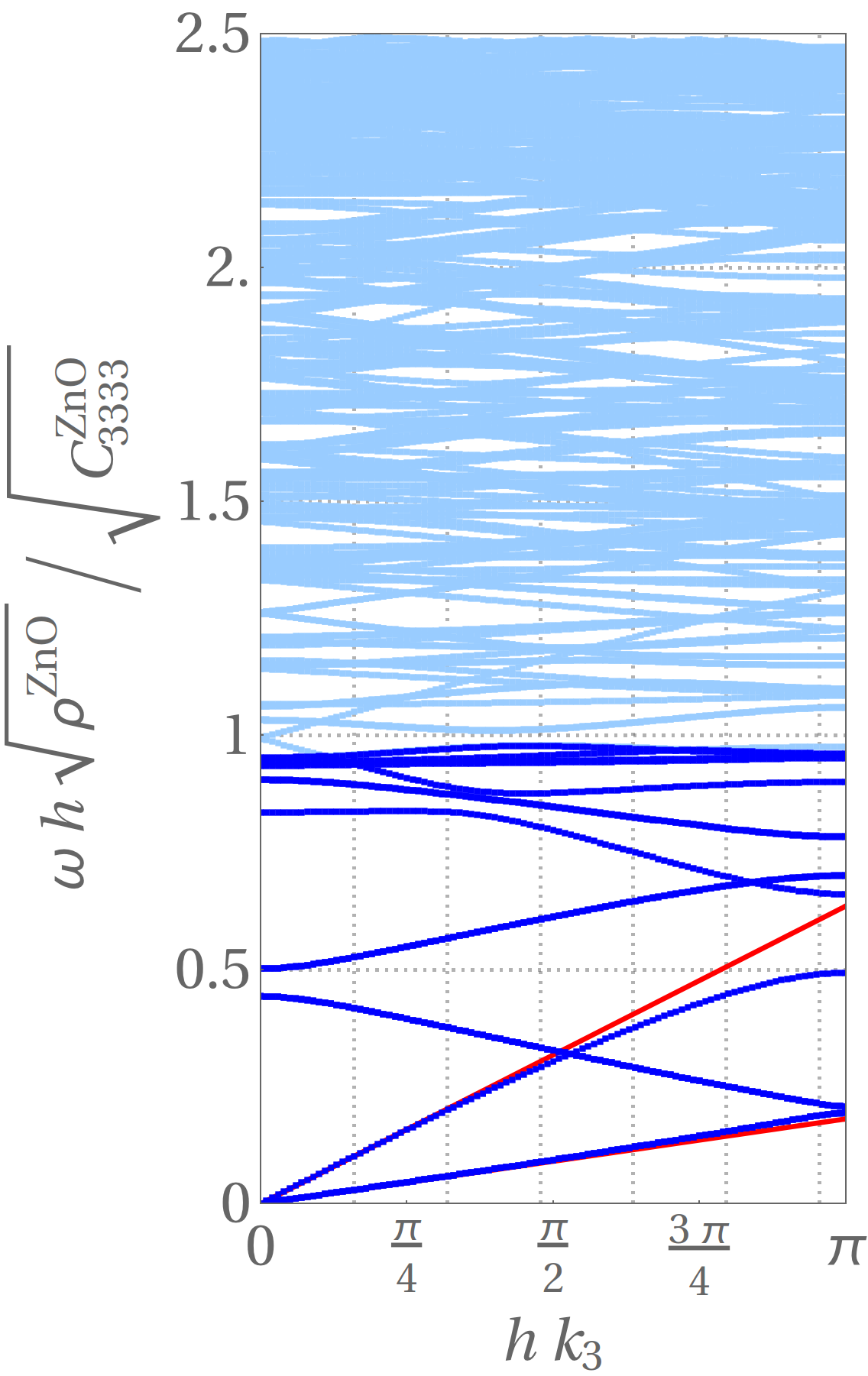}
                    \put (20,90) {$(c)$}
    \end{overpic}
\caption{Floquet-Bloch spectra for unit vector of propagation $\textbf{n}_3$ and different densities $\delta$ of the piezoelectric composite material. The dimensionless angular frequency $\omega \, h \sqrt{\rho^{ZnO}}/\sqrt{C^{ZnO}_{3333}}$ is plotted  versus the dimensionless wave number $k_3 h$. (a)  $\delta$=0.57; (b)  $\delta$=0.415; (c) $\delta$=0.10.}
    \label{Floquet1}
\end{figure}
The values of the dimensionless angular frequency $\omega \, h \sqrt{\rho^{ZnO}}/\sqrt{C^{ZnO}_{3333}}$,  against the dimensionless wave number $k_3h$, related to the unit vector of propagation  $\textbf{n}_3$ parallel to  $\textbf{e}_3$, are shown in Figure \ref{Floquet1} for three densities $\delta$ equal to 0.57, 0.415 and 0.10, respectively. The piezoelectric nanogenerator exhibits a high spectral density for all the densities. The first 15 branches of the spectrum are plotted in dark blue, while the other ones in light blue. In particular, in Figure \ref{Floquet1}(a) the Floquet-Bloch spectrum for 
 density $\delta$=0.57 is plotted. Only ten curves are visible, since multiple roots are found by virtue
of material symmetry characterizing the Periodic Cell. As expected, three acoustic branches departs from 
zero angular frequency (in the Figure only two are distinguishable, since two of them are coincident). The remaining 8 curves are the optical branches.
The Floquet-Bloch spectrum
exhibits one point of crossing between acoustic and optical branches and also a point of crossing
between different optical branches. Four partial band gaps are detected.
In Figure \ref{Floquet1}(b), instead, the Floquet-Bloch spectrum  for the material with density $\delta$= 0.415 is plotted. 
The spectrum is qualitatively similar to the previous one, in Figure \ref{Floquet1}(a).  The first three acoustic branches and the first two optical branches (two coincident curves) are very similar to the corresponding ones for $\delta$=0.57. On the other hand all the remaining branches are pulled down, resulting in curves closer together (more dense spectrum) and  partial band gaps basically characterized by lower amplitudes.
As a consequence, the relative position of the band gaps
is shifted towards lower frequencies. 
Moreover, in Figure \ref{Floquet1}(c) the Floquet-Bloch spectrum for the material with density $\delta$=0.1 is shown.  The same trend of Figure \ref{Floquet1}(b) is confirmed and emphasized. Again the first three acoustic branches and the first two optical branches remain almost unchanged, but all the others are shifted towards smaller angular frequencies.  A very dense spectrum is observed.
 This analysis suggests that, by focusing on a given frequency interval, for instance $0<\omega \, h \sqrt{\rho^{ZnO}}/\sqrt{C^{ZnO}_{3333}} $ $<1$,
 as the density $\delta$ decreases, the microstructured piezoelectric device exhibits better performances, since the spectrum is more dense. A considerably broader passband width is, indeed, found.  The acoustic branches of the Floquet-Bloch spectrum that characterize the heterogeneous material are compared against the dispersion curves obtained adopting
 the first order homogenization theory (red curves in Figure \ref{Floquet1}).  A good agreement is found, confirming the capability of the first order homogenization theory to satisfactorily reproducing the lowest (acoustic) branches of the
Bloch spectrum for a wide range of wavelengths. As expected,
the model at the first order only describes non-dispersive waves exhibiting a
linear dependence between the angular frequency  and the
wave-number. \\
\begin{figure}[hbtp]
  \centering
    \begin{overpic}[width=0.32\textwidth]{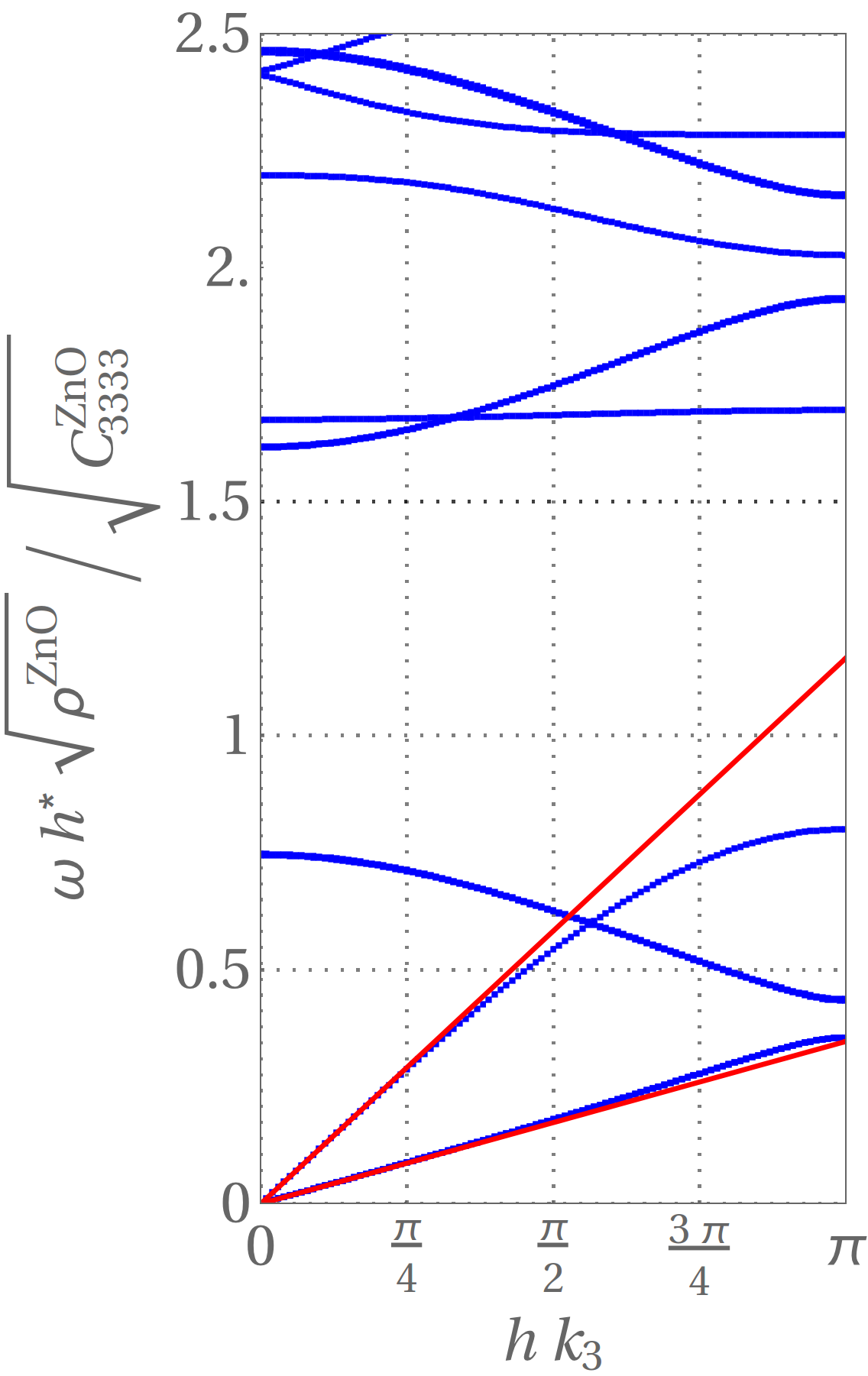}
         \put (20,90) {$(a)$}
    \end{overpic}\,\,\,
       \begin{overpic}[width=0.32\textwidth]{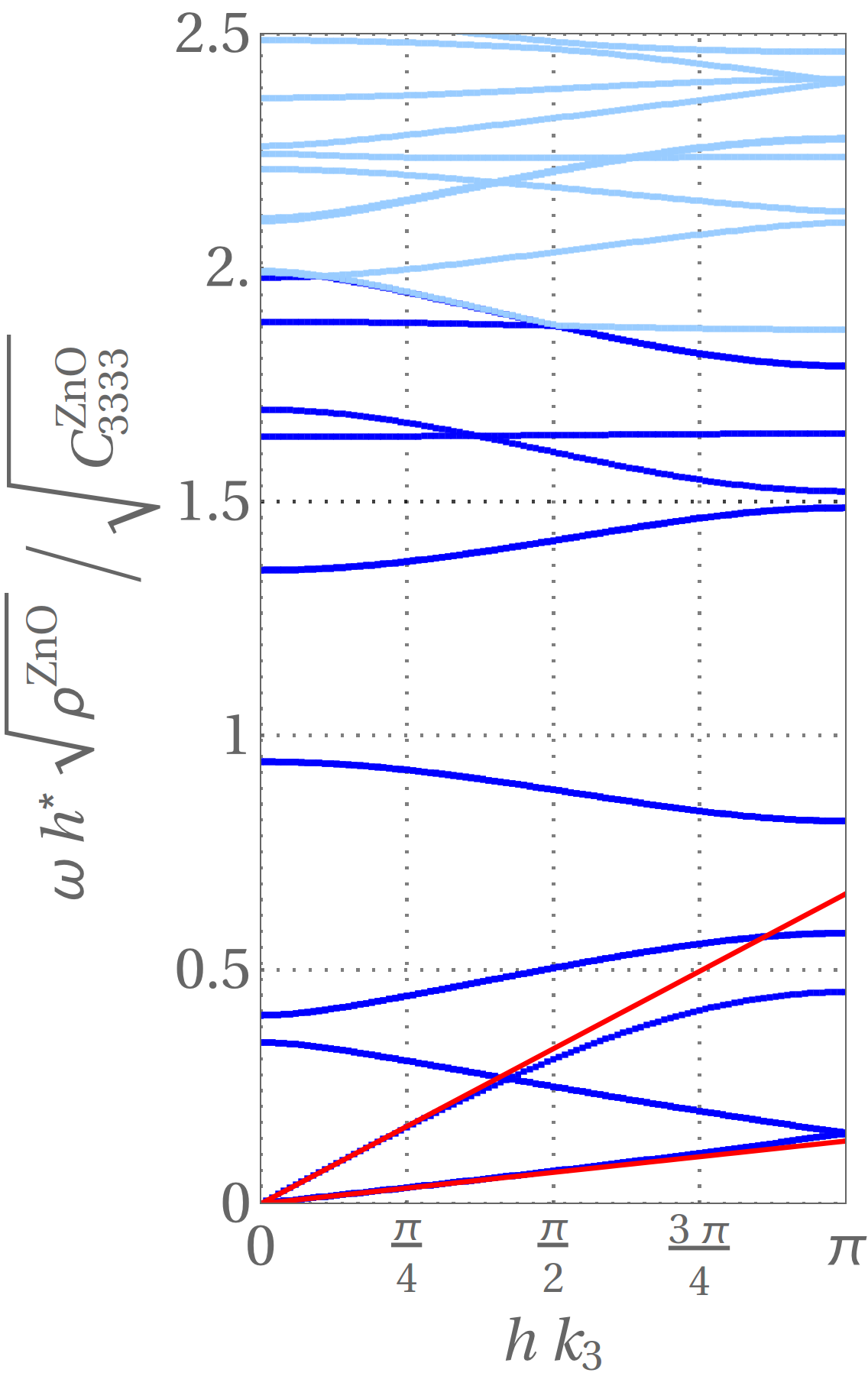}
                \put (20,90) {$(b)$}
    \end{overpic}\,\,\,
           \begin{overpic}[width=0.32\textwidth]{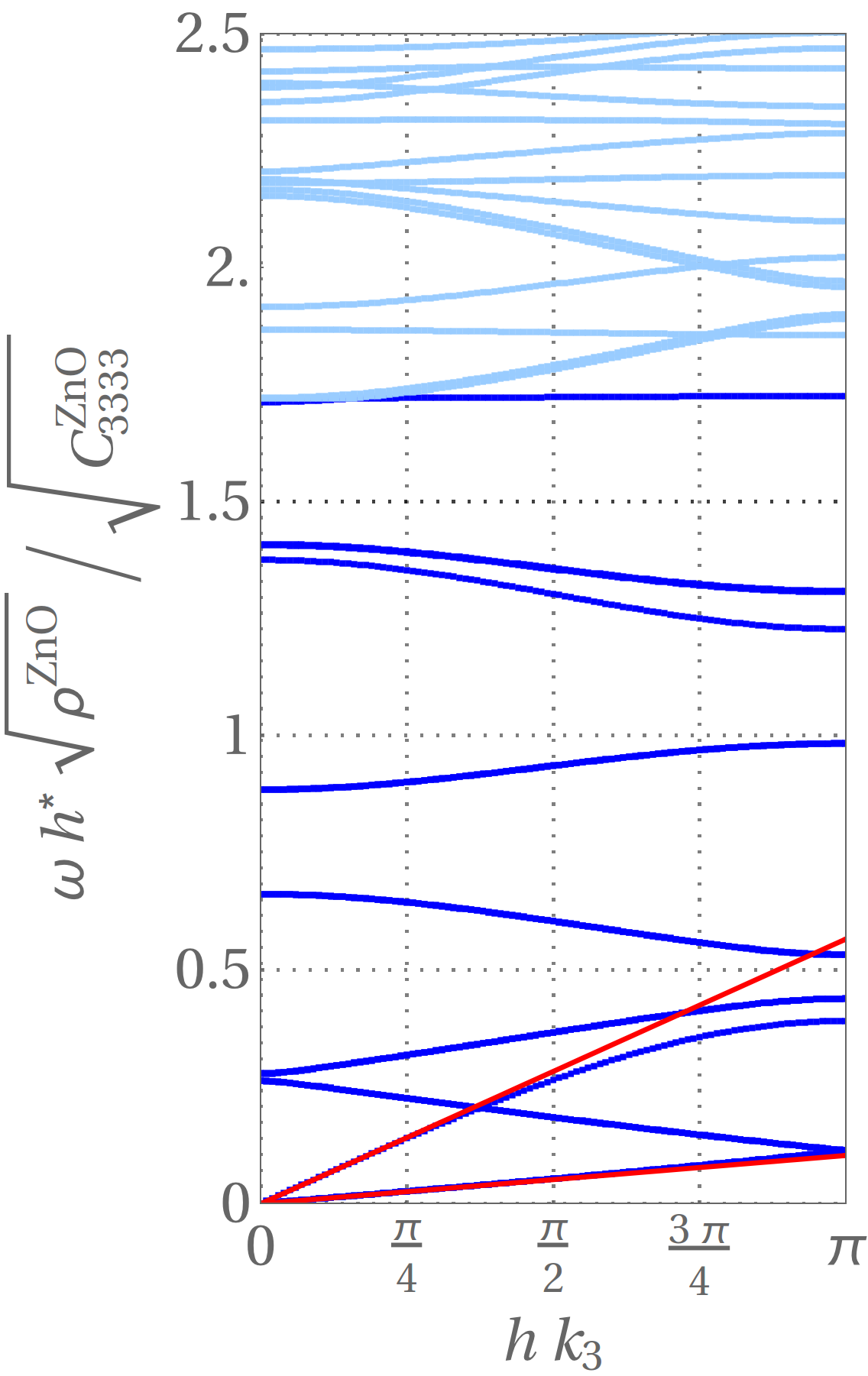}
                    \put (20,90) {$(c)$}
    \end{overpic}
\caption{Floquet-Bloch spectra for unit vector of propagation $\textbf{n}_3$ for different ratios 
between the height $h$ of the ZnO nanorod and the reference height $h^*$=1100 nm. The dimensionless angular frequency $\omega \, h^* \sqrt{\rho^{ZnO}}/\sqrt{C^{ZnO}_{3333}}$ is plotted versus versus the dimensionless wave number $k_3 h$. (a)  $h$=0.5 $h^*$; (b) $h$=1.5 $h^*$; (c) $h$=2 $h^*$.}
    \label{Floquet2}
\end{figure}
 \noindent A further investigation concerns the Floquet-Bloch spectrum for waves characterized by unit vector of propagation $\textbf{n}_3$, studying the influence of the height of the nanorods, considering the parameter $\alpha=h/h^*$, as the height $h$ varies  with $h^*$=1100 nm. A density $\delta$=0.415 is considered. In this case, the  dimensionless angular frequency $\omega \, h \sqrt{\rho^{ZnO}}/$ $\sqrt{C^{ZnO}_{3333}}$,  against the dimensionless wave number $k_3h$,  are shown in Figure \ref{Floquet2} for three heights $h$ equal to 550 nm, 1650 nm and 2200 nm, respectively. The first 15 branches of the spectrum are plotted in dark blue, while the others are plotted in light blue. It is remarked that, in the considered frequency range, 
 the spectral density tends to increase as the height $h$ increases.
In Figure \ref{Floquet2}(a), the case with $h$=0.5 $h^*$ is taken into account. A wide partial band gap is found with central dimensionless  frequency around 1.2, while a second partial band gap is characterized by  lower amplitude and higher central angular frequency. Again various crossing points are observed. Figures \ref{Floquet2}(b), and \ref{Floquet2}(c) shows the cases with $h$=1.5 $h^*$ and $h$=2 $h^*$, respectively. As the height increases, a higher number of partial band gaps, characterized by lower frequencies, are observed. The acoustic and optical branches tend to be shifted towards lower angular frequencies. Also in this case, the red curves represent the acoustic branches of the Floquet-Bloch spectrum obtained via a first order homogenization approach. Such linear dispersion functions are in good agreement with the the lowest (acoustic) branches of the
Bloch spectrum, characterizing the heterogeneous material, for a wide range of wavelengths.\\
A similar investigation, concerning both the influence of the density $\delta$ and of the height $h$, is performed considering the unit vector of propagation $\textbf{n}_1$ parallel to $\textbf{e}_1$. Also in this case, the first 15 branches of the spectrum are plotted in dark blue, while the others are plotted in light blue.  In Figure \ref{Floquet3}, three densities $\delta$ equal to 0.57, 0.415 and 0.10, respectively, are taken into account. Very dense spectra are here observed, where no partial band gaps are detected irrespective of the considered density.  It is confirmed that the spectrum density increases as the density $\delta$ decreases. Focusing on  Figure \ref{Floquet3}(a), various crossing points and a veering point (with dimensionless frequency about 2) are found. Three distinct acoustic branches  are observed, well approximated by the dispersion functions (red linear curves) obtained by the first order piezoelectric continuum. It is, moreover, noted that the first two acoustic branches tend to overlap each other as the density $\delta$ decreases. It is noteworthy that in the long wave regime the phase and group velocity, associated with the acoustic branches, tend to decrease as the density $\delta$ decreases. \\
 \begin{figure}[hbtp]
  \centering
    \begin{overpic}[width=0.32\textwidth]{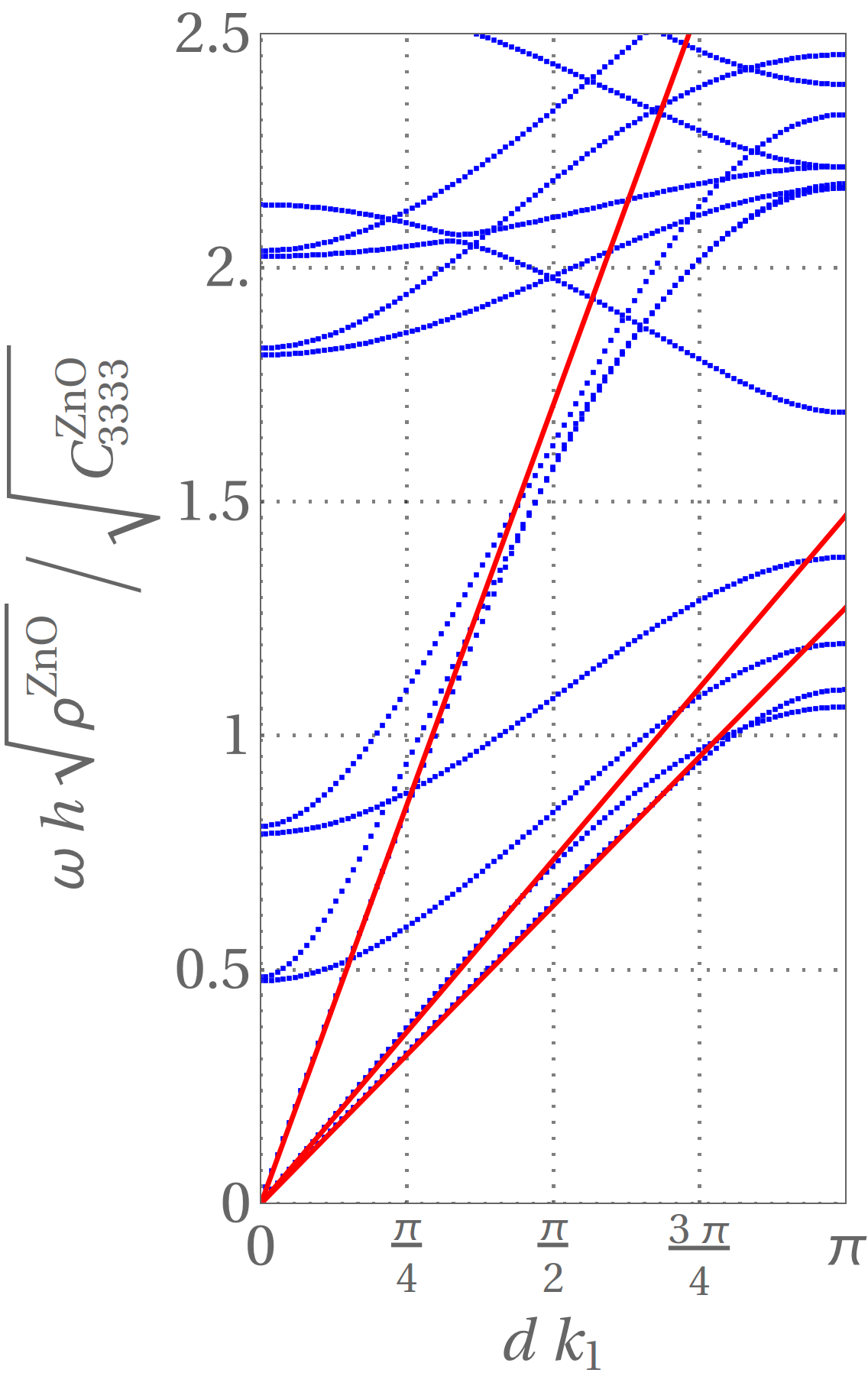}
         \put (20,90) {$(a)$}
    \end{overpic}\,\,\,
       \begin{overpic}[width=0.32\textwidth]{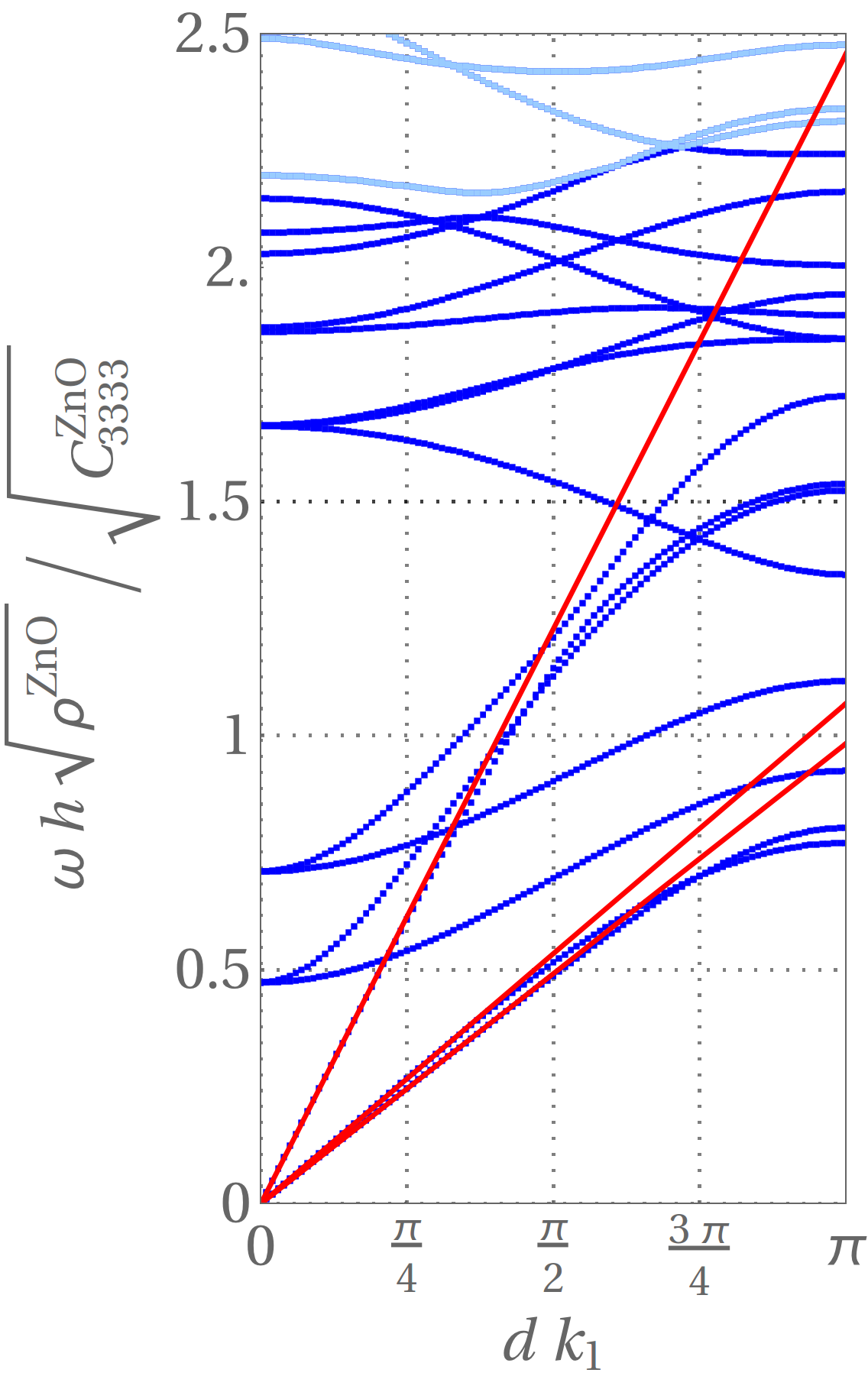}
                \put (20,90) {$(b)$}
    \end{overpic}\,\,\,
           \begin{overpic}[width=0.32\textwidth]{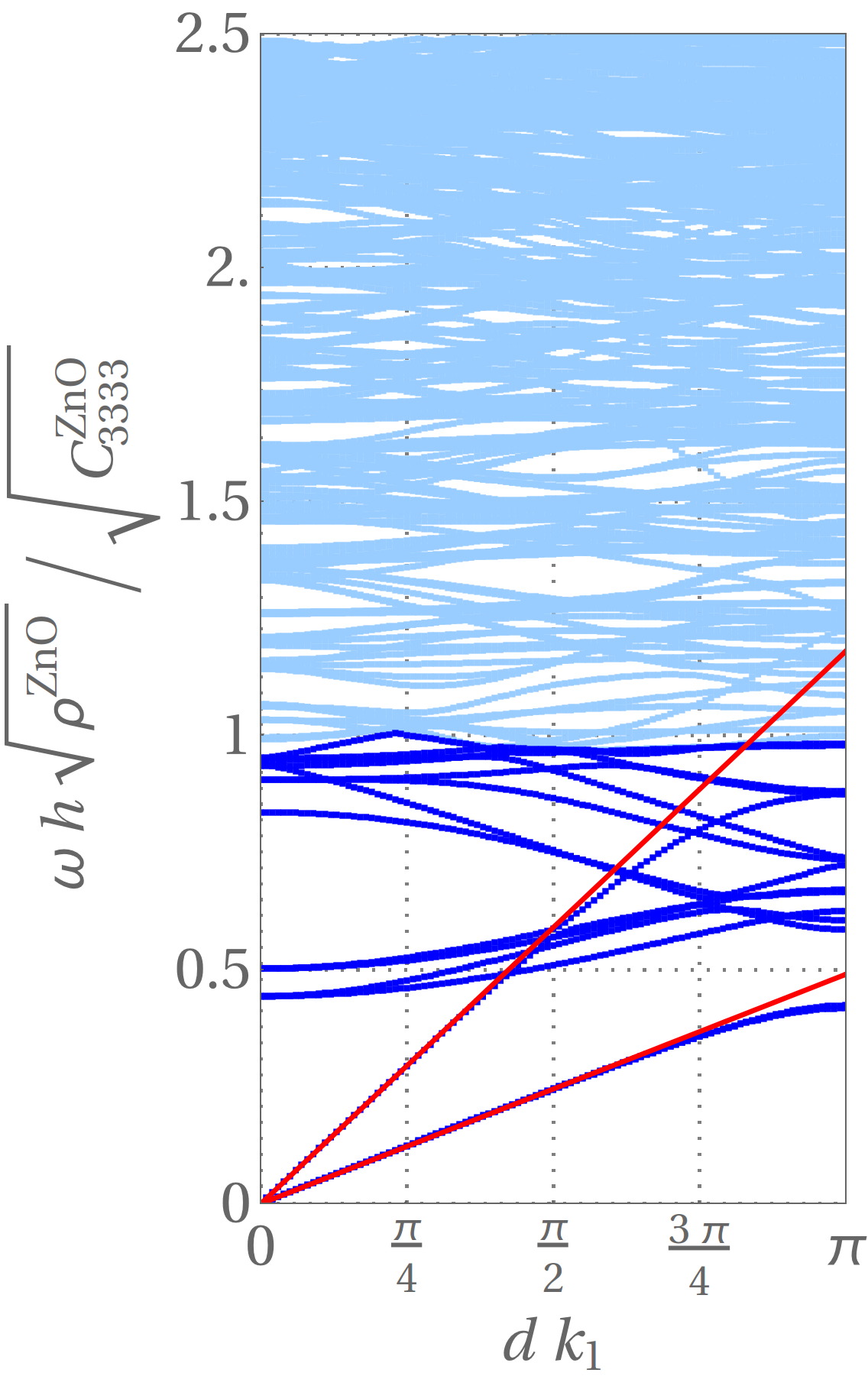}
                    \put (20,90) {$(c)$}
    \end{overpic}
\caption{Floquet-Bloch spectra for unit vector of propagation $\textbf{n}_1$ and different densities $\delta$ of the piezoelectric composite material. The dimensionless angular frequency $\omega \, h \sqrt{\rho^{ZnO}}/\sqrt{C^{ZnO}_{3333}}$ is plotted versus versus the dimensionless wave number $k_1 d$. (a)  $\delta$=0.57; (b)  $\delta$=0.415; (c) $\delta$=0.10.}
    \label{Floquet3}
\end{figure}
\begin{figure}[hbtp]
  \centering
    \begin{overpic}[width=0.32\textwidth]{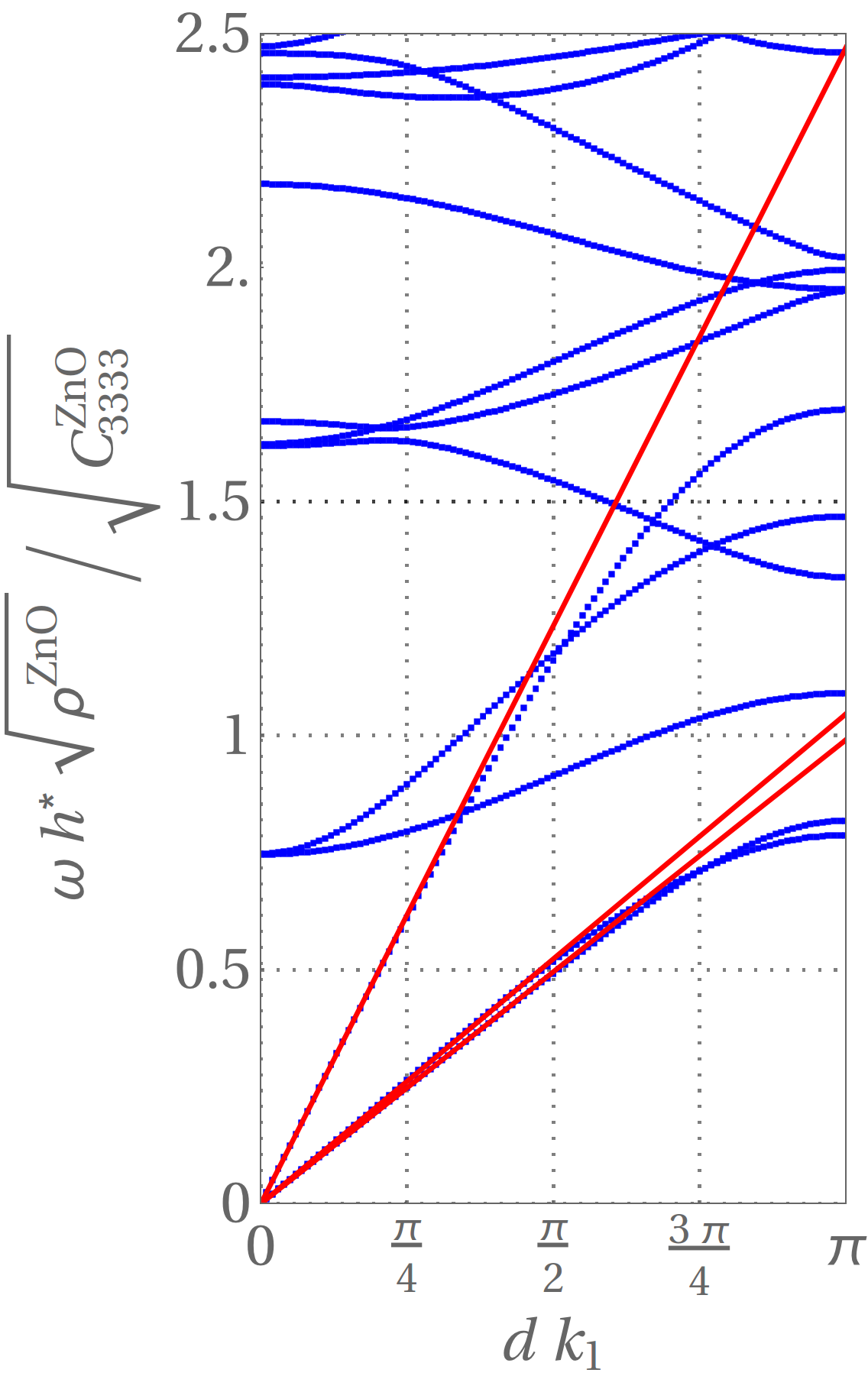}
         \put (20,90) {$(a)$}
    \end{overpic}\,\,\,
       \begin{overpic}[width=0.32\textwidth]{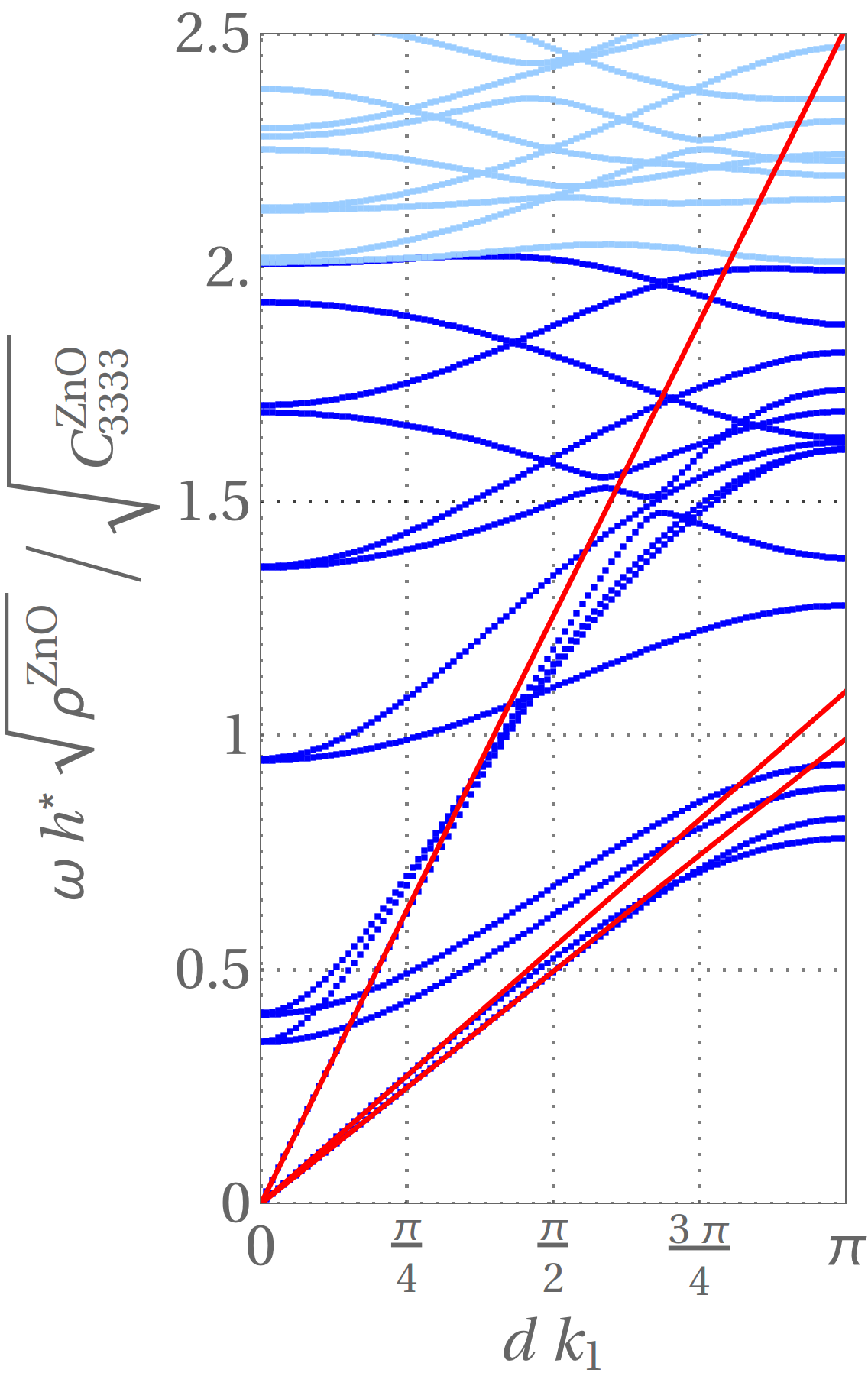}
                \put (20,90) {\colorbox{white}{$(b)$}}
    \end{overpic}\,\,\,
           \begin{overpic}[width=0.32\textwidth]{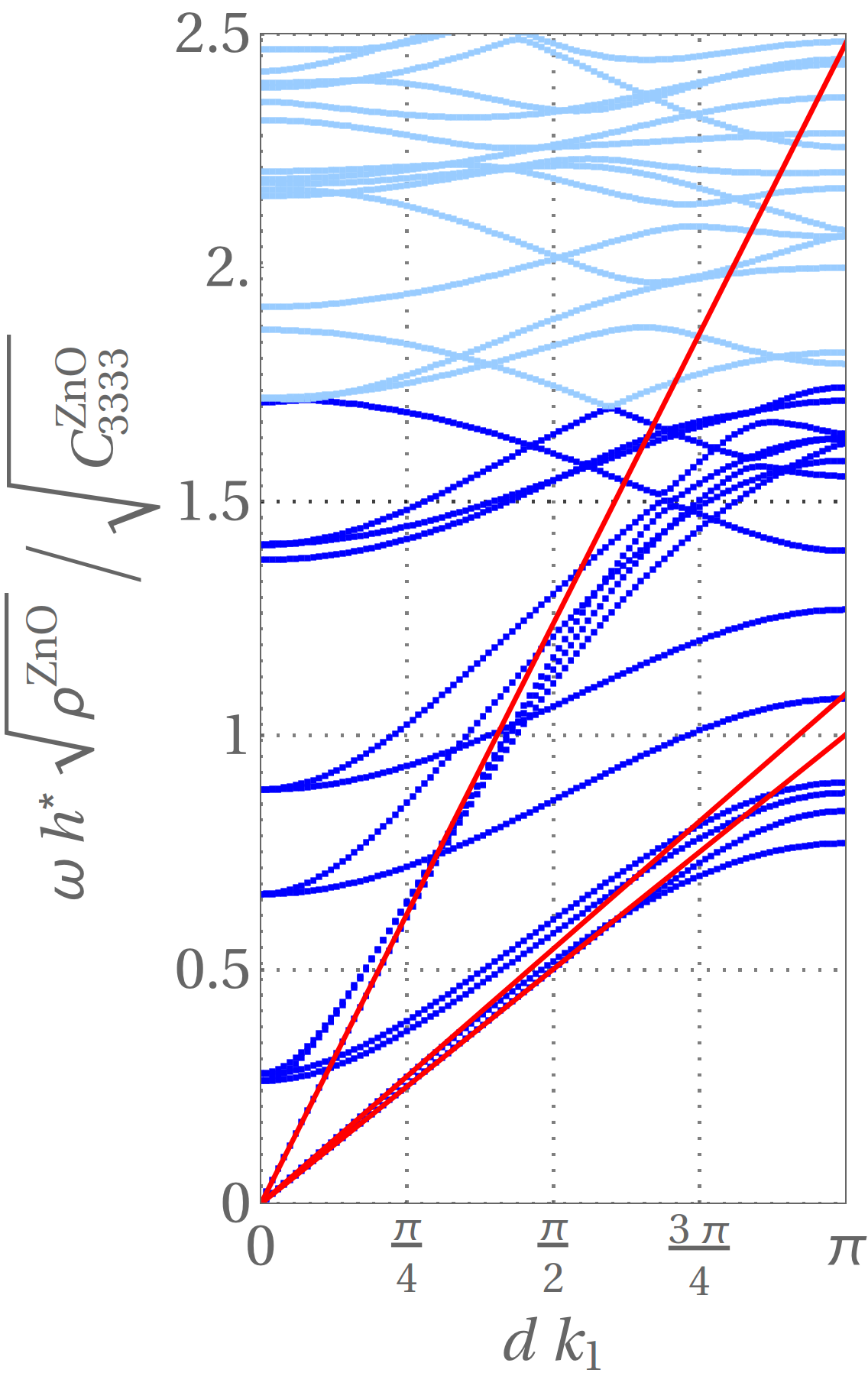}
                    \put (20,90) {\colorbox{white}{$(c)$}}
    \end{overpic}
\caption{Floquet-Bloch spectra for unit vector of propagation $\textbf{n}_1$ for different ratios 
between the height $h$ of the ZnO nanorod and the reference height $h^*$=1100 nm. The dimensionless angular frequency $\omega \, h^* \sqrt{\rho^{ZnO}}/\sqrt{C^{ZnO}_{3333}}$ is plotted versus versus the dimensionless wave number $k_1 d$. (a)  $h$=0.5 $h^*$; (b) $h$=1.5 $h^*$; (c) $h$=2 $h^*$.}
    \label{Floquet4}
\end{figure}
\noindent Analogous considerations are valid for the Figure \ref{Floquet4}, in which the  dimensionless angular frequency $\omega \, h \sqrt{\rho^{ZnO}}/$ $\sqrt{C^{ZnO}_{3333}}$,  against the dimensionless wave number $k_1d$,  are shown, for three heights $h$ equal to 550 nm, 1650 nm and 2200 nm, respectively, adopting $\delta$=0.415.  Also in this case, indeed, no partial band gaps are detected, and the spectrum density increases as the height $h$ increases. Various crossing and points are found. Finally, three distinct acoustic branches are observed, well approximated by the red linear curves, representing the dispersion functions of a first order homogeneous equivalent continuum. \\
 \begin{figure}[hbtp]
  \centering
    \begin{overpic}[width=0.47\textwidth]{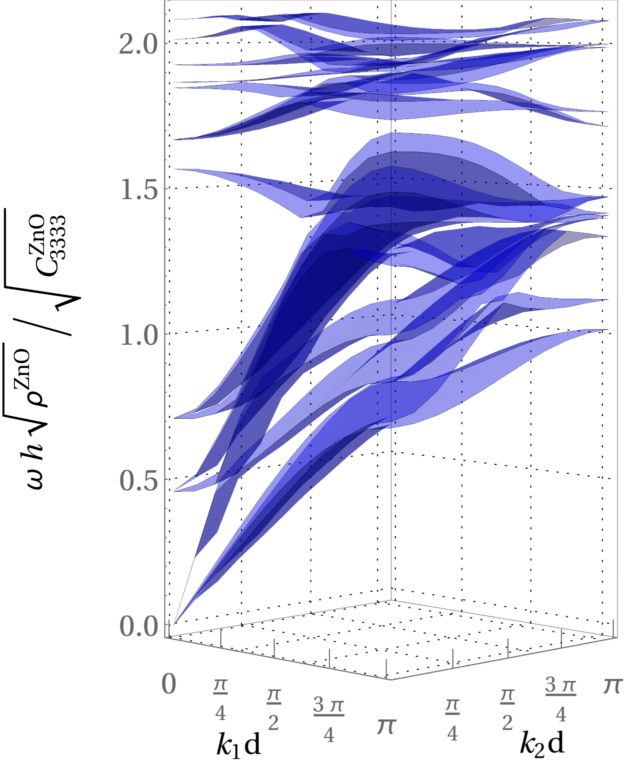}
    \put (73,90) {$(a)$}
    \end{overpic}\,\,\,
     \begin{overpic}[width=0.47\textwidth]{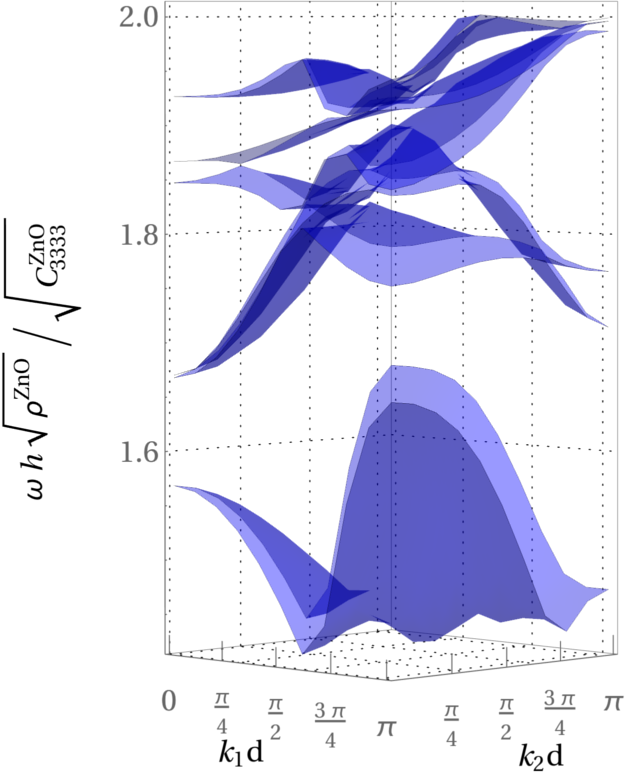}
         \put (73,90) {$(b)$}
    \end{overpic}\,\,\,
\caption{Axonometric view of the Floquet-Bloch spectra for unit vector of propagation $\textbf{n}$ laying on the plane with normal $\textbf{n}_3$. (a) First 15 surfaces; (b) zoomed view of the partial band gap.}
    \label{Floquetd}
\end{figure}
 \noindent A more comprehensive description of the acoustic characteristics of the piezoelectric
nanoscopic material is given by the dispersion surfaces in the Brillouin zone, representing the dimensionless angular frequency $\omega \, h \sqrt{\rho^{ZnO}}/\sqrt{C^{ZnO}_{3333}}$ 
of waves characterized by unit vector of propagation $\textbf{n}=cos \theta \, \textbf{n}_1+sen \theta \, \textbf{n}_2$, with $0 \le \theta \le \pi/2$. In  Figure \ref{Floquetd}(a), an axonometric view is show for the case with density $\delta$= 0.415.
In the domain of considered wave vectors, a high spectral density is observed, since the acoustic surfaces intersect the first optical ones and, moreover, the optical surfaces at higher frequencies intersect
each other over and over. One partial band gap is detected between the eight and ninth surfaces.\\
In particular, in Figure  \ref{Floquetd}(a) a zoomed view of the  surfaces defining the partial band gap is shown. As a consequence, in this range the propagation of waves characterized by $k_3=0$ is inhibited.

\clearpage
\section{Final Remarks}
Hybrid piezoelectric nanogenerators, made of 
Zinc Oxyde nanorods, embedded into a polymeric matrix and growth on a flexible polymeric supports, are investigated. Due to the high density and nearly regular distribution of the nanorods an equivalent periodic topology can be considered. The resulting device is made of a microstructured composite material, whose global response is strongly influenced by the microstructure, i.e. by the geometry and materials properties of each constituents and by their collaborative behaviour.\\
In such case,
a very detailed description could be obtained via  micromechanical approaches, resulting in often too cumbersome analyses. For this reason, multi-scale techniques can be alternatively used, being a very valuable tool to gather the overall behaviour.
In this framework, a  dynamic asymptotic homogenization approach, for periodic piezoelectric composites, is proposed. 
By exploiting this homogenization approach, a rigorous constitutive characterization of the periodic piezoelectric microstructured material is carried out, with the aim of optimizing their performances by analysing the behaviour of the material/device in the space of the physical-mechanical parameters. 
Among different parameters, focus is on the influence of both the height of the nanorods and of the density ( defined as ratio between the volume occupied by the nanorods and the volume of the Periodic Cell), since 
during the synthesis process of the samples it is possible to directly modify those parameters.
Under the assumption of considering an equivalent homogenized material at the macroscopic scale, also analytical solutions are found, providing benchmark comparative solutions. Almost perfect match is found by comparing the macroscopic analytic results with those obtained by the heterogeneous model via up-scaling relations, by confirming the optimal accuracy of the proposed homogenization scheme.\\
The behaviour of three piezoelectric microstructured nanogenerators, characterized by different working principles, is studied as a set of geometrical parameters changes. Both extension and bending nanogenerators are taken into account, considering either extension along the nanorods axis, or orthogonally to it. The influence of either the height of nanorods and their density on the overall behaviour of such devices is analysed. Concerning the  microstructured extension nanogenerators, both cases of uniform surface load and load with fixed resultant have been investigated. In the first case, maximum values of the potential differences are observed for low values of the density, in the range 0.1-0.2. In the second case, instead, the maximum potential differences correspond to higher values of 0.4-0.5. As expected, the variation of the nanorods height determine a grossly linear variation of the potential difference in the considered nanogenerator. In all the analysed case, it is noted that devices with free lateral faces exhibit better performances in terms of potential difference, with respect to the corresponding ones with constrained faces. With reference to microstructured bending nanogenerators, it is noted that an improved response is achieved by adding a further top layer, of increasing thickness, on the initial composite material made of two equal layers sandwiching the periodically distributed nanorods, embedded into the polymeric matrix. Moreover, for a fixed thickness of the additional top layer, the best response in terms of potential difference is observed for the highest values of the nanorods density. As regards the microstructured transversal extension nanogenerators, it emerges that the response, in terms of potential difference, is not affected by the thickness of the additional top layer. Moreover, for a fixed thickness, again the best response  is observed for the highest values of the nanorods density.\\
With reference to the study in the dynamic regime, the free wave propagation in such a periodic material is analysed. Considering waves travelling along the nanorords axis, it turns out that as the material density  decreases and the height of nanorods increases, more dense spectra are found. Moreover, various partial partial band gaps are detected for higher values of density, suggesting that lower values of the density enables better functioning as nanogenerators.
As the nanorods height increases, a higher number of partial band gaps, characterized by lower frequencies, are observed. The acoustic and optical branches tend to be shifted towards lower angular frequencies.
Finally, with reference to waves travelling in the plane normal to the nanorods axis, very dense spectra are here observed, where no partial band gaps, in the low frequency range, are detected irrespective of the considered density and height (few partial band gaps are found in the high frequency and characterized by small amplitudes).  It is confirmed that the spectrum density increases as the density decreases and the height increases. 
The acoustic branches of the Floquet-Bloch spectrum, characterizing the heterogeneous material, are compared against the dispersion curves obtained adopting
 the first order homogenization theory.  A good agreement is found, confirming the capability of the first order homogenization theory to satisfactorily reproducing the lowest (acoustic) branches of the
Bloch spectrum for a wide range of wavelengths.

\section*{Acknowledgement}
 The authors thankfully acknowledge financial support by Regione Puglia under the Future in
Research Program ”Development of next generation NEMS for energy harvesting”-NSUX1F1, and  National Group of Mathematical Physics
 (GNFM-INdAM).

\bibliographystyle{apalike}
\bibliography{BIBLIO}

\begin{thebibliography}{}

\bibitem[Aboudi et~al., 2001]{Aboudi2001}
Aboudi, J., Pindera, M., and Arnold, S. (2001).
\newblock Linear thermoelastic higher-order theory for periodic multiphase
  materials.
\newblock {\em Journal of Applied Mechanics}, 68(5):697--707.

\bibitem[Addessi et~al., 2013]{Addessi2013}
Addessi, D., De~Bellis, M.~L., and Sacco, E. (2013).
\newblock Micromechanical analysis of heterogeneous materials subjected to
  overall cosserat strains.
\newblock {\em Mechanics Research Communications}, 54:27 -- 34.

\bibitem[Addessi et~al., 2016]{Addessi2016}
Addessi, D., De~Bellis, M.~L., and Sacco, E. (2016).
\newblock A micromechanical approach for the cosserat modeling of composites.
\newblock {\em Meccanica}, 51(3):569--592.

\bibitem[Ahmed et~al., 2017]{Ahmed2017}
Ahmed, A., Saadatnia, Z., Hassan, I., Zi, Y., Xi, Y., He, X., Zu, J., and Wang,
  Z.~L. (2017).
\newblock Self-powered wireless sensor node enabled by a duck-shaped
  triboelectric nanogenerator for harvesting water wave energy.
\newblock {\em Advanced Energy Materials}, 7(7):1601705.

\bibitem[Allaire, 1992]{allaire1992homogenization}
Allaire, G. (1992).
\newblock Homogenization and two-scale convergence.
\newblock {\em SIAM Journal on Mathematical Analysis}, 23(6):1482--1518.

\bibitem[Almadhoun et~al., 2014]{almadhoun2014}
Almadhoun, M.~N., Hedhili, M.~N., Odeh, I.~N., Xavier, P., Bhansali, U.~S., and
  Alshareef, H.~N. (2014).
\newblock Influence of stacking morphology and edge nitrogen doping on the
  dielectric performance of graphene-polymer nanocomposites.
\newblock {\em Chemistry of Materials}, 26(9):2856--2861.

\bibitem[Andrianov et~al., 2008]{andrianov2008higher}
Andrianov, I.~V., Bolshakov, V.~I., Danishevs'~kyy, V.~V., and Weichert, D.
  (2008).
\newblock Higher order asymptotic homogenization and wave propagation in
  periodic composite materials.
\newblock In {\em Proceedings of the Royal Society of London A: Mathematical,
  Physical and Engineering Sciences}, volume 464, pages 1181--1201.

\bibitem[Askari et~al., 2019]{Askari2019}
Askari, H., Hashemi, E., Khajepour, A., Khamesee, M.~B., and Wang, Z.~L.
  (2019).
\newblock Tire condition monitoring and intelligent tires using nanogenerators
  based on piezoelectric, electromagnetic, and triboelectric effects.
\newblock {\em Advanced Materials Technologies}, 4(1):1800105.

\bibitem[Bacca et~al., 2013a]{Bacca2013a}
Bacca, M., Bigoni, D., Corso, F.~D., and Veber, D. (2013a).
\newblock Mindlin second-gradient elastic properties from dilute two-phase
  cauchy-elastic composites. part i: Closed form expression for the effective
  higher-order constitutive tensor.
\newblock {\em International Journal of Solids and Structures}, 50(24):4010 --
  4019.

\bibitem[Bacca et~al., 2013b]{Bacca2013b}
Bacca, M., Bigoni, D., Corso, F.~D., and Veber, D. (2013b).
\newblock Mindlin second-gradient elastic properties from dilute two-phase
  cauchy-elastic composites part ii: Higher-order constitutive properties and
  application cases.
\newblock {\em International Journal of Solids and Structures}, 50(24):4020 --
  4029.

\bibitem[Bacigalupo, 2014]{Bacigalupo2014}
Bacigalupo, A. (2014).
\newblock Second-order homogenization of periodic materials based on asymptotic
  approximation of the strain energy: formulation and validity limits.
\newblock {\em Meccanica}, 49(6):1407--1425.

\bibitem[Bacigalupo and Gambarotta, 2010]{BacigalupoGambarotta2010}
Bacigalupo, A. and Gambarotta, L. (2010).
\newblock Second-order computational homogenization of heterogeneous materials
  with periodic microstructure.
\newblock {\em ZAMM Journal of Applied Mathematics and Mechanics}, 90:796--811.

\bibitem[Bacigalupo and Gambarotta, 2012]{BACIGALUPO201216}
Bacigalupo, A. and Gambarotta, L. (2012).
\newblock Computational two-scale homogenization of periodic masonry:
  Characteristic lengths and dispersive waves.
\newblock {\em Computer Methods in Applied Mechanics and Engineering},
  213-216:16 -- 28.

\bibitem[Bacigalupo and Gambarotta, 2013]{Bacigalupo2013}
Bacigalupo, A. and Gambarotta, L. (2013).
\newblock A multi-scale strain-localization analysis of a layered strip with
  debonding interfaces.
\newblock {\em International Journal of Solids and Structures}, 50(13):2061 --
  2077.

\bibitem[Bacigalupo and Gambarotta, 2014]{BacigalupoGambarotta2014}
Bacigalupo, A. and Gambarotta, L. (2014).
\newblock Second-gradient homogenized model for wave propagation in
  heterogeneous periodic media.
\newblock {\em International Journal of Solids and Structures},
  51(5):1052--1065.

\bibitem[Bacigalupo et~al., 2016]{BacigalupoMorini2016}
Bacigalupo, A., Morini, L., and Piccolroaz, A. (2016).
\newblock Multiscale asymptotic homogenization analysis of thermo-diffusive
  composite materials.
\newblock {\em International Journal of Solids and Structures}, 85-86:15--33.

\bibitem[Bacigalupo et~al., 2017]{bacigalupo2017identification}
Bacigalupo, A., Paggi, M., Dal~Corso, F., and Bigoni, D. (2017).
\newblock Identification of higher-order continua equivalent to a cauchy
  elastic composite.
\newblock {\em Mechanics Research Communications},
  doi.org/10.1016/j.mechrescom.2017.07.002.

\bibitem[Bakhvalov and Panasenko, 1984]{Bakhvalov1984}
Bakhvalov, N. and Panasenko, G. (1984).
\newblock {\em Homogenization: Averaging Processes in Periodic Media}.
\newblock Kluwer Academic Publishers, Dordrecht-Boston-London.

\bibitem[Berger et~al., 2005]{BERGER200553}
Berger, H., Kari, S., Gabbert, U., Rodríguez-Ramos, R., Bravo-Castillero, J.,
  and Guinovart-Díaz, R. (2005).
\newblock A comprehensive numerical homogenisation technique for calculating
  effective coefficients of uniaxial piezoelectric fibre composites.
\newblock {\em Materials Science and Engineering: A}, 412(1):53 -- 60.
\newblock International Conference on Recent Advances in Composite Materials.

\bibitem[Bigoni and Drugan, 2007]{Bigoni2007}
Bigoni, D. and Drugan, W.~J. (2007).
\newblock Analytical derivation of {C}osserat moduli via homogenization of
  heterogeneous elastic materials.
\newblock {\em J Appl Mech}, 74:741--753.

\bibitem[Biswas and Poh, 2017]{biswas2017micromorphic}
Biswas, R. and Poh, L.~H. (2017).
\newblock A micromorphic computational homogenization framework for
  heterogeneous materials.
\newblock {\em Journal of the Mechanics and Physics of Solids}, 102:187--208.

\bibitem[Bloch, 1928]{Bloch1928}
Bloch, F. (1928).
\newblock Über die quantenmechanik der elektronen in kristallgittern.
\newblock {\em Z. Phys.}, 52:555--600.

\bibitem[Boutin, 1996]{Boutin1996}
Boutin, C. (1996).
\newblock Microstructural effects in elastic composites.
\newblock {\em International Journal of Solids and Structures}, 33:1023--1051.

\bibitem[Brillouin, 1960]{Brillouin1960}
Brillouin, L. (1960).
\newblock {\em Wave Propagation and Group Velocity}.
\newblock New York: Academic Press.

\bibitem[Briscoe and Dunn, 2015]{BRISCOE201515}
Briscoe, J. and Dunn, S. (2015).
\newblock Piezoelectric nanogenerators – a review of nanostructured
  piezoelectric energy harvesters.
\newblock {\em Nano Energy}, 14:15 -- 29.
\newblock Special issue on the 2nd International Conference on Nanogenerators
  and Piezotronics (NGPT 2014).

\bibitem[Choi et~al., 2017]{CHOI2017462}
Choi, M., Murillo, G., Hwang, S., Kim, J.~W., Jung, J.~H., Chen, C.-Y., and
  Lee, M. (2017).
\newblock Mechanical and electrical characterization of pvdf-zno hybrid
  structure for application to nanogenerator.
\newblock {\em Nano Energy}, 33:462 -- 468.

\bibitem[De~Bellis and Addessi, 2011]{DeBellis-Addessi11}
De~Bellis, M.~L. and Addessi, D. (2011).
\newblock A {C}osserat based multi--scale model for masonry structures.
\newblock {\em International Journal for Multiscale Computational Engineering},
  9(5):543--563.

\bibitem[De~Bellis and Bacigalupo, 2017]{de2017auxetic}
De~Bellis, M.~L. and Bacigalupo, A. (2017).
\newblock Auxetic behavior and acoustic properties of microstructured
  piezoelectric strain sensors.
\newblock {\em Smart Materials and Structures}, 26(8):085037.

\bibitem[Deraemaeker and Nasser, 2010]{DeraemaekerNasser2010}
Deraemaeker, A. and Nasser, H. (2010).
\newblock Numerical evaluation of the equivalent properties of macro fiber
  composite (mfc) transducers using periodic homogenization.
\newblock {\em International Journal of Solids and Structures},
  47(24):3272--3285.

\bibitem[Eftekhari, 2011]{eftekhari2011}
Eftekhari, A. (2011).
\newblock {\em Nanostructured conductive polymers}.
\newblock John Wiley \& Sons.

\bibitem[Fan et~al., 2016]{FAN2016}
Fan, F.~R., Tang, W., and Wang, Z.~L. (2016).
\newblock Flexible nanogenerators for energy harvesting and self-powered
  electronics.
\newblock {\em Advanced Materials}, 28(22):4283--4305.

\bibitem[Fantoni et~al., 2017]{fantoni2017multi}
Fantoni, F., Bacigalupo, A., and Paggi, M. (2017).
\newblock Multi-field asymptotic homogenization of thermo-piezoelectric
  materials with periodic microstructure.
\newblock {\em International Journal of Solids and Structures}, 120:31--56.

\bibitem[Fantoni et~al., 2018]{FANTONI2018319}
Fantoni, F., Bacigalupo, A., and Paggi, M. (2018).
\newblock Design of thermo-piezoelectric microstructured bending actuators via
  multi-field asymptotic homogenization.
\newblock {\em International Journal of Mechanical Sciences}, 146-147:319 --
  336.

\bibitem[Fish and Chen, 2001]{fish2001higher}
Fish, J. and Chen, W. (2001).
\newblock Higher-order homogenization of initial/boundary-value problem.
\newblock {\em Journal of engineering mechanics}, 127(12):1223--1230.

\bibitem[Floquet, 1883]{Floquet1883}
Floquet, G. (1883).
\newblock Sur les équations différentielles linéaires à coefficients
  périodiques.
\newblock {\em Annales de l'École Normale Supérieure}, 12:47--88.

\bibitem[Forest and Sab, 1998]{FOREST1998}
Forest, S. and Sab, K. (1998).
\newblock Cosserat overall modeling of heterogeneous materials.
\newblock {\em Mechanics Research Communications}, 25(4):449 -- 454.

\bibitem[Ga{\l}ka et~al., 1996]{galka1996some}
Ga{\l}ka, A., Telega, J.~J., and Wojnar, R. (1996).
\newblock Some computational aspects of homogenization of thermopiezoelectric
  composites.
\newblock {\em Comp. Assisted Mech. Eng. Sci}, 3(2):133--154.

\bibitem[Gambin and Kr\"oner, 1989]{GambinKroner1989}
Gambin, B. and Kr\"oner, E. (1989).
\newblock Higher order terms in the homogenized stress‐strain relation of
  periodic elastic media. physica status solidi (b).
\newblock {\em International Journal of Engineering Science}, 151(2):513--519.

\bibitem[Huang and Zhang, 2004]{huang2004}
Huang, C. and Zhang, Q. (2004).
\newblock Enhanced dielectric and electromechanical responses in high
  dielectric constant all-polymer percolative composites.
\newblock {\em Advanced Functional Materials}, 14(5):501--506.

\bibitem[H{\"u}tter, 2017]{hutter2017homogenization}
H{\"u}tter, G. (2017).
\newblock Homogenization of a cauchy continuum towards a micromorphic
  continuum.
\newblock {\em Journal of the Mechanics and Physics of Solids}, 99:394--408.

\bibitem[Jin et~al., 2016]{jin2016self}
Jin, L., Chen, J., Zhang, B., Deng, W., Zhang, L., Zhang, H., Huang, X., Zhu,
  M., Yang, W., and Wang, Z.~L. (2016).
\newblock Self-powered safety helmet based on hybridized nanogenerator for
  emergency.
\newblock {\em ACS nano}, 10(8):7874--7881.

\bibitem[Kaczmarczyk et~al., 2008]{kaczmarczyk2008scale}
Kaczmarczyk, {\L}., Pearce, C.~J., and Bi{\'c}ani{\'c}, N. (2008).
\newblock Scale transition and enforcement of rve boundary conditions in
  second-order computational homogenization.
\newblock {\em International Journal for Numerical Methods in Engineering},
  74(3):506--522.

\bibitem[Kanout\'e et~al., 2009]{Kanoute2009}
Kanout\'e, P., Boso, D., Chaboche, J., and Schrefler, B. (2009).
\newblock Multiscale methods for composites: a review.
\newblock {\em Archives of Computational Methods in Engineering}, 16(1):31--75.

\bibitem[Kouznetsova et~al., 2004]{KouznetsovaGeers2004}
Kouznetsova, V., Geers, M., and Brekelmans, W. (2004).
\newblock Multi-scale second-order computational homogenization of multi-phase
  materials: a nested finite element solution strategy.
\newblock {\em Computer Methods in Applied Mechanics and Engineering},
  193(48):5525--5550.

\bibitem[Lesi{\v{c}}ar et~al., 2014]{lesivcar2014second}
Lesi{\v{c}}ar, T., Tonkovi{\'c}, Z., and Sori{\'c}, J. (2014).
\newblock A second-order two-scale homogenization procedure using $ $ c-1
  macrolevel discretization.
\newblock {\em Computational mechanics}, 54(2):425--441.

\bibitem[Li et~al., 2017]{LI2017}
Li, M., Porter, A.~L., and Wang, Z.~L. (2017).
\newblock Evolutionary trend analysis of nanogenerator research based on a
  novel perspective of phased bibliographic coupling.
\newblock {\em Nano Energy}, 34:93 -- 102.

\bibitem[Li et~al., 2011]{li2011micro}
Li, X., Zhang, J., and Zhang, X. (2011).
\newblock Micro-macro homogenization of gradient-enhanced cosserat media.
\newblock {\em European Journal of Mechanics-A/Solids}, 30(3):362--372.

\bibitem[Liu et~al., 2018]{liu2018shape}
Liu, R., Kuang, X., Deng, J., Wang, Y.-C., Wang, A.~C., Ding, W., Lai, Y.-C.,
  Chen, J., Wang, P., Lin, Z., et~al. (2018).
\newblock Shape memory polymers for body motion energy harvesting and
  self-powered mechanosensing.
\newblock {\em Advanced Materials}, 30(8):1705195.

\bibitem[McCarthy et~al., 2016]{MCCARTHY2016355}
McCarthy, J., Watkins, S., Deivasigamani, A., and John, S. (2016).
\newblock Fluttering energy harvesters in the wind: A review.
\newblock {\em Journal of Sound and Vibration}, 361:355 -- 377.

\bibitem[Mindlin, 1974]{Mindlin1974}
Mindlin, R. (1974).
\newblock Equations of high frequency vibrations of thermopiezoelectric crystal
  plates.
\newblock {\em International Journal of Solids and Structures}, 10(6):625--637.

\bibitem[M{\"u}hlich et~al., 2012]{muhlich2012estimation}
M{\"u}hlich, U., Zybell, L., and Kuna, M. (2012).
\newblock Estimation of material properties for linear elastic strain gradient
  effective media.
\newblock {\em European Journal of Mechanics-A/Solids}, 31(1):117--130.

\bibitem[Opoku et~al., 2015]{OPOKU2015858}
Opoku, C., Dahiya, A.~S., Oshman, C., Cayrel, F., Poulin-Vittrant, G., Alquier,
  D., and Camara, N. (2015).
\newblock Fabrication of zno nanowire based piezoelectric generators and
  related structures.
\newblock {\em Physics Procedia}, 70:858 -- 862.
\newblock Proceedings of the 2015 ICU International Congress on Ultrasonics,
  Metz, France.

\bibitem[Peerlings and Fleck, 2004]{PeerlingsFleck2004}
Peerlings, R. and Fleck, N. (2004).
\newblock Computational evaluation of strain gradient elasticity constants.
\newblock {\em International Journal for Multiscale Computational Engineering},
  2(4).

\bibitem[Pettermann and Suresh, 2000]{PETTERMANN20005447}
Pettermann, H.~E. and Suresh, S. (2000).
\newblock A comprehensive unit cell model: a study of coupled effects in
  piezoelectric 1–3 composites.
\newblock {\em International Journal of Solids and Structures}, 37(39):5447 --
  5464.

\bibitem[Reccia et~al., 2018]{RECCIA201839}
Reccia, E., Bellis, M. L.~D., Trovalusci, P., and Masiani, R. (2018).
\newblock Sensitivity to material contrast in homogenization of random particle
  composites as micropolar continua.
\newblock {\em Composites Part B: Engineering}, 136:39 -- 45.

\bibitem[Saadatnia et~al., 2017]{saadatnia2017modeling}
Saadatnia, Z., Asadi, E., Askari, H., Zu, J., and Esmailzadeh, E. (2017).
\newblock Modeling and performance analysis of duck-shaped triboelectric and
  electromagnetic generators for water wave energy harvesting.
\newblock {\em International Journal of Energy Research}, 41(14):2392--2404.

\bibitem[Salvadori et~al., 2014]{SALVADORI2014114}
Salvadori, A., Bosco, E., and Grazioli, D. (2014).
\newblock A computational homogenization approach for li-ion battery cells:
  Part 1 – formulation.
\newblock {\em Journal of the Mechanics and Physics of Solids}, 65:114 -- 137.

\bibitem[Smyshlyaev and Cherednichenko, 2000]{Smyshlyaev2000}
Smyshlyaev, V. and Cherednichenko, K. (2000).
\newblock On rigorous derivation of strain gradient effects in the overall
  behaviour of periodic heterogeneous media.
\newblock {\em Journal of the Mechanics and Physics of Solids},
  48(6):1325--1357.

\bibitem[Stassi et~al., 2015]{STASSI2015}
Stassi, S., Cauda, V., Ottone, C., Chiodoni, A., Pirri, C.~F., and Canavese, G.
  (2015).
\newblock Flexible piezoelectric energy nanogenerator based on zno nanotubes
  hosted in a polycarbonate membrane.
\newblock {\em Nano Energy}, 13:474 -- 481.

\bibitem[Tran et~al., 2012]{Tran2012}
Tran, T., Monchiet, V., and Bonnet, G. (2012).
\newblock A micromechanics-based approach for the derivation of constitutive
  elastic coefficients of strain-gradient media.
\newblock {\em International Journal of Solids and Structures}, 49(5):783--792.

\bibitem[Trovalusci et~al., 2017]{TROVALUSCI2017164}
Trovalusci, P., Bellis, M. L.~D., and Masiani, R. (2017).
\newblock A multiscale description of particle composites: From lattice
  microstructures to micropolar continua.
\newblock {\em Composites Part B: Engineering}, 128:164 -- 173.

\bibitem[Trovalusci et~al., 2015]{TROVALUSCI2015396}
Trovalusci, P., Ostoja-Starzewski, M., Bellis, M. L.~D., and Murrali, A.
  (2015).
\newblock Scale-dependent homogenization of random composites as micropolar
  continua.
\newblock {\em European Journal of Mechanics - A/Solids}, 49:396 -- 407.

\bibitem[Wang et~al., 2015]{wang2015}
Wang, D., You, F., and Hu, G.-H. (2015).
\newblock Graphene/polymer nanocomposites with high dielectric performance:
  Interface engineering.
\newblock In {\em Graphene-Based Polymer Nanocomposites in Electronics}, pages
  49--65. Springer.

\bibitem[Wang et~al., 2005]{wang2005}
Wang, J.-W., Shen, Q.-D., Bao, H.-M., Yang, C.-Z., and Zhang, Q. (2005).
\newblock Microstructure and dielectric properties of {P}({VDF}-{T}r{FE}-{CFE})
  with partially grafted copper phthalocyanine oligomer.
\newblock {\em Macromolecules}, 38(6):2247--2252.

\bibitem[Wang, 2004]{Wang_2004}
Wang, Z.~L. (2004).
\newblock Zinc oxide nanostructures: growth, properties and applications.
\newblock {\em Journal of Physics: Condensed Matter}, 16(25):R829--R858.

\bibitem[Wang et~al., 2017]{wang2017toward}
Wang, Z.~L., Jiang, T., and Xu, L. (2017).
\newblock Toward the blue energy dream by triboelectric nanogenerator networks.
\newblock {\em Nano Energy}, 39:9--23.

\bibitem[Wang and Song, 2006]{wang2006piezoelectric}
Wang, Z.~L. and Song, J. (2006).
\newblock Piezoelectric nanogenerators based on zinc oxide nanowire arrays.
\newblock {\em Science}, 312(5771):242--246.

\bibitem[Yang et~al., 2017]{Dechao2017}
Yang, D., Qiu, Y., Jiang, Q., Guo, Z., Song, W., Xu, J., Zong, Y., Feng, Q.,
  and Sun, X. (2017).
\newblock Patterned growth of zno nanowires on flexible substrates for enhanced
  performance of flexible piezoelectric nanogenerators.
\newblock {\em Applied Physics Letters}, 110(6):063901.

\bibitem[Yang, 2004]{yang2004introduction}
Yang, J. (2004).
\newblock {\em An introduction to the theory of piezoelectricity}, volume~9.
\newblock Springer Science \& Business Media.

\bibitem[Yi et~al., 2005]{yi2005zno}
Yi, G.-C., Wang, C., and Park, W.~I. (2005).
\newblock Zno nanorods: synthesis, characterization and applications.
\newblock {\em Semiconductor Science and Technology}, 20(4):S22.

\bibitem[Zah and Miehe, 2013]{ZAH2013487}
Zah, D. and Miehe, C. (2013).
\newblock Computational homogenization in dissipative electro-mechanics of
  functional materials.
\newblock {\em Computer Methods in Applied Mechanics and Engineering}, 267:487
  -- 510.

\bibitem[Zhang et~al., 2007]{ZhangZhang2007}
Zhang, H., Zhang, S., Bi, J., and Schrefler, B. (2007).
\newblock Thermo‐mechanical analysis of periodic multiphase materials by a
  multiscale asymptotic homogenization approach.
\newblock {\em International Journal for Numerical Methods in Engineering},
  69(1):87--113.

\bibitem[Zhang et~al., 2016]{zhang2016wearable}
Zhang, N., Chen, J., Huang, Y., Guo, W., Yang, J., Du, J., Fan, X., and Tao, C.
  (2016).
\newblock A wearable all-solid photovoltaic textile.
\newblock {\em Advanced Materials}, 28(2):263--269.

\end{thebibliography}

\clearpage

\section*{Appendix A: Solutions of the hierarchical  differential problems }
The following hierarchical differential problems, expressed in terms of the sensitivities $u_k^{(j)}$, $\phi^{(j)}$ of both microscopic displacement  $u_k$ and potential $\phi$   fields, can be obtained from the asymptotic expansion of the field equations (\ref{eq.12})  by collecting the terms at the same order of $\varepsilon$.\\
More specifically, the recursive problem the at the order $\varepsilon^{-2}$ reads
\begin{align}
\begin{split}
&  \left(C_{ijkl}^{m} u^{(0)}_{k,l}\right)_{,j} + \left(e_{ijk}^{m}  \phi^{(0)}_{,k} \right)_{,j} = f_i^{(0)} (\textbf{x},t), \label{eq.A1}\\
&\left(e_{kli}^{m} u^{(0)}_{k,l}\right)_{,i} - \left(\beta_{il}^{m}  \phi^{(0)}_{,l} \right)_{,i} = g^{(0)} (\textbf{x},t).
\end{split}
\end{align}
The solvability condition in the class of the $Q$-periodic functions implies that  both source terms vanish, i.e. $f_i^{(0)} (\textbf{x},t)=0$
and $g^{(0)} (\textbf{x},t)=0$. It follows that the solution of the differential problem is independent on the fast variable $\boldsymbol{\xi}$
  and it takes the form reported in  (\ref{Phi0}).\\
  Analogously, by exploiting the solution of the differential problem (\ref{eq.A1}), the recursive problem the at the order $\varepsilon^{-1}$ is
\begin{align}
\begin{split}
& \left(C_{ijkl}^{m} u_{k,l}^{(1)}\right)_{,j}+ C_{ijkl,j}^{m} \frac{\partial U_k}{\partial x_l}  +\left(e_{ijk}^{m} \phi_{,k}^{(1)} \right)_{,j}+e_{ijk,j}^{m} \frac{\partial \Phi}{\partial x_k} = f_i^{(1)} (\textbf{x},t), \label{eq.A2}\\
& \left(e_{kli}^{m} u^{(1)}_{k,l} \right)_{,i}+ e_{kli,i}^{m}   \frac{\partial U_k}{\partial x_l}- \left(\beta_{il}^{m} \phi^{(1)}_{,l}\right)_{,i}-\beta_{il,i}^{m} \frac{\partial \Phi}{\partial x_l}= g^{(1)} (\textbf{x},t) .
\end{split}
\end{align}
Here, the solvability condition in the class of the $Q$-periodic functions implies that
\begin{align}
\begin{split}
& f_i^{(1)} (\textbf{x},t) = \langle C_{ijkl,j}^{m}\rangle \frac{\partial U_k}{\partial x_l}+ \langle e_{ijk,j}^{m}\rangle \frac{\partial \Phi}{\partial x_k}=0, \label{eq.A3}\\
& g^{(1)} (\textbf{x},t)= \langle e_{kli,i}^{m}\rangle \frac{\partial U_k}{\partial x_k} -\langle \beta_{il,i}^{m}\rangle \frac{\partial \Phi}{\partial x_l}=0 ,
\end{split}
\end{align}
since the components of the constitutive tensors at the microscopic scale are $Q$-periodic. In this case, the solution of the differential problem takes the form reported in  (\ref{Phi1}).\\
Similarly,  the recursive problem the at the order $\varepsilon^{0}$ is obtained by exploiting the solution of both the differential problems (\ref{eq.A1}) and (\ref{eq.A2}), and it reads
\begin{align}
\begin{split}
& \left(C_{ijkl}^{m} u_{k,l}^{(2)}\right)_{,j}+
\left[  \left(C_{ijkq_2}^{m} N_{kpq_1}^{(1)}\right)_{,j} + C_{iq_1 p q_2}^{m} + C_{iq_2kl}^{m} N^{(1)}_{kpq_1,l}+\left(e_{ijq_2}^{m} \widetilde{W}_{pq_1}^{(1)} \right)_{,j} + e_{iq_2k}^{m}\widetilde{W}_{pq_1,k}^{(1)} \right]  \frac{\partial^2 U_p}{\partial x_{q_1}\partial x_{q_2}} +\\ &+ \left(e_{ijk}^{m} \phi_{,k}^{(2)} \right)_{,j}+ \left[ \left(C_{ijkq_2}^{m} \widetilde{N}_{kq_1}^{(1)}\right)_{,j} + C_{iq_2kl}^{m} \widetilde{N}_{kq_1,l}^{(1)}+\left(e_{ijq_2}^{m} {W}_{q_1}^{(1)} \right)_{,j} +e_{iq_1q_2}^{m}+e_{iq_2k}^{m} {W}_{q_1,k}^{(1)}\right] \frac{\partial^2 \Phi}{\partial x_{q_1}\partial x_{q_2}}+\\
&- \rho^m \ddot{U}_i = f_i^{(2)} (\textbf{x},t), \label{eq.A4}
\end{split}
\end{align}
\begin{align}
\begin{split}
&\left(e_{kli}^{m} u_{k,l}^{(2)}\right)_{,i}+
\left[  \left(e_{kq_2i}^{m} N_{kpq_1}^{(1)}\right)_{,i} + e_{kl q_2}^{m} N^{(1)}_{kpq_1,l}+ e_{p q_2 q_1}^{m}-\left(\beta_{iq_2}^{m} \widetilde{W}_{pq_1}^{(1)} \right)_{,i}-\beta_{q_2l}^{m}  \widetilde{W}_{pq_1,l}^{(1)} \right]  \frac{\partial^2 U_p}{\partial x_{q_1}\partial x_{q_2}} +\\ &- \left(\beta_{il}^{m} \phi_{,l}^{(2)} \right)_{,i}+ \left[ \left(e_{kq_2i}^{m} \widetilde{N}_{kq_1}^{(1)}\right)_{,i} + e_{klq_2}^{m} \widetilde{N}_{kq_1,l}^{(1)}-\left(\beta_{iq_2}^{m} {W}_{q_1}^{(1)} \right)_{,i} -\beta_{q_1q_2}^{m}-\beta_{q_2l}^{m} {W}_{q_1,l}^{(1)}\right] \frac{\partial^2 \Phi}{\partial x_{q_1}\partial x_{q_2}}=\\
&= g^{(2)} (\textbf{x},t) . \nonumber
\end{split}
\end{align}
Finally, the solvability condition in the class of the $Q$-periodic functions implies that
\begin{align}
\begin{split}
& f_i^{(2)} (\textbf{x},t) = \langle C_{iq_1pq_2}^{m} + C_{iq_2kl}^{m} N^{(1)}_{kpq_1,l}+ e^m_{iq_2k} \widetilde{W}^{(1)}_{,k}\rangle \frac{\partial^2 U_p}{\partial x_{q_1} \partial x_{q_2}}+ \\
& +\langle C_{iq_2kl}^{m} \widetilde{N}^{(1)}_{kq_1,l}+ e_{iq_1q_2}^{m}+e^m_{iq_2k} {W}^{(1)}_{q_1,k}\rangle \frac{\partial^2 \Phi}{\partial x_{q_1} \partial x_{q_2}}-\langle \rho^m \rangle \ddot{U}_i, \label{eq.A5}\\
& g^{(2)} (\textbf{x},t)= \langle e_{klq_2}^{m} N^{(1)}_{kpq_1,l}+e_{pq_2q_1}^{m}-\beta_{q_2l} \widetilde{W}^{(1)}_{pq_1,l} \rangle \frac{\partial^2 U_p}{\partial x_{q_1} \partial x_{q_2}}+\\
&+\langle e_{klq_2q}^{m}\widetilde{N}^{(1)}_{kq_1,l}- \beta_{q_1q_2}^{m}-\beta_{q_2l}^{m} {W}^{(1)}_{q_1,l}\rangle \frac{\partial^2 \Phi}{\partial x_{q_1} \partial x_{q_2}} ,
\end{split}
\end{align}
The solution of the differential problem takes the form reported in (\ref{Phi2}).

\section*{Appendix B: Average field equation of infinite order }
By plugging the down-scaling relations (\ref{polynExp}) in the field equations (\ref{eq:1}), i.e. by reassembling the recursive differential problems at different $\varepsilon$ orders (\ref{eq.A1}), (\ref{eq.A2}) and (\ref{eq.A4}), the following average field equation of infinite order are obtained as
\begin{align}
\begin{split}
&n^{(2,0)}_{ipq_1q_2}  \frac{\partial^2 U_p}{\partial x_{q_1} \partial x_{q_2}}+\widetilde{n}^{(2,0)}_{iq_1q_2}  \frac{\partial^2 \Phi}{\partial x_{q_1} \partial x_{q_2}} -n_{ip}^{(2,0)} \ddot{U_p} +\mathcal{O}\left( {\bf{\varepsilon}} \right)+b_i (\textbf{x})= 0, \label{eq.B1}\\
& \widetilde{w}^{(2,0)}_{pq_1q_2}  \frac{\partial^2 U_p}{\partial x_{q_1} \partial x_{q_2}}- w^{(2,0)}_{q_1q_2}  \frac{\partial^2 \Phi}{\partial x_{q_1} \partial x_{q_2}}  +\mathcal{O}\left( {\bf{\varepsilon}} \right)- \rho_e (\textbf{x})= 0,
\end{split}
\end{align}
where the components of the constant global constitutive tensors, involved in the governing equations, take the form
\begin{align}
\begin{split}
&n^{(2,0)}_{ipq_1q_2} = \frac{1}{2} \left \langle C_{iq_1pq_2}^{m} + C_{iq_2kl}^{m} N^{(1)}_{kpq_1,l}+ e^m_{iq_2k} \widetilde{W}^{(1)}_{,k}+
C_{iq_2pq_1}^{m} + C_{iq_1kl}^{m} N^{(1)}_{kpq_2,l}+ e^m_{iq_1k} \widetilde{W}^{(1)}_{,k} \right \rangle, \label{eq.B2}\\
&\widetilde{n}^{(2,0)}_{iq_1q_2}= \frac{1}{2} \left \langle C_{iq_2kl}^{m} \widetilde{N}^{(1)}_{kq_1,l}+ e_{iq_1q_2}^{m}+e^m_{iq_2k} {W}^{(1)}_{q_1,k} +
 C_{iq_1kl}^{m} \widetilde{N}^{(1)}_{kq_2,l}+ e_{iq_2q_1}^{m}+e^m_{iq_1k} {W}^{(1)}_{q_2,k}\right \rangle,\\
& n^{(2,0)}_{ip} = \left \langle \rho^m \right \rangle
 \delta_{ip},\\
& \widetilde{w}^{(2,0)}_{pq_1q_2}=\frac{1}{2} \left \langle e_{klq_2}^{m} N^{(1)}_{kpq_1,l}+e_{pq_2q_1}^{m}-\beta_{q_2l} \widetilde{W}^{(1)}_{pq_1,l} +
e_{klq_1}^{m} N^{(1)}_{kpq_2,l}+e_{pq_1q_2}^{m}-\beta_{q_1l} \widetilde{W}^{(1)}_{pq_2,l}
\right \rangle,\\
& w^{(2,0)}_{q_1q_2} =\frac{1}{2} \left \langle \beta_{q_1q_2}^{m}+\beta_{q_2l}^{m} {W}^{(1)}_{q_1,l}-e_{klq_2q}^{m}\widetilde{N}^{(1)}_{kq_1,l} +
\beta_{q_2q_1}^{m}+\beta_{q_1l}^{m} {W}^{(1)}_{q_2,l}-e_{klq_1q}^{m}\widetilde{N}^{(1)}_{kq_2,l}
 \right \rangle.
\end{split}
\end{align}
In order to obtain a formal solution of equations (\ref{eq.B1}), the macroscopic variables $U_k$ and $\Phi$ are asymptotically expanded as follows
\begin{align}
\begin{split}
&{U_k}\left( {{\bf{x}}, t} \right) =\sum\limits_{j = 0}^{ + \infty } {{\varepsilon ^j}} {U_k^{(j)}}\left( {{\bf{x}}, t} \right),\\
&{\Phi}\left( {{\bf{x}}, t} \right) =\sum\limits_{j = 0}^{ + \infty } {{\varepsilon ^j}} {\Phi^{(j)}}\left( {{\bf{x}}, t} \right).
\end{split}
\label{eq.B3}
\end{align}
By plugging (\ref{eq.B3}) into the (\ref{eq.B1}), the asymptotic expansion of the average field equation of infinite order takes the form
\begin{align}
\begin{split}
&n^{(2,0)}_{ipq_1q_2}  \left(\frac{\partial^2 U_p^{(0)}}{\partial x_{q_1} \partial x_{q_2}}+ \varepsilon \frac{\partial^2 U_p^{(1)}}{\partial x_{q_1} \partial x_{q_2}}+ \varepsilon^2 \frac{\partial^2 U_p^{(2)}}{\partial x_{q_1} \partial x_{q_2}}+... \right)+\widetilde{n}^{(2,0)}_{iq_1q_2}  \left(\frac{\partial^2 \Phi^{(0)}}{\partial x_{q_1} \partial x_{q_2}}+ \varepsilon \frac{\partial^2 \Phi^{(1)}}{\partial x_{q_1} \partial x_{q_2}}+ \right.\\
& \left.+ \varepsilon^2 \frac{\partial^2 \Phi^{(2)}}{\partial x_{q_1} \partial x_{q_2}}+...  \right) -n_{ip}^{(2,0)} \left( \ddot{U_p}^{(0)}+\varepsilon \ddot{U_p}^{(1)}+\varepsilon^2 \ddot{U_p}^{(2)}  \right) + ...+b_i (\textbf{x})= 0, \label{eq.B4}\\
& \widetilde{w}^{(2,0)}_{pq_1q_2}  \left(\frac{\partial^2 U_p^{(0)}}{\partial x_{q_1} \partial x_{q_2}}+ \varepsilon \frac{\partial^2 U_p^{(1)}}{\partial x_{q_1} \partial x_{q_2}}+ \varepsilon^2 \frac{\partial^2 U_p^{(2)}}{\partial x_{q_1} \partial x_{q_2}}+... \right)- w^{(2,0)}_{q_1q_2}  \left(\frac{\partial^2 \Phi^{(0)}}{\partial x_{q_1} \partial x_{q_2}}+ \varepsilon \frac{\partial^2 \Phi^{(1)}}{\partial x_{q_1} \partial x_{q_2}}+ \right.\\
& \left.+ \varepsilon^2 \frac{\partial^2 \Phi^{(2)}}{\partial x_{q_1} \partial x_{q_2}}+...  \right)  +...- \rho_e (\textbf{x})= 0.
\end{split}
\end{align}
In particular, by collecting the terms in (\ref{eq.B4}) at the same order of $\varepsilon$, an infinite set of macroscopic hierarchical differential problems, expressed in terms of the sensitivities $U_p^{(j)}$, $\Phi^{(j)}$ of both macroscopic displacement  $U_p$ and potential $\Phi$   fields can be determined. 
Namely, the recursive problem at the macroscopic scale of order $\varepsilon^0$ reads
\begin{align}
\begin{split}
&n^{(2,0)}_{ipq_1q_2}  \frac{\partial^2 U_p^{(0)}}{\partial x_{q_1} \partial x_{q_2}}+\widetilde{n}^{(2,0)}_{iq_1q_2}  \frac{\partial^2 \Phi^{(0)}}{\partial x_{q_1} \partial x_{q_2}} -n_{ip}^{(2,0)} \ddot{U_p}^{(0)}+b_i (\textbf{x})= 0, \label{eq.B5}\\
& \widetilde{w}^{(2,0)}_{pq_1q_2} \frac{\partial^2 U_p^{(0)}}{\partial x_{q_1} \partial x_{q_2}}- w^{(2,0)}_{q_1q_2}  \frac{\partial^2 \Phi^{(0)}}{\partial x_{q_1} \partial x_{q_2}}- \rho_e (\textbf{x})= 0,
\end{split}
\end{align}
while the generic recursive problem at the macroscopic scale of order $\varepsilon^m$ with $m \in \mathbb{Z}$, $m\ge 1$, is found as
\begin{align}
\begin{split}
&n^{(2,0)}_{ipq_1q_2}  \frac{\partial^2 U_p^{(m)}}{\partial x_{q_1} \partial x_{q_2}}+\widetilde{n}^{(2,0)}_{iq_1q_2}  \frac{\partial^2 \Phi^{(m)}}{\partial x_{q_1} \partial x_{q_2}} -n_{ip}^{(2,0)} \ddot{U_p}^{(m)}+s_i^{(m)}(\textbf{x},t)  = 0, \label{eq.B6}\\
& \widetilde{w}^{(2,0)}_{pq_1q_2} \frac{\partial^2 U_p^{(m)}}{\partial x_{q_1} \partial x_{q_2}}- w^{(2,0)}_{q_1q_2}  \frac{\partial^2 \Phi^{(m)}}{\partial x_{q_1} \partial x_{q_2}}-v^{(m)}(\textbf{x},t) = 0,
\end{split}
\end{align}
where $s_i^{(m)}$ and $v^{(m)}$ are known  $\mathfrak{L}$-periodic fields, playing the role of source terms. Such fields depend both on higher order constant tensors, that appear in the terms at orders equal or higher than  $\varepsilon^m$ of the average field equation of infinite order  (\ref{eq.B1}), and on sensitivities $U_k^{(j)}$ and $\Phi^{(j)}$ descending from macroscopic hierarchical differential problems  of order lower than $\varepsilon^m$.\\
Due to the $\mathfrak{L}$-periodicity of the source terms, the solution of equations (\ref{eq.B5}) and (\ref{eq.B6}) are in turn $\mathfrak{L}$-periodic. More specifically, the uniqueness of such solution is guaranteed by the fulfilment of the following normalization conditions $1/|\mathfrak{L}| \int_{\mathfrak{L}} U_p^{(m)} d \textbf{x}=0$ and $1/|\mathfrak{L}| \int_{\mathfrak{L}} \Phi^{(m)} d \textbf{x}=0$, where $|\mathfrak{L}|=\eta L^3$.\\
It is worth noting that the formal structure of equation (\ref{eq.B6}) is the same as the one of (\ref{eq.B5}), i.e. the governing equation of a first order piezoelectric continuum \citep{Mindlin1974}. Nevertheless, while the source terms in (\ref{eq.B5}) are body forces and free charge densities acting on the heterogeneous medium, the terms in (\ref{eq.B6})
 are auxiliary source terms that take into account nonlocal effects occurring in equation (\ref{eq.B1}), that asymptotically approximate the governing equations of the heterogeneous piezoelectric material, i.e. equation (\ref{eq:1}).

\end{document}